\documentclass{article}
\usepackage{graphicx} 
\usepackage{xcolor}
\usepackage{amsthm}
\usepackage{amsmath}
\usepackage{amsfonts}
\usepackage{dsfont}
\usepackage{natbib}
\usepackage{pdflscape}
\usepackage{enumitem}
\usepackage{comment}
\theoremstyle{definition}
\usepackage{hyperref}
\usepackage[colorinlistoftodos]{todonotes} 
\usepackage{amssymb} 

\newtheorem{assumption}{Assumption}

\newtheorem{theorem'}[ccounter]{Theorem}

\newtheorem{lemma'}[ccounter]{Lemma}
\newtheorem{corollary'}[ccounter]{Corollary}
\newtheorem{conjecture'}[ccounter]{Conjecture}
\newtheorem{proposition'}[ccounter]{Proposition}
\newtheorem{definition'}[ccounter]{Definition}
\newtheorem{discussion'}[ccounter]{Internal Discussion}

\newtheorem{remark'}[ccounter]{Remark}
\newtheorem{assumption'}[cccounter]{Assumption}

\usepackage{algorithm}  
\usepackage{algpseudocode}
\renewcommand{\algorithmiccomment}[1]{\bgroup\hfill//~#1\egroup}

\usepackage{float}       

\newcommand{\arate}{\tau_\ar}

\newcommand{\A}{\mathcal{A}}

\newcommand{\arhj}{\widehat{\tau}^{\textnormal{hj}}_n}

\newcommand{\arhjk}{\widehat{\tau}^{\textnormal{hj}}_{n,k}}

\newcommand{\arhjs}{\widehat{\tau}^{\textnormal{hj}*}}
\newcommand{\ha}{\textnormal{hj}}

\newcommand{\T}{\mathcal{T}}

\newcommand{\obs}{\textnormal{obs}}
\newcommand{\nmc}{n_{mc}}
\newcommand{\Prob}{\textnormal{P}}
\newcommand{\E}{\textnormal{E}}

\newcommand{\ar}{\mathcal{A}} 
\newcommand{\N}{\textnormal{N}}
\newcommand{\V}{\textrm{V}}

\newcommand{\dimA}{\widehat{\tau}^a_n}

\newcommand\AVar{\operatorname{V}}

\newcommand\VA{\operatorname{V}_n(A)}

\newcommand\VAn{\operatorname{V}_n(A^n)}

\renewcommand\P{\mathbb{P}}

\newcommand{\nAn}{n_{An}}
\newcommand{\ytA}{\overline{y}_A(1)}

\title{Randomization Inference  For the Always-Reporter Average Treatment Effect}
\date{\today}
\author{
Haoge Chang\thanks{Department of Economics, Columbia University, \href{mailto:hc3516@columbia.edu}{\nolinkurl{hc3516@columbia.edu}}}
\and
Zeyang Yu\thanks{Department of Politics, Princeton University, \href{mailto:arthurzeyangyu@princeton.edu}{\nolinkurl{arthurzeyangyu@princeton.edu}}}
}
\begin{document}

\maketitle
\begin{abstract}
    This article studies randomization inference for treatment effects in randomized controlled trials with attrition, where outcomes are observed for only a subset of units. We assume monotonicity in reporting behavior as in \cite{lee2009training} and focus on the average treatment effect for always-reporters (AR-ATE), defined as units whose outcomes are observed under both treatment and control.
Because always-reporter status is only partially revealed by observed assignment and response patterns, we propose a worst-case randomization test that maximizes the randomization p-value over all always-reporter configurations consistent with the data, with an optional pretest to prune implausible configurations. Using studentized Hajek- and chi-square–type statistics, we show the resulting procedure is finite-sample valid for the sharp null and asymptotically valid for the weak null. We also discuss computational implementations for discrete outcomes and integer-programming-based bounds for continuous outcomes.
\end{abstract}
\section{Introduction}

Sample attrition, in which some units’ outcomes are unobserved after randomization, is common in field experiments and can introduce selection bias if it is systematically related to potential outcomes. A popular framework for addressing the attrition problem, since \cite{lee2009training}, imposes a monotonicity assumption under which treatment affects reporting behavior in only one direction.\footnote{See also \cite{zhang2003estimation}.} Under this assumption, the observed outcome distributions imply bounds on the average treatment effect of always-reporters (AR-ATE hereafter) —units who would report outcomes regardless of treatment assignment.\footnote{Other common approaches to addressing the sample attrition problem include worst-case bounds \citep{horowitz2000nonparametric} and inverse-probability weighting under a conditional ignorability assumption \citep{robins1994estimation}.}

Existing methods for conducting inference for AR-ATE under monotonicity rely on asymptotic approximations \cite{lee2009training, imbens2004confidence, stoye2009more}. This paper proposes an alternative randomization-based testing procedure for inference on the AR-ATE. The primary advantage of our approach is that it delivers strong finite-sample guarantees under the sharp null hypothesis (e.g., no treatment effects for all always-reporters), while maintaining asymptotic validity when treatment effects may be heterogeneous. Specifically, our procedure is finite-sample valid for testing the sharp-null hypothesis and remains asymptotically valid under the weak null hypothesis that the average treatment effect for always-reporters is zero.

The principle that randomization inference based on properly chosen (e.g. studentizied) statistics delivers the dual guarantees as described above has long been known (see the related literature section below). This paper applies this principle to randomization inference in the presence of sample attrition. However, the structure of our problem differs substantially from previously studied settings. In particular, the target subpopulation (always-reporters) is unobservable, and the outcome distributions of always-reporters cannot be recovered exactly even under the sharp-null hypothesis.

To address this challenge, we adopt a worst-case p-value approach and consider a worst-case randomization test that maximizes the randomization p-value over all always-reporter configurations consistent with the observed data. Importantly, our model implies two balance conditions that can be tested under the sharp-null hypothesis: (i) balance of outcomes, as in standard randomization-inference settings, and (ii) balance in the number of always-reporters between the treatment and control groups. The second balance condition can be used to construct a two-step testing procedure or combined with the first condition to yield a chi-square–type test statistic. This stands in contrast to the existing literature, where the sharp null hypothesis typically implies only a balance condition on the outcomes. The second balance follows from the observation that, while always-reporters cannot be individually identified, random assignment implies that their expected counts are equal across the treatment and control groups.

The worst-case p-value approach delivers finite-sample valid inference under the sharp null hypothesis via a simple argument. Establishing asymptotic validity under mild conditions, however, requires more refined analysis than in the existing literature. In particular, we show that our procedure is uniformly asymptotically valid over a parameter region in which the differential attrition rate between treatment and control groups may be negligible. This empirically relevant region is generally ruled out by the analysis of existing procedures \cite{lee2009training}.\footnote{A notable exception is \cite{semenova2025generalized}, which proposes a pretesting-based approach.} Incorporating this region requires a refined conditional analysis, as it may induce nonstandard asymptotic distributions for the component of the statistic that tests balance in the number of always-reporters, depending on the particular sequence of parameters under consideration.

In addition to the statistical results, we provide computationally feasible algorithms for implementing the proposed statistical procedure. For outcomes with finite support (e.g., binary or categorical), we exploit the symmetry of randomization distribution to compute p-values via exhaustive enumeration. For continuous outcomes, we reformulate the p-value computation as a collection of smaller optimization problems, each solved using integer linear programming.

The remainder of the paper is organized as follows. Section~\ref{section:related_literature} discusses related literature, Section~\ref{section:setup} describes the setup and standard assumptions, Section~\ref{section:statistical_procedure} introduces our statistical procedures, Section~\ref{section:statistical_guarantees} states additional assumptions and establishes the statistical guarantees, Section~\ref{section:implementation} discusses computational implementation and Section\ref{section:simulation} presents simulation results.

\subsection{Related Literature}\label{section:related_literature}

Asymptotic inference methods for (trimming) bounds have been studied in \cite{imbens2004confidence,lee2009training,stoye2009more,samii2023generalizing,semenova2025generalized}. Randomization inference and design-based inference with sample attrition or missing outcomes has been examined in \cite{lin2017placement,ivanova2022randomization,heussen2024randomization,heng2025design,li2025randomization}. To our knowledge, none of the existing papers establish the dual guarantee in the monotone attrition framework as provided by our procedure. \citep{kline2025finite} also consider randomization-based procedures for weak null hypotheses and their approaches requires a user-specified apriori bounds on the outcome support.

In general, heterogeneity-robust permutation/randomization inference is studied in \cite{neuhaus1993conditional}, \cite{janssen1997studentized},  \cite{janssen1999testing}, \cite{janssen2003bootstrap}, 
\cite{chung2013exact}, \cite{canay2017randomization},
\cite{wu2021randomization}, \cite{zhao2021covariate}, \cite{cohen2022gaussian}, 
\cite{tuvaandorj2024robust} and \cite{aronow2024randomization}. The primary difference of our paper with respect to this literature has been discussed in the introduction. 

\section{Setup, Dataset and Statistical Framework}\label{section:setup}
\subsection{Notation and Identification Assumptions}
We consider an experiment with $n$ units, where each unit is randomized to the treatment group or the control group. Let $y_i(d)$ denote the potential outcome of unit $i$ under treatment status $d\in \{0,1\}$. Let $r_i(d)$ denote the reporting status of the $i$th unit when under treatment status $d$. We write $r_i(1) = 1$ if the $i$th unit is assigned to treatment and is present in the following-up survey, and we write $r_i(1)=0$ if it is not present. The reporting status $r_i(0)$ is defined analogously.

As in \cite{lee2009training}, we make the following assumption on the reporting status.
\begin{assumption}[Monotonicity]\label{a:monotonicity}
For all  $i \in [ n ] $, we have  $r_i(1) \ge r_i(0)$.
\end{assumption}
The monotonicity assumption rules out the existence of units who would report if untreated but not report if treated. We note that the direction of the monotonicity is not critical here. With slight modification, the same identification argument and statistical procedure work  if one assumes that $r_i(1)\leq r_i(0)$ for all $i\in[n]$. The key requirement is that treatment affects reporting status in the same direction for all units.

Given Assumption \ref{a:monotonicity}, an unit $i$ belongs to one of the three principal strata based on its potential reporting status. Specifically, it can be:
\begin{itemize}
\item an always-reporter if $r_i(1) = 1$ and $r_i(0) = 1$,
\item an if-reporter if $r_i(1) = 1$ and $r_i(0) = 0$, or
\item a never-reporter if $r_i(1) = 0$ and $r_i(0) = 0$.
\end{itemize}
We refer to the sets of always-reporters, if-reporters, and never-reporters as the principal reporting strata \cite{frangakis2002principal}. They are the groups defined with respect to the potential reporting status.

Let $\mathcal{A}$ denote the set of always-reporters, defined as:
\begin{equation}\label{def:always_reporter}
    \ar=\{i\in[n]: r_i(0)=1, r_i(1)=1\}.
\end{equation}
The parameter of interest is the average treatment effect (ATE) of the always-reporters, defined as:
\begin{equation}\label{def:always_reporter_ate}
\arate = \frac{1}{|\mathcal A|} \sum_{i \in \mathcal A} \big( y_i(1)  - y_i(0) \big),   
\end{equation}
where we implicitly assume that $|\mathcal{A}|\geq 1$ so that the quantity is well-defined. We refer to this parameter as the \emph{always-reporter average treatment effect} (AR-ATE).\footnote{For if-reporters, we never observe their control outcomes; for never-reporters, we observe neither their treated nor control outcomes. For the ATE parameters of these groups, it should be clear that we can do no better than worst-case bounds under a bounded outcome assumption. }  

Under Assumption~\ref{a:monotonicity} and using the argument in \cite{lee2009training}, one can establish that AR-ATE is partially identified and a sharp bound on AR-ATE can be derived.\footnote{For example, see Section 7.4 of \cite{gerber2012field}.  } 
In particular, the logic of \cite{lee2009training} implies
\begin{enumerate}[label=(\roman*)]
    \item the fraction of always-reporters, $\pi_{\mathrm{AR}}$, is identified by the reporting rate in the control group;
    \item the fraction of if-reporters, $\pi_{\mathrm{IR}}$, is identified by the differential reporting rate between the treated and control groups;
    \item the average untreated outcome for always-reporters is identified by the average outcomes among reporters in the control group;
    \item the fraction of if-reporters among the reporters in the treated group is $\pi_{\mathrm{IR}\mid r(1)=1}=\pi_{\mathrm{IR}}/\left(\pi_{\mathrm{AR}}+\pi_{\mathrm{IR}}\right)$
    \item the average treated outcome for always-reporters, which is partially identified, is bounded between the upper-($\pi_{\mathrm{IR}\mid r(1)=1}$) and lower-($\pi_{\mathrm{IR}\mid r(1)=1}$) trimmed means of the treated outcomes among reporters in the treated group with estimable sample analogues.\footnote{Let $\{x_{(i)}\}_{i=1}^n$ with $x_{(1)}\le \cdots \le x_{(n)}$ denote the order statistics of $n$ real numbers and let $s\in[0,1]$. The upper-$s$ trimmed mean of $\{x_{(i)}\}_{i=1}^n$ is $\frac{1}{n-\lfloor sn\rfloor}\sum_{i=1}^{,n-\lfloor sn\rfloor}x_{(i)}$ and the lower-$s$ trimmed mean of $\{x_{(i)}\}_{i=1}^n$ is $\frac{1}{n-\lfloor sn\rfloor}\sum_{i=\lfloor sn\rfloor+1}^{n}x_{(i)}$.} 
\end{enumerate}
For statistical inference on the identified set, \cite{lee2009training} uses asymptotic methods. In this paper, we retain the identification framework of \cite{lee2009training} but replace asymptotics methods with a randomization-based inferential approach.

\subsection{Experimental Design and Data Generating Process}
We assume that the experiments under consideration are completely randomized: among $n$ units, researchers assign $n_1$  units to treatment, chosen uniformly at random. For each unit $i$, we let $D_i$ be a random variable such that $D_i=1$ if unit $i$ is assigned to treatment, and $D_i=0$ otherwise. We note that $\Prob(D_i =1 ) = n_1/n$. We denote a random assignment vector following a complete randomization of $n_1$ out of $n$ units as $D=\left(D_i\right)_{i=1}^n\sim \textrm{CR}(n,n_1)$.


A sample of $n$ units is associated with an unknown potential-outcome-potential-reporting-status table $\left(\left(y_i(1),y_i(0),r_i(1),r_i(0)\right)\right)_{i=1}^n$. The units are randomly assigned to the treatment according to a realization of the assignment vector $D^{\obs}=(D^{\obs}_i)_{i=1}^n \sim \textrm{CR}(n,n_1)$. We observe reporting status for all units given by $R_i=r_i\left(D_i\right)$. For reported units with $R_i=1$, we observe their outcomes $Y_i=y_i\left(D_i\right)\in \mathbb{R}$. For units with $R_i=0$, we do not observe their outcomes and we denote their outcomes as $\mathbf{NA}$.
The observed dataset hence consists of a set of outcome--assignment--reporting-status triples 
\begin{equation}\label{eqn:data}
\mathcal{D} = \{(Y_i,D^{\obs}_i,R_i)\}_{i=1}^n\in \left(\mathbb{R}\cup \mathbf{NA}\right)^n \times \{0,1\}^n \times \{0,1\}^n.
\end{equation}
We will also write $\mathcal{D}=\left(Y,D^\obs,R\right)$ where $Y=(Y_i)_{i=1}^n$, $D^\obs=(D^\obs_i)_{i=1}^n$ and $R=(R_i)_{i=1}^n$ if needed.



We adopt the finite-population (design-based) framework, treating the potential outcomes and potential reporting status 
as fixed parameters \citep{imbens2015causal}. 
The sole source of randomness in our model is the vector of random treatment assignments, and statistical uncertainties are evaluated exclusively with respect to this randomness.
\subsection{Problem Statement}
With the setup above, we now state our problem. Given a sample of $n$ units and the set of always-reporters $\ar$ as defined in (\ref{def:always_reporter}), we are interested in designing statistical inferential procedures that are valid as a test (but with different guarantees) for both the sharp-null hypothesis:
\begin{equation}\label{eqn:sharp_null}
    \textrm{H}_0^s: y_i(1)=y_i(0), \forall i\in\mathcal{A},
\end{equation}
and the weak-null hypothesis:
\begin{equation}
      \textrm{H}_0^w: \frac{1}{|\ar|}\sum_{i\in\ar}y_i(1)= \frac{1}{|\ar|}\sum_{i\in\ar}y_i(0).   
\end{equation}
We shall propose below procedures that are finite-sample valid for testing the sharp-null hypothesis and asymptotically valid for the weak-null hypothesis.

\section{Statistical Procedure}\label{section:statistical_procedure}

We first give a heuristic explanation for our inferential procedure.  We denote $A_i=1$ if the unit $i$ is an always-reporter and $A_i=0$ if otherwise.  Denote the binary vector of always-report indicators as  $A=\left(A_i\right)_{i=1}^n\in\{0,1\}^n$. We note that $A$ is unknown to the researchers and is only partially revealed by the realized assignments (e.g. see discussion in Section \ref{subsection:principal_strata}). 

If we know the set of always-reporters, randomization inference under the sharp-null hypothesis is straightforward. For example, we can use the (possibly studentized) absolute value of the difference-in-means statistic
\begin{equation}
    \widehat{\tau}(\tilde{D},Y,A)=\left|\frac{\sum_{i=1}^n \tilde{D}_iA_iY_i}{\sum_{i=1}^n \tilde{D}_iA_i} - \frac{\sum_{i=1}^n \left(1-\tilde{D}_i\right)A_iY_i}{\sum_{i=1}^n \left(1-\tilde{D}_i\right)A_i}\right|, 
\end{equation}
as our test statistic, where $\tilde{D}=(\tilde{D}_i)_{i=1}^n \in \{0,1\}^n$ is an arbitrary treatment assignment, $Y=\left(Y_i\right)_{i=1}^n$ is the vector of (possibly missing) observed outcomes, and $A$ is the binary vector of always-report indicators defined above.  We can obtain the randomization distribution of the statistics under the sharp-null hypothesis and reject the sharp-null hypothesis (\ref{eqn:sharp_null}) if the p-value $p(A)$ is less than a pre-specified level $\alpha$. A standard argument \cite{imbens2015causal} shows that this test is a finite-sample valid level-$\alpha$ test for testing the sharp-null hypothesis. However, this procedure is infeasible because it relies on the knowledge of the set of always-reporters, which is unknown in practice.   

To address this difficulty, we take a worst-case approach: we calculate worst-case p-value as the maximum across all randomization p-values based on sets of always-reporters consistent with the observed data. We shall explain below how to construct such sets of always-reporters in Section \ref{subsection:principal_strata}. For now, let $\mathbb{A}(D^\obs,R)$ be the set of sets of always-reporters that are consistent with the observed data. Formally, the worse-case p-value is defined as
\begin{equation}\label{eqn:worst_case_p_value}
    p^{\textrm{worst}}=\sup_{ A \in \mathbb{A}(D^\obs,R)} p(A).
\end{equation}
For statistical testing, we will reject the sharp-null hypothesis if $ p^{\textrm{worst}}\leq \alpha$. 

The statistical inferential algorithm is detailed in Algorithm \ref{alg:inf}. Given a sample size $n$, we denote an arbitrary test statistic as:
\begin{equation}\label{eqn:test_statistics}
    \mathcal{T}(\cdot,\cdot,\cdot): \left(\mathbb{R}\cup \mathbf{NA}\right)^n\times \{0,1\}^n \times \{0,1\}^n \to [-\infty,\infty],
\end{equation}
which is a function mapping (possibly missing) observed outcomes, treatment assignments and always-reporters indicators to the extended real line. We discuss the choice of test statistics in Section \ref{subsection:statistics}.
\begin{algorithm}[ht]
\caption{Randomization Inference for AR-ATE}\label{alg:inf}
    \begin{algorithmic}[1]
        \State \textbf{Input}: dataset $   \mathcal{D}=(Y,D^{\obs},R)$, where $Y=\left(Y_i\right)_{i=1}^n\in \left(\mathbb{R}\cup \mathbf{NA}\right)^n$, $D^{\obs}=\left(D_i^{\textrm{obs}}\right)_{i=1}^n\in \{0,1\}^n$ and $R=\left(R_i\right)_{i=1}^n\in \{0,1\}^n$, test statistics $\mathcal{T}(\cdot,\cdot,\cdot)$ as defined in (\ref{eqn:test_statistics}), pre-test significance level $\beta \in [0,1]$.
        \State \textbf{Step 1}: Compute the set of possible reporting tables $\mathbb{A}\left(D,R\right)$ as (\ref{eqn:compatible_reporting_ps}). 
        \State \textbf{Step 2}: Prune $\mathbb{A}\left(D,R\right)$ with Algorithm \ref{alg:prune} at the significance level $\beta$.
       \State \textbf{Step 3}: For each possible reporting table $A=\left(A_i\right)_{i=1}^n\in \mathbb{A}\left(D,R\right)$, calculate the p-value:
    \begin{equation}\label{eqn:pA}
        p(A)=\E_{D\sim \textrm{CR}(n,n_1)}\left[\mathds{1}\{\mathcal{T}(Y,D,A) \geq  \mathcal{T}(Y,D^{\obs},A) \}\right],
    \end{equation}
    where $Y$, $D^{\obs}$ and $A$ are fixed and the expectation is taken with respect to the random binary vector $D=\left(D_i\right)_{i=1}^n $, which is generated by complete randomization (selecting $n_1$ of $n$ units without replacement).
     \State \textbf{Step 3}: Collect the $p$-values for all possible reporting tables $\{p(A)\}_{A\in \mathbb{A}\left(D,R\right)}$ and output 1 (reject) if $\max_{A\in \mathbb{A}(D,R)}p(A)\leq \alpha-\beta$, otherwise output 0 (fail to reject).
    \end{algorithmic}
\end{algorithm}
\begin{remark'}
    In practice, one does not need to iterate over all possible reporting tables. One can terminate the algorithm and output $0$ as soon as a single p-value is above $\alpha-\beta$. 
\end{remark'}

In what follows, Section \ref{subsection:principal_strata} describes the construction of all reporting tables compatible with the observed assignments and reporting statuses. Section \ref{subsection:statistics} introduces several test statistics.

\subsection{Assign units to principal reporting strata}\label{subsection:principal_strata}

To construct the worst-case p-value, we need to first construct the set of sets of always-reporters that are consistent with the observed data. Recall that $R_i$ is the reporting status of the $i$th unit and $D_i$ is its treatment assignment. 

Conditioning on treatment assignments, it is possible to assign some subjects to principal reporting strata based on realized assignments and reporting statuses:
\begin{itemize}
    \item If $D_i=0$ and $R_i=1$, the unit must be an always-reporter;
    \item If $D_i=0$ and $R_i=0$, the unit could be an if-reporter or a never-reporter, and it can not be an always-reporter;
    \item If $D_i=1$ and $R_i=1$, the unit could be an if-reporter or an always-reporter;
    \item If $D_i=1$ and $R_i=0$, the unit must be a never reporter. 
\end{itemize}
The toy example in Table \ref{tab:reporting_strata} illustrates the attribution procedure. Note that the always-reporters in the control group can be identified exactly. The ambiguity comes from the treated units with $D_i=1$ and $R_i=1$, which are a mixture of always-reporters and if-reporters. Hence the set of sets of always reporters that are consistent with the data can be described as:

\begin{equation}\label{eqn:compatible_reporting_ps}
    \mathbb{A}\left(D,R\right) =\left\{A\in \{0,1\}^n: \begin{array}{ll}
        A_i=1, \text{ if } D_i=0, R_i=1\\
        A_i=0, \text{ if } D_i=0, R_i=0\\
        A_i=0, \text{ if } D_i=1, R_i=0\\
        A_i\in \{0,1 \} , \text{ if } D_i=1, R_i=1\\
    \end{array}\right\}
\end{equation}

\begin{table}[h]
    \centering
    \begin{tabular}{|c|c|c|c|c|c|c|}
    \hline 
    index     & $r_i(1)$ & $r_i(0)$ & $D_i$ & $R_i$ & AR? & True Principal Stratum   \\
    \hline
    1     & 1 & 1 & 1 & 1 & ? & always-reporter   \\
    2     & 1 & 0 & 1 & 1 & ? & if-reporter   \\
    3     & 0 & 0 & 1 & 0 & NO & never-reporter   \\
    4     & 1 & 1 & 0 & 1 & YES & always-reporter   \\
    5     & 1 & 0 & 0 & 0 & NO & if-reporter   \\
    6     & 0 & 0 & 0 & 0 & NO & never-reporter   \\
    \hline 
    \end{tabular}
    \caption{The column \textit{index} lists the indices for units. The column $r_i(1)$ indicates the reporting status if treated. The column $r_i(0)$ indicates the reporting status if untreated. The column $D_i$ shows the observed treatment assignments. The column $R_i$ shows the observed reporting status. The column $AR?$ indicates whether the unit is an always-reporter. The column \textit{True Principal Stratum} denotes the true (but unknown) principal reporting stratum that the unit belongs.}
    \label{tab:reporting_strata}
\end{table}
We shall call each $A\in\mathbb{A}(D,R)$ a \textit{reporting table}, which is a vector of always-reporter indicators. 
\subsubsection{Refine sets of always-reporters with pretesting}\label{section:pretest}
Based on the observed missing pattern, we can pretest the number of always-reporters in the data. Because the treatment assignment is randomized independent of the reporting status, the fractions of always-reporters should on average be the same. This argument allows one to prune probabilistically improbable tables in $\mathbb{A}(D,R)$ that contain too many or too few always-reporters. This is an application of thr Bergers-Boo procedure in our setting \cite{berger1994p}. For example, given a table $A\in\mathbb{A}\left(D,R\right)$, define the test-statistic:
\begin{equation}\label{eqn:dima}
   \dimA(D,A)=\frac{1}{n_1}\sum_{i=1}^n D_iA_i - \frac{1}{n_0}\sum_{i=1}^n \left(1-D_i\right)A_i,
\end{equation}
and we reject a table if the p-value based on its randomization distribution is smaller than $\beta$. A detailed procedure is included in Algorithm \ref{alg:prune}.\footnote{Note that the randomization distribution of $\widehat{\textrm{DIM}}(D,A)$ depends only on the number of always-reporters in $A$ because of the symmetry of complete randomization. Hence, all tables with the same number of always-reporters yield identical rejection decisions. This observation can substantially reduce the computational burden.}  Other pre-testing procedures, such as the one based on tail bounds or test inversions \cite{rigdon2015randomization} can also be employed.

\begin{algorithm}
\caption{ Prune $\mathbb{A}\left(D^\obs,R\right)$ based on the number of always reporters at level $\beta$ }\label{alg:prune}
\begin{algorithmic}[1]
\State \textbf{Input}: $\{(R_i,D^\obs_i)\}_{i=1}^n$, compatible reporting tables $\mathbb{A}\left(D^\obs,R\right)$ as defined in (\ref{eqn:compatible_reporting_ps}), significance level $\beta\in(0,1)$
\State \textbf{Step 1}: Calculate $n_{1\ar}=\sum_{i=1}^n R_iD_i$.
\For{ $i \gets 0:n_{1\ar}$ }
    \State Select arbitrary $A\in \mathbb{A}\left(D,R\right)$ such that $\sum_{i=1}^n D^\obs_iA_i=i$.
    \State Given the selected $A$, calculate
        \begin{equation}
        p(i)=\E_{D\sim \textrm{CR}(n,n_1)}\left[\mathds{1}\{\dimA(D^\obs,A)\geq  \dimA(D,A) \}\right],
    \end{equation}
      where $A$ is fixed and the expectation is taken with respect to the random binary vector $D=\left(D_i\right)_{i=1}^n $, which is generated by complete randomization (selecting $n_1$ of $n$ units without replacement).
\EndFor
\State \textbf{Step 2}:  Calculate the set $N_A=\{i\in [0,n_{1\ar}]: p(i)\geq \beta\}$.
\State \textbf{Return} the pruned set
\begin{equation}
    \mathbb{A}^{\textrm{prune}}\left(D^\obs,R\right)=\Big\{A\in \mathbb{A}\left(D^\obs,R\right): \sum_{i=1}^n D^\obs_iA_i \in N_A\Big\}. 
\end{equation}
\end{algorithmic}
\end{algorithm}

\subsection{Test Statistics}\label{subsection:statistics}
\subsubsection{The Studentized-Hajek Statistic}
Given a reporting table $A$ that is consistent with the data, define the Hajek estimator $\arhj(Y,D,A)$: 
    \begin{equation}\label{eqn:hajek_a}
        \arhj\left(Y,D,A\right)=\frac{\sum_{i=1}^n D_iA_iY_i}{\sum_{i=1}^n D_iA_i} -\frac{\sum_{i=1}^n \left(1-D_i\right)A_iY_i}{\sum_{i=1}^n \left(1-D_i\right)A_i} , 
    \end{equation}
and its variance estimator:
\begin{equation}\label{hajek:var}
    \widehat{\AVar}^{\ha}_n \left( Y,D,A \right)= \frac{\sum_{i=1}^n D_iA_i\left( Y_i-\widehat{\tau}^1_A\right)^2}{\left(\sum_{i=1}^n D_iA_i\right)^2}  +  \frac{\sum_{i=1}^n (1-D_i)A_i\left( Y_i-\widehat{\tau}^0_A\right)^2}{\left(\sum_{i=1}^n \left(1-D_i\right)A_i\right)^2} 
\end{equation}
where,
\begin{equation}\label{def:avg}
\widehat{\tau}^1_{A}= \frac{\sum_{i=1}^n D_iA_iY_i}{\sum_{i=1}^n D_iA_i}, \widehat{\tau}^0_{A} = \frac{\sum_{i=1}^n \left(1-D_i\right)A_iY_i}{\sum_{i=1}^n \left(1-D_i\right)A_i}.
\end{equation}
We define the absolute value of the studentized-Hajek statistic as:
    \begin{equation}\label{eqn:t0}
   \T^0_n\left(Y,D,A\right) =\left| \frac{\arhj\left(Y,D,A\right)}{\sqrt{\widehat{\AVar}^{\ha}_n \left( Y,D,A \right)}}\right|.
    \end{equation}

\subsubsection{The Chi-square Statistics}
Apart from the balance of outcomes among always-reporters, we can include the balance of the number of always-reporters between the treated and control groups, leading to variations of Wald statistics. The variance of the statistic $ \dimA$, $\VA$, is defined as \begin{equation}\label{eqn:va}
   \VA= \frac{n^2}{n_1n_0(n-1)} \left( \frac{1}{n}\sum_{i=1}^n A_i - \left(\frac{1}{n}\sum_{i=1}^nA_i\right)^2  \right).
     \end{equation}
We now define two test statistics: 
\begin{equation}\label{eqn:t1}
  \T^1_n(Y,D,A)= \left(\T^0_n\left(Y,D,A\right)\right)^2 +   \left(\frac{\dimA(D,A)}{ \sqrt{\VA}}\right)^2,
\end{equation}
and
\begin{equation}\label{eqn:t2}
    \T^2_n\left(Y,D,A\right) = \left(\T^0_n\left(Y,D,A\right)\right)^2 +   \left(\left\lfloor\frac{\dimA(D,A)}{ \sqrt{\VA}}\right\rfloor_-\right)^2 ,
\end{equation}
where we define $\lfloor x\rfloor_-=\max\{0,-x\}$ and $0/0=0$.

\section{Statistical Guarantees}\label{section:statistical_guarantees}
\subsection{Randomization-based Procedures}
Let $\Pi_n$ be a distribution of random assignment variables $\{D_i\}_{i=1}^n$ implementing a completely randomized design. We first define a general parameter space encoding Assumption \ref{a:monotonicity}.
\begin{equation}\label{eqn:Theta_n}
    \Theta_n=\Big\{ \left(\{\left(y_i(1),y_i(0),r_i(1),r_i(0)\right)\}_{i=1}^n,\Pi_n\right):r_i(1)\geq r_i(0), \forall i\in[n]\Big\}.
\end{equation}
Note that each element $\theta_n\in \Theta_n$ is a combination of potential outcomes, potential reporting status, and an experimental design. Each element $\theta_n$ completely determines distributions of our test statistics. For the asymptotic guarantee of testing the weak-null hypothesis, we need the following assumptions on our parameter space:
\begin{assumption}[Parameter Space]\label{a:theta}
Let $n\in \mathbb{N}$, and constants $\delta  \in [-1,1) $, $B>0$ and $s\in (0,1]$ be given. 

Given a set of potential outcomes and reporting status $\{(y_i(1),y_i(0),r_i(1),r_i(0))\}_{i=1}^n$, define the always-reporter indicator for each  unit $i$ as $A_i=1$ if $r_i(1)=r_i(0)=1$ and $A_i=0$ otherwise. Define the set of always-reporters as $\mathcal{A}=\{i: A_i=1\}$. Let $n_A=\sum_{i=1}^n A_i$ denote the number of always-reporters. The following conditions hold:
\begin{enumerate}[label=(\roman*)]
\item For the given $s$, $n_A\geq sn$.\label{a:theta1}
\item
For the given $\delta$, we have \label{a:theta2}
\begin{equation}
 {\sigma_{y(1), y(0), A}} \geq - \delta\cdot {\sigma_{y(1),A} \cdot \sigma_{ y(0),A}} ,   
\end{equation}
where $\sigma_{y(1),y(0),A}$ is the always-reporter-group correlation of adjusted potential outcomes, 
\begin{equation}
    \sigma_{y(1),y(0),A}=\frac{1}{n_A-1}\sum_{i\in \A}
    \left(y_i(1)-\mu_{y(1),A}\right)\left(y_i(0)-\mu_{y(0),A}\right),
\end{equation}
where  $\mu_{y(a),A}=n_A^{-1}\sum_{i\in \A}y_i(a)$ for $a\in\{0,1\}$,  and $\sigma^2_{ y(1),A}$ and $\sigma^2_{y(0),A}$ are the always-reporter-group potential outcome variances,
\begin{equation*}
  \sigma^2_{y(a),A}=\frac{1}{n_A-1}\sum_{i\in\A}\left(y_i(a)-\mu_{y(a),A}\right)^2,
\end{equation*}
for $a\in\{0,1\}$. $\sigma^2_{y(1),A}$ and $\sigma^2_{y(0),A}$ are positive. If $n_A=1$, all variances and covariances are defined to be zero.
\item For each $a \in \{ 0, 1 \}$ and the given $B$, \label{a:theta3}
\begin{equation}\label{fourthmoment}
    \left(\frac{1}{n_A}\sum_{i\in \A} |y_i(a)- \mu_{y(a),A}|^{4}\right)^{1/4}\leq  B  \sigma_{y(a),A}.
\end{equation}
\end{enumerate}
\end{assumption}
\begin{remark'}
Condition (i) assumes that the always-reporters take an non-negligible share of the sample. 
Condition (ii) rules out extreme correlations between the potential treatment and control outcomes of the compliers, a common assumption in the finite-population setting. 
Condition (iii) is a standard technical condition that arises when applying the central limit theorem for triangular arrays, and may be thought of as excluding heavy-tailed or sparse data. 
\end{remark'}

\begin{assumption}[Completely Randomized Experiment]\label{a:cr}    Given $n\in \mathbb{N}$ and a constant $0<r\leq1/2$, we have $n_1\in [rn,(1-r)n]$.
\end{assumption}

We denote the parameter space of potential outcomes, potential reporting statuses, and experimental designs satisfying Assumptions 1, 2 and 3 with $n$ units and constants $\delta$, $r$, and $B$ as $\Theta_n\left(\delta,s,r,B\right)\subset \Theta_n $.
For the weak-null hypothesis, we further define the parameter space:
\begin{equation}\label{defn:theta_w}
    \Theta^w_n(\delta,s,r,B)
    = \left\{ \theta \in \Theta_n(\delta,s,r,B) : 
    \frac{1}{n_A}\sum_{i\in\A}y_i(1) = \frac{1}{n_A}\sum_{i\in\A}y_i(0)  \right\}.
\end{equation}
For the sharp-null hypothesis, we define the parameter space:
\begin{equation}\label{defn:theta_s}
      \Theta^s_n
    = \Big\{\theta\in\Theta_n: y_i(1)=y_i(0),\forall i\in \A\Big\}.   
\end{equation}
Note that the parameter space for the sharp-null hypothesis needs not satisfy Assumption \ref{a:theta} and Assumption \ref{a:cr}. 

The following theorem states that using test statistics $\T_0$, $\T_1$ and $\T_2$  introduced in (\ref{eqn:t0}),  (\ref{eqn:t1}), and (\ref{eqn:t2}), the statistical procedure as in Algorithm \ref{alg:inf} provides finite-sample valid level-$\alpha$ test against the sharp-null hypothesis and asymptotically-valid  level-$\alpha$ test against the weak-null hypothesis.
\begin{theorem'}\label{thm:main}
Consider the statistical procedure in Algorithm \ref{alg:inf} with test statistics $\T_n^0$, $\T_n^1$ and $\T_n^2$ defined in (\ref{eqn:t0}),  (\ref{eqn:t1}), and (\ref{eqn:t2}). Fix a significance level $\alpha\in(0,0.25]$ and a pre-testing level $\beta\in [0,\alpha)$. Let $\Theta^s_n$ be the parameter space of the sharp-null hypothesis defined in (\ref{defn:theta_s}). We have
\begin{equation}\label{thm:finite_sample_guarantee}
    \sup_{\theta_n^s\in \Theta_n^{\textrm{s}}}\Prob_{\theta_n^s}\left( p^{\textrm{worst}}\leq \alpha\right)\leq \alpha.
\end{equation}
Fix $\delta <1$, $s\in(0,1]$, $r \in (0,1/2]$, and $B > 0$. Let $\Theta^w_n(\delta,s,r,B)$ be the parameter space of the weak-null hypothesis defined in (\ref{defn:theta_w}). We have
\begin{equation}\label{thm:asymptotic_guarantee}
    \limsup_{n\to\infty}\sup_{\theta^w_n\in \Theta^w_n(\delta,s,r,B)}\Prob_{\theta^w_n}\left( p^{\textrm{worst}}\leq \alpha\right)\leq \alpha
\end{equation}
 \end{theorem'}

\begin{remark'}
Our asymptotic uniform guarantee includes cases where the sample consists almost entirely of always-reporters and few (potentially zero) if-reporters. Other asymptotic methods generally rule out this portion of the parameter space \citep{lee2009training} or rely on problem-specific tuning parameters \citep{SEMENOVA2025106055}. Including this part of the parameter space necessitates a more involved mathematical analysis.
\end{remark'}
\subsection{Asymptotic-Distribution-Based Procedures}

We note that, as an intermediate step in the proof of Theorem \ref{thm:main}, we derived the asymptotic distributions (up to the slackness introduced by the variance bounds)  of the statistics $\mathcal{T}_n^0$, $\mathcal{T}_n^1$, and $\mathcal{T}_n^2$ under the true but unknown always-reporter vector $A$. This result permits inference based on asymptotic critical values. Procedures based on critical values obtained from asymptotic approximations do not provide the finite-sample guarantee as in \eqref{thm:finite_sample_guarantee}; they retain only the asymptotic guarantee as in \eqref{thm:asymptotic_guarantee}. However, as we discuss in Section \ref{section:implementation}, procedures based on asymptotic critical values are typically much easier to compute in practice. 

The inferential algorithm based on asymptotic critical values is presented in Algorithm~\ref{alg:asy}. We first define three functions of
\(n_A^1=\sum_{i=1}^n D_iA_i\), given \(n_1\), \(n_0\), and \(n_A\):
\begin{align}
    & g_0(n_A^1) = 0,\qquad 
    g_1(n_A^1) = \V_n^{-1}(A)\left(\frac{n_A^1}{n_1} - \frac{n_A-n_A^1}{n_0}\right)^2, \\
    & g_2(n_A^1) = \V_n^{-1}(A)\left(\left\lfloor\frac{n_A^1}{n_1} - \frac{n_A-n_A^1}{n_0}\right\rfloor_{-}\right)^2,
\end{align}
where $\V_n(A)$ is defined in \eqref{eqn:va} and we adopt the convention that $0/0=0$.
These three functions correspond to the three statistics for balance in the number of always-reporters in test statistics \eqref{eqn:t0}, \eqref{eqn:t1}, and \eqref{eqn:t2}.
\begin{algorithm}[ht]
\caption{Asymptotic Inference for the AR-ATE}\label{alg:asy}
\begin{algorithmic}[1]
  \State \textbf{Input:} Dataset $\mathcal{D}=(Y,D^{\obs},R)$ as in \eqref{eqn:data}; test statistics $\mathcal{T}_i(\cdot,\cdot,\cdot)$, $i\in\{0,1,2\}$, as in \eqref{eqn:t0}, \eqref{eqn:t1}, and \eqref{eqn:t2}; significance level $\alpha\in(0,1)$; pre-test significance level $\beta\in[0,\alpha)$.
  \State \textbf{Step 1:} Create the set of compatible reporting tables $\mathbb{A}(D,R)$ as in \eqref{eqn:compatible_reporting_ps}.
  \State \textbf{Step 2:} Prune $\mathbb{A}(D,R)$ using Algorithm \ref{alg:prune} at significance level $\beta$.
  \State \textbf{Step 3:} Initialize $Rej\gets 1$. (1 = reject, 0 = fail to reject)
  \State \textbf{Step 4:} Enumerate the possible sizes of always-reporter vectors:
  \[
    \mathrm{Cardinality}\text{-}A
    =
    \left\{
      \sum_{i=1}^n A_i \,:\,
      A=(A_i)_{i=1}^n \in \mathbb{A}(D,R)
    \right\}.
  \]
  \For{each $k \in \mathrm{Cardinality}\text{-}A$}
    \State Compute the $(1-\alpha+\beta)$-quantile $q^k_{i,1-\alpha+\beta}$ of $Z+g_i(n_A^1)$, where
    $Z\sim \mathrm{N}(0,1)$,
    $n_A^1=\sum_{i=1}^n D_iA_i$,
    $D=(D_i)_{i=1}^n \sim \mathrm{CR}(n,n_1)$, and $Z$ is independent of $D$.
    \State Compute $
      \mathcal{T}_i^{\max,k}=
      \max_{A\in \mathbb{A}(D,R)\,:\,\sum_{i=1}^n A_i=k}
      \mathcal{T}_n^i\!\left(Y,D^{\obs},A\right).$  
    \If{$\mathcal{T}_i^{\max,k}\le q^k_{i,1-\alpha+\beta}$}
      \State Set $Rej\gets 0$.
    \EndIf
  \EndFor
  \State \textbf{Step 5:} \textbf{return} $Rej$.
\end{algorithmic}
\end{algorithm}

\begin{theorem'}\label{thm:asy}
Consider the statistical procedure in Algorithm \ref{alg:asy}. Fix a significance level $\alpha\in(0,0.25]$ and a pre-testing level $\beta\in [0,\alpha)$. Fix $\delta <1$, $s\in(0,1]$, $r \in (0,1/2]$, and $B > 0$. Let $\Theta^w_n(\delta,s,r,B)$ be the parameter space of the weak-null hypothesis defined in (\ref{defn:theta_w}). Define the event
\begin{equation*}
    \mathcal{E}_n=\{ (Y,D^\obs,R): \text{ Algorithm } \ref{alg:asy} \text{ returns } Rej=1\}.
\end{equation*}
Then we have,
\begin{equation}\label{thm:asymptotic_guarantee}
    \limsup_{n\to\infty}\sup_{\theta^w_n\in \Theta^w_n(\delta,s,r,B)}\Prob_{\theta^w_n}\left( \mathcal{E}_n\right)\leq \alpha.
\end{equation}
 \end{theorem'}
\begin{remark'}
We remark that the limiting distribution of $g_1(n_A^1)$ can depend on the sequence of always-reporter vectors. We give two simple examples here and other sequences are also possible and may yield different limits. If a sequence $\left(A^n\right)_{n=1}^{\infty}$ satisfies, for a constant $c<1$,
\[
\frac{1}{n}\sum_{i=1}^n A_i^n \leq c, \quad \forall n,
\]
then $g_1(n_A^1)$ has a limiting chi-squared distribution with one degree of freedom. In contrast, if $\sum_{i=1}^n A_i^n=1$ for all $n$, then $g_1(n_A^1)$ has a point mass at $0$ as its limiting distribution. Both sequences are permitted by our parameter space. A similar comment applies to $g_2(n_A^1)$. Both the randomization-based procedure in Algorithm \ref{alg:inf} and the asymptotic procedure in Algorithm \ref{alg:asy} are agnostic to the particular sequence of always-reporter vectors, and both deliver uniformly asymptotically valid inference.
\end{remark'}

\section{Implementations}\label{section:implementation}

\subsection{Overview}

In randomization inference, computing exact p-values by enumerating all possible treatment assignments is typically infeasible. Instead, p-values are usually approximated via Monte Carlo simulation, using a sufficiently large number of independent random assignments to approximate the randomization distribution.

Let $n_{mc}$ denote the number of Monte-Carlo simulations. For each simulation $s\in \{1,...,n_{mc}\}$, denote the $s$th  simulated assignment variables as $D^s=\left(D^s_i\right)_{i=1}^n$. The computation of the worst-case p-value problem based on Monte-Carlo draws can be represented as:
\begin{equation}\label{eq:p_worst_mc}
   p^{\textrm{worst,mc}}= \max_{A\in \mathbb{A}(D^\obs,R)}\sum_{s=1}^{\nmc}\mathds{1}\left\{ \mathcal{T}^i_n\left(Y,D_s,A\right) \geq \mathcal{T}^i_n\left(Y,D^{\obs},A\right) \right\},
\end{equation}
with $\mathcal{T}_n^i, i\in\{0,1,2\}$ defined in \eqref{eqn:t0}, \eqref{eqn:t1} and \eqref{eqn:t2} respectively, and $\mathbb{A}(D^\obs,R)$ defined in \eqref{eqn:compatible_reporting_ps}. 

The optimization problem \eqref{eq:p_worst_mc} can be challenging to solve practically. If the number of reporting treated units is \(r_1=\sum_{i=1}^n D_i^{\obs}R_i\), then the set \(\mathbb{A}(D^\obs,R)\) will generically contain \(2^{r_1}\) candidate always-reporter tables. Pre-testing the number of always-reporters in the treated group can shrink the search space, but typically not enough to make brute-force enumeration practical for moderately sized datasets---for example, even when \(r_1=50\).

We discuss two computation approaches in this section, designed for different outcome data variable types. Section \ref{outcome:finite_support} considers the case where the observed outcome variables are discrete and have a small support. For example, this is the case when the outcome variables are binary, categorical or count-valued. In such a setting, 
exhaustive enumeration is feasible even with a large sample size, after exploiting the symmetry of complete randomization. 

Section \ref{section:IP}  considers a second approach which can be used with continuous outcomes. It decomposes \eqref{eq:p_worst_mc} into smaller problems, each solvable using (linear) integer programming techniques. The resulting p-value, $p^{\textrm{worst,IP,mc}}$, has the guarantee that $p^{\textrm{worst,IP,mc}}\geq p^{\textrm{worst,mc}}$, hence enabling valid yet possibly conservative inference. In our simulations, we did not observe a major loss of power due to such relaxations, provided each subproblem is tight enough. See Lemma \ref{lemma:v}.

We use the following notations through out the section. Given a dataset with a sample size $n$ and a vector of always-reporter indicators $A=\left(A_i\right)_{i=1}^n$, let $n_A=\sum_{i=1}^n A_i$ denote the number of always-reporters. Denote the random assignment variables $\left(D_i\right)_{i=1}^n\sim \textrm{CR}(n,n_1)$. We define the random variables $n_{A,1}=\sum_{i=1}^n D_iA_i$ and $n_{A,0}=\sum_{i=1}^n \left(1-D_i\right)A_i$. 
\subsection{Outcomes with a small number of support points}\label{outcome:finite_support}
Datasets with discrete outcomes (e.g., binary, categorical or count-valued) are common. With such dataset, exhaustive enumeration is feasible even with a large sample size, after exploiting the symmetry of complete randomization. 

Suppose that the support of the observed outcomes, $\mathcal{S}$, has cardinality $K$. We enumerate the support as $\mathcal{S}=\{v_1,...,v_K\}$. For the randomization distribution under the sharp-null hypothesis, we define the number of always-reporters with outcome $v_k$ as $n_A^k=\sum_{i=1}^n A_i\mathds{1}\{Y_i=v_k\}$, $k\in [K]$, and the number of always-reporters assigned to treatment and control groups with outcome $k$ as $n_{A}^{1,k}=\sum_{i=1}^n D_iA_i\mathds{1}\{Y_i=v_k\}$ and $n_{A}^{0,k}=\sum_{i=1}^n \left(1-D_i\right)A_i\mathds{1}\{Y_i=v_k\}$, where $D=\left(D_i\right)_{i=1}^n$ is a generic assignment vector.

Given $n_A$ and $\{n_{A}^k\}_{k\in [K]}$, the distribution of the test statistics \eqref{eqn:t0}, \eqref{eqn:t1} and \eqref{eqn:t2} are completely determined by the distribution of $\{n_{A}^{1,k}\}_{k\in [K]}$. For example, the squared studentized-Hajek statistics \eqref{eqn:t0} can be written as a function of $n_A$, $\{n_{A}^k\}_{k\in [K]}$,  $n_{A,1}$ and $\{n_{A}^{1,k}\}_{k\in [K]}$:
\begin{align*}
  \bigl(\T_n^0(Y,D,A)\bigr)^2=\frac{ \left(  n_{A,1}^{-1}\sum_{k=1}^K v_k n_{A}^{1,k}-n_{A,0}^{-1}\sum_{k=1}^K v_k n_{A}^{0,k}\right)^2 }{ n_{A,1}^{-1}\widehat{\sigma}^2_1 +n_{A,0}^{-1} \widehat{\sigma}^2_0}, 
\end{align*}
where, for $a\in \{0,1\}$,
\begin{align*}
    \widehat{\sigma}^2_a = \frac{1}{n_{A,a}}\sum_{k=1}^Kn_{A}^{a,k}\left( v_k -\frac{1}{n_{A,a}}\sum_{k=1}^K v_k n_{A}^{a,k}\right)^2,
\end{align*}
with $n_{A,1}=\sum_{k=1}^Kn_{A}^{1,k}$, $n_{A,0}=n_A-n_{A,1}$, and $n_A^{0,k}=n_A^k-n_A^{1,k}$. The probability mass function of $\{n_{A}^{1,k}\}_{k\in [K]}$ is
\begin{equation*}
 \textrm{pmf}\left(\{n_{A}^{1,k}\}_{k\in [K]}\right) = \binom{n}{n_1}^{-1}\binom{n-n_A}{n_1-n_A^1}\prod_k \binom{n_A^k}{n_A^{1,k}},
\end{equation*}
which depends on $\{n,n_1,\{n_{A}^k\}_{k=1}^K\}$. Hence, the randomization distributions of test statistics \eqref{eqn:t0}, \eqref{eqn:t1} and \eqref{eqn:t2} under the sharp-null hypothesis are completely determined by the parameters $\{n,n_1,\{n_{A}^k\}_{k=1}^K\}$.

Given an always-reporters vector $A=\left(A_i\right)_{i=1}^n\in \mathbb{A}(D^\obs,R)$, we denote the vector of outcome counts as $c\left(A\right)=\left(n_{A}^k\right)_{k=1}^K$. By the argument above, the randomization-based $p$-value associated with a given always-reporter table depends on $A$ only through the outcome counts vector $c(A)$, together with the sample size $n$ and the number of treated units $n_1$:
\begin{equation}\label{eqn:count_vector_element}
    p(A)=\widetilde{p}\bigl(c(A),n,n_1\bigr).
\end{equation}
Let $\mathbb{C}(D^\obs,R)$ denote all possible counts vector that is compatible with the data:
\begin{equation}\label{eqn:conut_vector}
    \mathbb{C}(D^\obs,R)=\{ c\left(A\right),A\in \mathbb{A}(D^\obs,R)\}.
\end{equation}
We have the reduction,
\begin{equation}
        p^{\textrm{worst}}=\sup_{ A \in \mathbb{A}(D^\obs,R)} p(A)=\sup_{ c \in \mathbb{C}(D^\obs,R)} \widetilde{p}(c,n,n_1).
\end{equation}
If the support of observed outcomes has cardinality $K$, the worst-case cardinality of $\mathbb{C}(D^\obs,R)$ is upper-bounded by $n^K$. Provided that we consider a regime where $K$ does not increase with $n$, an exhaustive search algorithm has a time complexity that is polynomial in $n$. When $K$ is small, for example, $K=2$ for binary outcomes, a simple implementation would lead to a practical algorithm. A pseudo algorithm is included as Algorithm \ref{alg:12152025_2} in Appendix Section \ref{section:algorithms}. 

\subsection{Integer Programming (IP) Approach for outcomes with a large number of support points}\label{section:IP}
When the observed outcome data have large support, the reduction in Section~\ref{outcome:finite_support} does not yield practically efficient algorithms.
To optimize over a large search space, we employ integer programming techniques.

Because $\mathcal{T}_n^0$, $\mathcal{T}_n^1$, and $\mathcal{T}_n^2$ are nonlinear functions of the always-reporter indicators $A$, a direct integer-programming formulation is challenging. In particular, after factorization, $\mathcal{T}_n^0$ and $\mathcal{T}_n^1$ can be expressed as sums of ratios of polynomials in $A$ of degree up to 10, and $\mathcal{T}_n^2$ is further complicated by the floor operator $\lfloor \cdot \rfloor_{-}$. Standard linearization techniques in IP are not easily applicable even with small instances.

To address the computational challenge, we decompose \eqref{eq:p_worst_mc} into a collection of smaller subproblems that are substantially easier to solve in our simulations. Moreover, the subproblems can be solved independently, making this step readily distributable across computing resources.  The decomposition proceeds in two steps. First, we rewrite \eqref{eqn:pA} as a set of subproblems with a simpler algebraic structure (for example, involving lower-degree polynomials). Second, we further partition these subproblems by the number of always-reporters \(n_A\). The two-step reductions are useful because they decompose the original problem into integer programming subproblems with only quadratic (second-order) polynomials, eliminating non-polynomial features such as floor functions. For details, see the discussion after Theorem \ref{thm:reduction}.

\subsubsection{Step 1: Decomposing \eqref{eqn:pA} into subproblems with a simpler algebraic structure}\label{section:Step1}
Let $\mathcal{T}_n$ denote, generically, any one of the test statistics $\mathcal{T}_n^0$, $\mathcal{T}_n^1$, and $\mathcal{T}_n^2$. 
Let $L,U\in[0,\infty]$ be two (possibly infinity-valued) scalars such that
\begin{equation}\label{eqn:LU}
    L \le \min_{A\in \mathbb{A}(D^\obs,R)} \mathcal{T}_n(Y,D^\obs,A)
    \le \max_{A\in \mathbb{A}(D^\obs,R)} \mathcal{T}_n(Y,D^\obs,A)
    \le U .
\end{equation}
Relatively tight bounds $L$ and $U$ are easy to obtain, either analytically or via numerical procedures. See Algorithm \ref{alg:6}, Algorithm \ref{alg:7} and Section \ref{section:boundstd}.

Consider a partition of the interval $[L,U]$ with increasingly ordered endpoints $\{t_i\}_{i=0}^I$, where $t_0=L$ and $t_I=U$. For each subinterval $[t_{i-1},t_i]$, consider the following value associated with each always-reporter vector $A$:
\begin{equation}\label{eqn:vAtt}
v(A,t_{i-1},t_i)=
\begin{cases}
\displaystyle \E_{D}\!\left[\mathds{1}\!\left\{\T_n(Y,D,A)\ge t_{i-1}\right\}\right],
& \text{if } \T_n(Y,D^{\obs},A)\le t_i,\\[0.5em]
-\infty, & \text{otherwise,}
\end{cases}
\end{equation}
where $D\sim \textrm{CR}(n,n_1)$. Also consider the optimization problem:
\begin{equation}\label{eqn:inner_problem}
    v_i=\max_{A\in\mathbb{A}(D^\obs,R)}v(A,t_{i-1},t_i).
\end{equation}
Lemma~\ref{lemma:v} suggests that the collection of optimal values of \eqref{eqn:inner_problem} across different pairs of consecutive endpoints is informative about $p^{\textrm{worst}}$.

\begin{lemma'}\label{lemma:v}
Given endpoints $\{t_i\}_{i=0}^I$ with $t_0=L$ and $t_I=U$ satisfying \eqref{eqn:LU}, let $A^{i*}$ denote an optimal always-reporter vector for problem~\eqref{eqn:inner_problem} on the interval $[t_{i-1},t_i]$. We have the following inequalities:
\begin{equation*}
 p^{\textrm{worst}} \leq \max_{i\in [I]}v_i,
\end{equation*}
\begin{equation*}
 v_i \leq  p^{\textrm{worst}} +\E_{D\sim \textrm{CR}(n,n_1)}\left[\mathds{1}\{\T_n(Y,D,A^{i*})\in [t_{i-1},t_i)\}\right], \forall i\in [I].
\end{equation*}
In particular, let $i^*$ be the optimal index of the problem $ \max_{i\in [I]}v_i$. We have
\begin{equation}\label{eqn:bound38}
p^{\textrm{worst}} \leq \max_{i\in [I]}v_i \leq  p^{\textrm{worst}} +\E_{D\sim \textrm{CR}(n,n_1)}\left[\mathds{1}\{\T_n(Y,D,A^{i^*})\in [t_{i^*-1},t_i^*)\}\right].
\end{equation}
\end{lemma'}

From a practical perspective, this lemma has three implications. 
First, if $\max_{i\in[I]} v_i \le \alpha$, then it certifies that $p^{\mathrm{worst}} \le \alpha$. 
Second, provided that the randomization distributions induced by the optimizing always-reporter vectors are not overly concentrated on the interval $[t_{i^*-1},\, t_{i^*})$, the quantity $\max_{i\in[I]} v_i$ should be close to the true $p^{\mathrm{worst}}$.\footnote{Ideally, we would like to show that, as the maximum interval length tends to zero, the additional term on the right-hand side of \eqref{eqn:bound38} also tends to zero. If the underlying distribution were continuous, this would follow from an application of the dominated convergence theorem. In our setting, however, the randomization distributions are discrete with point masses, so such arguments are not directly applicable. We are not aware of a simple refinement that yields a useful bound for this term.} 
Third, it transforms the task of comparing two ratios of polynomials into comparing a ratio of polynomials with a scalar. This reduction simplifies the computation problem. After Step~2 below, we can clear the denominators (variance estimators) in the test statistics associated with outcome balance by multiplying them on both sides, yielding inequalities in which both sides are polynomials of degree at most $2$.

  We note that the lemma and the accompanying discussion remain valid, after replacing $p^{\mathrm{worst}}$ with $p^{\mathrm{worst,mc}}$, when the expectation (average) is taken over simulated assignments, as opposed to the exact expectation under the complete randomization distribution.

\subsubsection{Step 2: Split by the number of always-reporters}\label{section:step2}

To make the discussion simpler, we introduce some additional notations. For a given always-reporter vector $A$, we index the always-reporters by \(a\in[n_A]\) via a bijection
\(\pi_A:[n_A]\to [n]\). Without loss of generality, we label indices for both \(i\) and \(a\) so
that the first $r_0=\sum_{i=1}^n (1-D_i^{\mathrm{obs}})R_i$ 
elements correspond to always-reporters assigned to the control group. We write \(\pi_A(a)=i\) when unit \(i\)
is labeled as the \(a\)-th always-reporter. Our indexing convention implies $\pi_A(i)=i$ for $i\leq r_0$ and all $A\in\mathbb{A}(D^\obs,R)$.

The original assignment variables \(D=(D_i)_{i\in[n]}\) and outcome variables \(Y=(Y_i)_{i\in[n]}\) indexed by $i$ then induce
assignment variables and outcome variables indexed by \(a\),
\[
\widetilde{D}^A_a := D_{\pi_A(a)},\quad \widetilde{Y}_a := Y_{\pi_A(a)},\quad a\in[n_A].
\]
We also introduce matching variables $\{x^A_{ai}\}_{a\in [n_A],i\in[n]}$, where $x^A_{ai}=1$ if $\pi_A(i)=a$ and $x^A_{ai}=0$ otherwise. By our indexing convention, we have $x^A_{ai}=1$ if $i=a$ and $i\leq r_0$ for every always-reporter table $A\in \mathbb{A}(D^{\obs},R)$. With the matching variables, the outcome for the $a$th always-reporter can be expressed as $Y_a=\sum_{i=1}^n x^A_{ai}Y_i$.

Each always-reporter table $A$ induces the assignment variables $\tilde{D}^A=\left(\tilde{D}^A_a\right)_{a\in [n_A]}$ and matching variables $\{x^A_{ai}\}_{a\in [n_A],i\in [n]}$. It is important to note that due to the symmetry of complete randomization, $\tilde{D}^A$ has the same distribution for all always-reporter tables with the same number of always-reporters. Motivated by this fact, we write $\tilde{D}^A$ as $\tilde{D}^{n_A}$. We denote the distribution of $\tilde{D}^{n_A}$ as $\mathcal{L}(n,n_1,n_A)$.


We now note that \(\mathcal{T}_n^0\), \(\mathcal{T}_n^1\), and \(\mathcal{T}_n^2\) depend on \((Y,D,A)\) only through \(Y\), the induced assignment vector \(\tilde{D}^{A}\), and the matching variables \(x^{A}=\{x^{A}_{ai}\}_{a,i}\). Define
\[
n_{A,1}=\sum_{a=1}^{n_A} D_a
\qquad\text{and}\qquad
n_{A,0}=\sum_{a=1}^{n_A} (1-D_a).
\]
For example, \(\mathcal{T}_n^0\) in \eqref{eqn:t0} can be written as
\begin{equation}\label{eqn:tildeTn0}
 \mathcal{T}_n^0(Y,D,A)
=\widetilde{\mathcal{T}}_n^0(Y,D^A,x^A)
=\frac{\widehat{\mu}_n^2(Y,D^A,x^A)}{\widehat{\sigma}^{2,\ha}_n\!\left(Y,D^A,x^A\right)},   
\end{equation}
where
\[
\widehat{\mu}_n(Y,D^A,x^A)
=\widehat{\mu}^1_n(Y,D^A,x^A)-\widehat{\mu}^0_n(Y,D^A,x^A),
\]
with
\[
\widehat{\mu}^1_n(Y,D^A,x^A)=\frac{1}{n_{A,1}}\sum_{a,i} D_a x_{ai} Y_i,
\quad
\widehat{\mu}^0_n(Y,D^A,x^A)=\frac{1}{n_{A,0}}\sum_{a,i} (1-D_a) x_{ai} Y_i,
\]
and
\[
\widehat{\sigma}^{2,\ha}_n\!\left(Y,D^A,x^A\right)
=\widehat{v}^1_n(Y,D^A,x^A)+\widehat{v}^0_n(Y,D^A,x^A),
\]
where
\[
\widehat{v}^1_n(Y,D^A,x^A)
=\frac{1}{n_{A,1}^2}\sum_{a,i} D_a x_{ai} Y_i^2
-\frac{1}{n_{A,1}}\Big(\widehat{\mu}^1_n(Y,D^A,x^A)\Big)^2,
\]
\[
\widehat{v}^0_n(Y,D^A,x^A)
=\frac{1}{n_{A,0}^2}\sum_{a,i} (1-D_a) x_{ai} Y_i^2
-\frac{1}{n_{A,0}}\Big(\widehat{\mu}^0_n(Y,D^A,x^A)\Big)^2.
\]

Define the space of matching variables associated with all always-reporter tables with $k$ always-reporters as
\begin{equation}\label{eqn:assignment_variables}
\begin{split}
    &x(k,D^\obs,R)=\\
    &\left\{ \{x_{ai}\}_{a\in [k],i\in [n]}: \begin{array}{ll}
            x_{ai}\in \{0,1\} & \forall a\in [k],i\in [n],\\
        x_{ai}=1, &\forall a=i, a\leq r_0\\
        \sum_{i}x_{ai}=1,& \forall a\in [k],\\
         \sum_{a}x_{ai}\leq 1, &\forall i\in [n],\\
        x_{ai}\in \{0,1\}, &\forall a\in [k], D_i^\obs=1, R_i=1,\\
        x_{ai}=0,& \forall a\in [k], D_i^\obs=1, R_i=0
    \end{array}\right\}.    
\end{split}
\end{equation}
We note that the test statistics $\T_n^1$ and $\T_n^2$ can be re-expressed, in a similar fashion, as
\begin{equation}\label{eqn:tildeTn12}
    \widetilde{\T}_n^1=\widetilde{\T}_n^0+g_1(n_{A,1}), \text{ and, } \widetilde{\T}_n^2=\widetilde{\T}_n^0+g_2(n_{A,1}).
\end{equation}
The functions $g_1$ and $g_2$ are defined in \eqref{eqn:2025102385} and \eqref{eqn:2025102386} in Appendix~\ref{section:asydist}. We omit their explicit forms for brevity.
The only point we use is that, for a fixed always-reporter size $n_A$, both $g_1$ and $g_2$ depend on the data only through $n_{A,1}$.

Lemma \ref{lemma:change_of_variable}  summarizes the discussion above.
\begin{lemma'}\label{lemma:change_of_variable}
Let $\widetilde{\T}_n$ denote one of the test statistics $\widetilde{\T}^0_n$, $\widetilde{\T}^1_n$, or $\widetilde{\T}^2_n$ defined in \eqref{eqn:tildeTn0} and \eqref{eqn:tildeTn12}. Given an always-reporter table $A=(A_i)_{i=1}^n$, observed outcomes $Y$, and observed assignments $D^{\obs}$, recall the p-value $p(A)$ defined in \eqref{eqn:pA} and the quantity $v(A,t_{i-1},t_i)$ defined in \eqref{eqn:vAtt} with endpoints $t_{i-1}$ and $t_i$.


Define the p-value associated with the matching variables $x=\{x_{ai}\}_{a,i}$ by 
\begin{equation}\label{eqn:px}
\widetilde{p}(x)
=
\E_{D^{n_A}}\!\left[
\mathds{1}\!\left\{
\widetilde{\mathcal{T}}_n\!\left(Y,\widetilde{D}^{n_A},x\right)
\ge
\widetilde{\mathcal{T}}_n\!\left(Y,\widetilde{D}^{n_A}_\obs,x\right)
\right\}
\right]    
\end{equation}

and, with endpoints $t_{i-1}$ and $t_i$, define
\begin{equation}\label{eqn:vxtt}
\widetilde{v}(x,t_{i-1},t_i)=
\begin{cases}
\displaystyle 
\E_{\widetilde{D}^{n_A}}\!\left[
\mathds{1}\!\left\{\widetilde{\mathcal{T}}_n\!\left(Y,\widetilde{D}^{n_A},x\right)\ge t_{i-1}\right\}
\right],
& \text{if } \widetilde{\mathcal{T}}_n\!\left(Y,\widetilde{D}^{n_A}_\obs,x\right)\le t_i,\\[0.5em]
-\infty, & \text{otherwise,}
\end{cases}
\end{equation}
where $\widetilde{D}^{n_A}\sim \mathcal{L}(n,n_1,n_A)$ and $\widetilde{D}^{n_A}_\obs$ is the assignment vector indexed by $a$ that is induced by the observed assignments.

Let $x^A=\{x^A_{ai}\}_{a,i}$ denote the matching variables induced by $A$. Then
\[
\widetilde{p}\!\left(x^A\right)=p(A)
\qquad\text{and}\qquad
\widetilde{v}\!\left(x^A,t_{i-1},t_i\right)=v\!\left(A,t_{i-1},t_i\right).
\]
\end{lemma'}
Lemma \ref{lemma:v} and Lemma \ref{lemma:change_of_variable} imply the following theorem.
\begin{theorem'}\label{thm:reduction}
Let $L,U\in[0,\infty]$ be two scalars satisfying \eqref{eqn:LU}, and consider a partition of the interval $[L,U]$ with increasingly ordered endpoints $\{t_i\}_{i=0}^I$, where $t_0=L$ and $t_I=U$. Recall the definition of $v_i$ from \eqref{eqn:inner_problem}. We have,
\begin{equation*}
    \max_{i\in [I]}v_i=\max_{k\in [n]}\max_{i\in [I]}\max_{x\in x(k,D^\obs,R)}v(x,t_{i-1},t_i) \geq p^{\textrm{worst}}.
\end{equation*}
\end{theorem'}
The theorem implies that the optimization problem on the left-hand side can be decomposed into smaller subproblems indexed by $k$, the number of always-reporters, and by $i$, the interval index. 

In addition, it converts the optimization over always-reporter indicators \(A=\{A_i\}_{i=1}^n\) into an optimization over the matching variables \(\{x_{ai}\}_{a,i}\). Importantly, the component of the test statistics \(\mathcal{T}^1_n\) and \(\mathcal{T}^2_n\) that assesses balance in the always-reporter indicators does not depend on \(\{x_{ai}\}_{a,i}\). To illustrate this point, consider the following IP formulation of the subproblem
\(\max_{x\in x(k,D^\obs,R)} v(x,t_{i-1},t_i)\) for some $k$ and $i$.
Let \(\{\tilde{D}^k_s\}_{s=1}^{n_{mc}}\) be \(n_{mc}\) Monte Carlo draws of the assignment vector (indexed by \(a\)) from \(\mathcal{L}(n,n_1,k)\). The subproblem can be written as
\begin{equation}
    \max_{\{x_{ai}\}_{a,i},\{I_s\}_{s=1}^{n_{mc}}} \sum_{s=1}^{n_{mc}} I_s,
\end{equation}
subject to
\begin{align*}
  &  \widehat{\mu}^2_n\!\left(Y,D^k_s,x\right)
  +\Bigl(g_i(n_{A,s}^1)-t_{i-1}\Bigr)\widehat{\sigma}^{2,\ha}_n\!\left(Y,D^k_s,x\right)
  \ \geq\ L_s\,(1-I_s), \qquad \forall s\in [n_{mc}],\\
  &  \widehat{\mu}^2_n\!\left(Y,D^k_{\obs},x\right)
  +\Bigl(g_i(n_{A,\obs}^1)-t_{i}\Bigr)\widehat{\sigma}^{2,\ha}_n\!\left(Y,D^k_{\obs},x\right)
  \ \leq\ 0,\\
  & x\in x(k,D^\obs,R), \qquad I_s\in \{0,1\}, \qquad \forall s\in [n_{mc}],
\end{align*}
where (i) the functions \(g_i(\cdot)\), $i\in \{0,1,2\}$, correspond to different test statistics $\T_n^0, \T_n^1$ and $\T_n^2$, and are defined in
\eqref{eqn:2025102385} and \eqref{eqn:2025102386} and; (ii) each \(L_s\) is any constant satisfying
\[
L_s \ \leq\ \min_{x\in x(k,D^\obs,R)}
\Bigl\{
\widehat{\mu}^2_n\!\left(Y,D^A_s,x\right)
+\bigl(g_i(n_{A,s}^1)-t_{i-1}\bigr)\widehat{\sigma}^{2,\ha}_n\!\left(Y,D^A_s,x\right)
\Bigr\},
\]
and the collection \(\{L_s\}_{s=1}^{n_{mc}}\) can be obtained either analytically or computationally; (iii) the indicator \(I_s\) must be set to zero whenever
\begin{equation}
\frac{\widehat{\mu}^2_n\!\left(Y,D^A_s,x\right)}{\widehat{\sigma}^{2,\ha}_n\!\left(Y,D^A_s,x\right)}
+g_i(n_{A,s}^1)<t_{i-1} ,
\footnote{We implicitly assume \(\widehat{\sigma}^{2,\ha}_n\!\left(Y,D^A_s,x\right)>0\) for all \(x\in x(k,D^\obs,R)\) and all \(s\in [n_{mc}]\). This condition is often satisfied empirically and can be verified quickly.}
\end{equation}
and is otherwise unconstrained.

We note that \(g_i(n_{A,s}^1)\) is independent of the decision variables and can therefore be treated as a known scalar for each \(s\). Moreover, both
\(\widehat{\mu}^2_n(Y,D^A_s,x)\) and \(\widehat{\sigma}^{2,\ha}_n(Y,D^A_s,x)\)
are quadratic functions of the matching variables \(x\). By contrast, obtaining a formulation with comparable structure in the original decision-variable space
\(A=(A_i)_{i=1}^n\) appears challenging.

We conclude by noting that the reduction in Section \ref{section:step2}, together with a technique analogous to that in Section \ref{section:Step1}, can be combined with a bisection search to solve Algorithm \ref{alg:asy}, which relies on critical values from the asymptotic distribution for inference. In practice, we find the resulting computations to be highly efficient. This approach is useful because it provides a fast heuristic that delivers a reasonable lower bound for the worst-case Monte Carlo $p$-value \(p^{\textrm{worst,mc}}\). The resulting solution can be used to warm-start the integer-programming method to accelerate computation, or to certify non-rejection at the chosen significance level. 

Readers can find a complete pseudo algorithm incorporating the discussions above in Algorithm \ref{alg:12152025_c}.

\section{Simulation Results}\label{section:simulation}
This section evaluates the finite-sample performance and computational cost of the proposed worst-case randomization tests.
We report (i) empirical rejection rates under the sharp-null scenario (test size), (ii) empirical rejection rates under an alternative with a positive AR-ATE (power), and (iii) runtime of the implementation.

Each simulated dataset contains $n=100$ units, with complete randomization assigning $n_1=50$ to treatment and $n_0=50$ to control. Units belong to one of three principal reporting strata under Assumption~\ref{a:monotonicity}.
We set the population shares to
\[
\pi_{\mathrm{AR}}=0.9,\qquad \pi_{\mathrm{IR}}=0.05,\qquad \pi_{\mathrm{NR}}=0.05.
\]

We generate potential outcomes for all units according to the following model:
\[
y_i(0) \sim \textrm{N}(0,1), \qquad y_i(1)=y_i(0)+\tau,
\]
where $\textrm{N}(0,1)$ denote a standard normal random variable and $\tau$ is the treatemtent effect. We consider two scenarios. $\tau=0$ and $\tau=1$.

We implement the worst-case randomization test in Algorithm~\ref{alg:inf} using the following the one-sided chi-square statistics $\mathcal T^2_n$ in \eqref{eqn:t2}. We choose the pruning step at $\beta=0.05$ and the nominal size at $\alpha=0.05$. The number of simulated randomization draws is 1000. All computations were performed on a Linux compute node equipped with an Intel Xeon Platinum 8268 CPU (2.90 GHz), providing 8 logical cores.

In the null scenario (ATE 
=0), the rejection rate is 
0.002. In most simulations, the algorithm terminates at the heuristic stage once a reporting configuration yields a randomization p-value above the rejection threshold, resulting in a median runtime of 35.26 seconds and a 90th percentile runtime of 72.11 seconds. There are rare cases where the null is rejected and the algorithm must exhaustively verify all admissible configurations; in two such instances this required roughly 10 hours.

In the alternative scenario (ATE = 1), the rejection rate is 0.9218. This scenario sees a median runtime of 308 seconds and a 90th percentile runtime of 7,080 seconds. Relative to the null case, a larger share of simulations require additional verification beyond the heuristic stage, resulting in longer runtimes and a substantially higher maximum computation time of approximately 75,000 seconds.

In general, we find that when the lower bound of the test statistic produced by the heuristic stage lies in the range of approximately 6–9, the algorithm may require substantial computation time to verify the rejection decision. By contrast, when the null hypothesis is not rejected or when the hypothesis is rejected with a sufficiently large test statistic (e.g., greater than or equal to 9), the algorithm typically terminates quickly.

\begin{table}[t]
\centering
\caption{Rejection Rates and Runtime for the Worst-Case Test Using $\mathcal T^2_n$}\label{tab:sim_T2}
\begin{tabular}{ccccc}
\hline
ATE 
& Rej. Rate 
& Median Time (s) 
& 90th Pct. Time (s) 
& Max Time (s) \\
\hline
$0$ & 0.002 &   35.26&   72.11 &   37956 \\
$1$ & 0.9218  & 308 & 7080 &  75068\\
\hline
\end{tabular}

\begin{flushleft}
\footnotesize
Notes: Simulations use $n=100$ with $n_1=50$ under complete randomization and principal strata shares $(\pi_{\mathrm{AR}},\pi_{\mathrm{IR}},\pi_{\mathrm{NR}})=(0.9,0.05,0.05)$.
Each design is evaluated using $1{,}000$ Monte Carlo repetitions and $1{,}000$ random treatment assignments per repetition.
The nominal test level is $\alpha=0.05$, and the pretest level is set to $\beta=0.005$.
Rejection rates are computed using the worst-case randomization test based on the statistic $\mathcal T^2_n$.
The reported runtimes correspond to the median, 90th percentile, and maximum wall-clock time per dataset.
\end{flushleft}
\end{table}

\appendix
\section{Proofs}
This section contains the proof of Theorem \ref{thm:main}, which is included in Section \ref{section:prooftheoremmain}. The proof requires multiple preliminary steps: 
\begin{enumerate}
    \item A study of asymptotic distributions of $\T^0_n$, $\T^1_n$ and $\T^2_n$ in Section \ref{section:asydist} and Section \ref{section:var_proof}. Th key theorem is Theorem \ref{thm:feasible_clt}.
    \item A study of randomization-based critical value in Section \ref{section:rand_cv}. The key theorem is Theorem \ref{thm:thmA24}.
\end{enumerate}
Auxiliary lemmas are included in Section \ref{section:auxiliary_lemma}. 

The key technical challenge is handling the random variable \(n_A^1=\sum_{i=1}^n D_iA_i\), the number of treated always-reporters. A CLT for \(n_A^1\) in the uniform sense does not hold under Assumption \ref{a:theta}; for instance, the sample may consist entirely of always-reporters. This rules out applying a combinatorial CLT to justify asymptotic validity as in \citep{wu2021randomization, aronow2024randomization}. Instead, we pursue a careful conditional analysis and invoke recent Berry--Esseen bounds for the combinatorial CLT developed by \citep{shi2022berry}.

\subsection{Notations}
Throughout the appendix, we use $Z\sim \N(0,1)$ to denote a standard normal variables with mean zero and unit variance. 

We write \(\mathrm{CR}(n,n_1),n\geq n_1\) for the distribution of a random vector \(D=(D_i)_{i=1}^n\in\{0,1\}^n\) that selects exactly \(n_1\) of the \(n\) units to have \(D_i=1\), with the remaining \(n_0=n-n_1\) units having \(D_i=0\), uniformly over all such assignments.

Unless stated otherwise, all probability measures $\mathbb{P}_n$ are taken under the complete randomization distribution of the assignment vector $D=(D_i)_{i=1}^n\sim \mathrm{CR}(n,n_1)$, with $n$ and $n_1$ understood from context. 

Quantities such as the number of always-reporters $n_A$ and the number of treated units $n_1$ are inherently dependent on $n$. Unless noted otherwise, we suppress their notational dependence on $n$.

\subsection{Asymptotic Distributions of $\T^0_n$, $\T^1_n$ and $\T^2_n$}\label{section:asydist}

Section \ref{section:beb} states a Berry-Esseen bound from \cite{shi2022berry} adapted to our setting. Section \ref{section:ap} states a probabilistic bound for the number of always-reporters in the treated and control group. Section \ref{section:conditional_hajek} studies the conditional properties of the Hajek estimator, conditioning on the number of always-reporters in the treated group.

\subsubsection{Berry-Esseen Bound for the Combinatorial Central Limit Theorem}\label{section:beb}
Theorem \ref{thm:beb} gives a Berry–Esseen bound for a two-arm completely randomized design, adapted from \cite{shi2022berry}.
\begin{theorem'}\label{thm:beb}
Given potential outcomes $\{(w_i(1),w_i(0))\}_{i=1}^n$ and a completely randomized design $\textrm{CR}(n,n_1)$, consider the difference-in-means estimator:
\begin{equation*}
    \widehat{\tau}_n=\frac{1}{n_1}\sum_{i=1}^n D_iw_i(1)-\frac{1}{n_0}\sum_{i=1}^n \left(1-D_i\right)w_i(0),
\end{equation*}
for $\tau_n=n^{-1}\sum_{i=1}^n \left(w_i(1)-w_i(0)\right)$. Denote the variance of $\widehat{\tau}_n$ as
\begin{equation*}
    \AVar_n=\AVar_n\left(\widehat{\tau}_n\right)=\frac{1}{n_1}v^1+\frac{1}{n_0}v^0 - \frac{1}{n}v^{01},
\end{equation*}
where,
    \begin{align*}
    v^1 &= \frac{1}{n-1}\sum_{i=1}^n \left( w_i(1)- \overline{w}(1)\right)^2, v^0 = \frac{1}{n-1}\sum_{i=1}^n\left( w_i(0)- \overline{w}(0)\right)^2,\\ 
    v^{01}&=\frac{1}{n-1}\sum_{i=1}^n \left(w_i(1)-w_i(0)-\left(\overline{w}(1)-\overline{w}(0)\right)\right)^2,
\end{align*}
with $\overline{w}(1)=n^{-1}\sum_{i=1}^n w_i(1)$ and $\overline{w}(0)=n^{-1}\sum_{i=1}^n w_i(0)$. 
Suppose we have 
\begin{equation*}
    \AVar_n\left(\widehat{\tau}_n\right) \geq c^{-2}\left( \frac{1}{n_1}v^1+\frac{1}{n_0}v^0\right)
\end{equation*}
for some $c\geq 1$. Then there exists a universal constant $C$, which may depend on $c$, such that
\begin{equation*}
    \sup_{t\in \mathbb{R}}\left| \P\left( \AVar_n^{-\frac{1}{2}} \left(\widehat{\tau}_n-\tau_n\right) \leq t  \right)-\P\left( Z\leq t\right)\right| \leq C \max_{a\in \{0,1\}} \max_{i\in [n]}\frac{\left|w_i(a)-\overline{w}(a)\right|}{ \sqrt{n_av^a}}
\end{equation*}
\end{theorem'}
\begin{proof}
    The result follows from Theorem 1-(ii) in \cite{shi2022berry}, with $F=(1,-1)$, $b=1$.
\end{proof}
It should be noted if we consider $-\widehat{\tau}_n$ instead of $\widehat{\tau}_n$, the same inequality holds, i.e.,
\begin{equation*}
    \sup_{t\in \mathbb{R}}\left| \P\left( \AVar_n^{-\frac{1}{2}} \left(\widehat{\tau}_n-\tau_n\right) \geq t  \right)-\P\left( Z\geq t\right)\right| \leq C \max_{a\in \{0,1\}} \max_{i\in [n]}\frac{\left|w_i(a)-\overline{w}(a)\right|}{ \sqrt{n_av^a}}
\end{equation*}
\subsubsection{Probabilistic Bounds on Always-Reporter Counts in Treatment and Control Groups}\label{section:ap}
Given a vector of always-reporter indicators $A=\left(A_i\right)_{i=1}^n$, recall the definition of $\dimA$ from (\ref{eqn:dima}):
\begin{align*}
     \dimA(D,A) = &\frac{1}{n_1}\sum_{i=1}^nD_iA_i - \frac{1}{n_0}\sum_{i=1}^n \left(1-D_i\right)A_i.
\end{align*}
Suppose $D\sim \textrm{CR}(n,n_1)$. The variance of $\dimA$ can be shown to be
\begin{equation}\label{eqn:variance_A2}
   \VA= \frac{n^2}{n_1n_0(n-1)} \left( \frac{1}{n}\sum_{i=1}^n A_i - \left(\frac{1}{n}\sum_{i=1}^nA_i\right)^2  \right).
\end{equation}
Lemma \ref{lemma:var_bound_A} provides an upper-bound on the variance term.
\begin{lemma'}\label{lemma:var_bound_A}
\begin{equation*}
    \max_{\{A_i\}_{i=1}^n\subset \{0,1\}^n} \frac{1}{n}\sum_{i=1}^n A_i - \left(\frac{1}{n}\sum_{i=1}^nA_i\right)^2 \leq \frac{1}{4} 
\end{equation*}
\end{lemma'}
\begin{proof}
The maximization problem reduces to maximizing $a - a^{2}$ over $a \in [0,1]$, where $a = \frac{1}{n}\sum_{i=1}^{n} A_i$. The maximum equals $1/4$ and is attained at $a = 1/2$.
\end{proof}

Lemma \ref{lemma:nonnegligible_share} states that, under mild regularity conditions, the numbers of always-reporters in the treated and control groups are non-negligible with probability tending to 1.
\begin{lemma'}\label{lemma:nonnegligible_share}
Consider a vector of always–reporter indicators $A=\left(A_i\right)_{i=1}^n$, and a completely randomized designs $D=\left(D_i\right)_{i=1}^n\sim \mathrm{CR}(n,n_{1})$. Let $n_{0}=n-n_{1}$ and $n_{A}=\sum_{i=1}^n A_i$. Suppose the following conditions hold:
\begin{enumerate}[label=(\roman*)]
    \item $n_{1}/n\in [r,1-r]$ with some $r\in (0,\frac{1}{2}]$; \label{lemma:ns1}
    \item $n_{A} \geq sn$ for some $s\in (0,1]$.\label{lemma:ns2}
\end{enumerate}
We have
\begin{equation*}
   \P_n\left( \sum_{i=1}^n D_iA_i\leq  \frac{n_{1}n_{A}}{2n}\right)\leq  \frac{1-r}{rs^2(n-1)},  
\end{equation*}
and,
\begin{equation*}
  \P_n\left( \sum_{i=1}^n \left(1-D_i\right)A_i\leq \frac{n_{0}n_{A}}{2n}\right)\leq  \frac{1-r}{rs^2(n-1)}.
\end{equation*}
\end{lemma'}
\begin{proof}
We have the following calculations
\begin{equation}\label{eqn:measure}
    \begin{split}
          & \P_n\left( \frac{\sum_{i=1}^n D_i A_i}{n_{1}} - \frac{\sum_{i=1}^n \left(1-D_i\right) A_i}{n_{0}}\leq -\frac{n_A}{2n_{0}} \right)\\
    = & \P_n\left( \frac{n\sum_{i=1}^n D_i A_i}{n_{1}n_{0}} - \frac{n_A}{n_{0}}\leq -\frac{n_A}{2n_{0}} \right) =  \P_n\left( \frac{n\sum_{i=1}^n D_i A_i}{n_{1}n_{0}}\leq \frac{n_A}{2n_{0}} \right) \nonumber\\
    = &\P_n\left(\sum_{i=1}^n D_i A_i\leq \frac{n_A n_{1}}{2n} \right).
    \end{split}
\end{equation}
We then have
\begin{align}
         & \P_n\left( \frac{\sum_{i=1}^n D_i A_i}{n_{1}} - \frac{\sum_{i=1}^n \left(1-D_i\right) A_i}{n_{0}}\leq -\frac{n_A}{2n_{0}} \right)\nonumber \\
 \leq  &\P_n\left( \left|\frac{\sum_{i=1}^n D_i A_i}{n_{1}} - \frac{\sum_{i=1}^n \left(1-D_i\right) A_i}{n_{0}}\right|\geq \frac{n_A}{2n_{0}} \right) \leq \frac{\VAn}{{n_A}^2\left(2n_{0}\right)^{-2}} \label{eqn:chebyshev1}\\
 \leq & \frac{n^2}{4n_{1}n_{0}(n-1)}\frac{1}{{n_A}^2\left(2n_{0}\right)^{-2}}= \frac{n^2n_0}{(n-1)n_{1}n_A^2}\leq \frac{\left(1-r\right)n^3}{2rs^2(n-1)n^3} \label{eqn:worst_case_bound}\\
 \leq & \frac{1-r}{rs^2(n-1)},\nonumber
\end{align}
where $\VAn$ is defined in (\ref{eqn:variance_A2}), (\ref{eqn:chebyshev1}) is by the Chebyshev inequality, and (\ref{eqn:worst_case_bound}) follows from Lemma \ref{lemma:var_bound_A} and premises \ref{lemma:ns1} and \ref{lemma:ns2}. The statement for $\sum_{i=1}^n \left(1-D_i\right)A_i$ can be proved analogously
\end{proof}

\subsubsection{Conditional Properties of the Hajek Estimator}\label{section:conditional_hajek}
The following lemma says that under a completely randomized assignment of all units, the assignments for the always-reporters, conditional on exactly $k$ of them being treated, are themselves completely randomized as well.
\begin{lemma'}\label{lemma:cr}
Given a vector of always–reporter binary indicators $A=\left(A_i\right)_{i=1}^n $, denote $\A=\{i\in[n]:A_i=1\}$ and $n_A=\sum_{i=1}^n A_i$. Suppose $(D_i)_{i=1}^n \sim \textrm{CR}(n,n_1)$, then nonnegative integer $k\leq n_A$,
\begin{equation*}
    (D_i)_{i\in\A}\bigg\vert\sum_{i=1}^n D_iA_i=k \sim \textrm{CR}(n_A,k).
\end{equation*}
\end{lemma'}
\begin{proof}
First, consider $\P_n \left( \sum_{i = 1}^{n} D_{i} A_{i} = k \right)$. We have
\begin{align*}
    \P_n\left( \sum_{i = 1}^{n} D_{i} A_{i} = k \right) &= \P_n\left( \sum_{i = 1}^{n} D_{i} A_{i} = k, \sum_{i = 1}^{n} D_{i} (1 - A_{i}) = n_{1} - k \right) \\
    &= \frac{\binom{n_{A}}{k} \binom{n - n_{A}}{n_{1} - k}}{\binom{n}{n_{1}}},
\end{align*}
where the first equality follows from the fact that $(D_i)_{i=1}^n \sim \textrm{CR}(n,n_1)$, and the second equality follows by inspection.

Next, consider $\P_{n} \left( (D_i)_{i\in\A}, \sum_{i = 1}^{n} D_{i} A_{i} = k \right)$. We have
\begin{align*}
    \P_{n} \left( (D_i)_{i\in\A}, \sum_{i = 1}^{n} D_{i} A_{i} = k \right) &= \P_{n} \left( (D_i)_{i\in\A}, \sum_{i = 1}^{n} D_{i} A_{i} = k, \sum_{i = 1}^{n} D_{i} (1 - A_{i}) = n_{1} - k \right) \\
    &= \frac{\binom{n - n_{A}}{n_{1} - k}}{\binom{n}{n_{1}}},
\end{align*}
where the first equality follows from the fact that $(D_i)_{i=1}^n \sim \textrm{CR}(n,n_1)$, and the second equality follows by inspection.

The desired result follows immediately.
\end{proof}

We note that the Hajek estimator conditioning on the event that $\sum_{i=1}^n D_iA_i=k$, with $1\leq k\leq n_A-1$, can be expressed as:
\begin{equation}\label{eqn:hajek_ak}
      \arhjk(Y,D,A)= \frac{1}{k}\sum_{i=1}^n D_iA_iY_i- \frac{1}{n_A-k}\sum_{i=1}^n \left(1-D_i\right)A_iY_i.
\end{equation}

Lemma \ref{lemma:conditional_mean_variance} collects the conditional mean and variance characterizations of the Hajek estimator. They immediately follow from Lemma \ref{lemma:cr} and a standard calculation \cite{imbens2015causal}.
\begin{lemma'}[Conditional Mean and Variance]\label{lemma:conditional_mean_variance}
Suppose we are given a vector of always-reporter indicators $A=\left(A_i\right)_{i=1}^n$,  outcomes $\{\left(y_i(1),y_i(0)\right)\}_{i=1}^n$, and a completely randomized design $D=\left(D_i\right)_{i=1}^n\sim \textrm{CR}(n,n_1)$. Consider the estimators $\arhj $ defined in (\ref{eqn:hajek_a}) and $\arhjk$ defined in (\ref{eqn:hajek_ak}). Define $\mathcal{A}=\{i\in[n]:A_i=1\}$ and $n_A=\sum_{i=1}^n A_i$. Suppose $n_A\geq 1$ and $1\leq k\leq n_A-1$.  We have:
\begin{enumerate}[label=(\roman*)]
    \item $\E\left[\arhj(Y,D,A)\Big|\sum_{i=1}^n D_iA_i=k\right]$ = $\E\left[\arhjk(Y,D,A)\Big|\sum_{i=1}^n D_iA_i=k\right]$
    \item $\AVar_n\left(\arhj\Big|\sum_{i=1}^n D_iA_i=k\right)=k^{-1}v^1_{A}+\left(n_A-k\right)^{-1}v^0_{A}-n_A^{-1}v^{01}_{A}$,
    where
\end{enumerate}
    \begin{align}
    v^1_{A} &= \frac{1}{n_A-1}\sum_{i\in \A} \left( y_i(1)- \overline{y}_{A}(1)\right)^2, v^0_{A}= \frac{1}{n_A-1}\sum_{i\in \A} \left( y_i(0)- \overline{y}_{A}(0)\right)^2,\label{eqn:va}\\ 
    v^{01}_{A}&=\frac{1}{n_A-1}\sum_{i\in \A} \left(y_i(1)-y_i(0)-\left(\overline{y}_{A}(1)-\overline{y}_{A}(0)\right)\right)^2,\label{eqn:va01}
\end{align}
with $\overline{y}_{A}(1)=n_A^{-1}\sum_{i\in \A}y_i(1)$ and  $\overline{y}_{A}(0)=n_A^{-1}\sum_{i\in \A}y_i(0)$. If $n_A=1$, all variances and covariances are defined to be zero.
\end{lemma'}
We shall write the conditional variance of the Hajek estimator as 
\begin{equation}\label{eqn:AVnk}
    \AVar_{n,k}\left(\arhj\right)=\AVar_n\left(\arhj\Big|\sum_{i=1}^n D_iA_i=k\right),
\end{equation}
viewing it as a function of $k$. In the remaining section, we let $\delta, s, r, B$ be the constants fixed in the statement of Theorem \ref{thm:main}, and use the shorthand notation $\Theta^w_n =\Theta^w_n (\delta, s, r , B)$. We typically denote an element in $\Theta_n^w$ by $\theta_n^w$. Lemma \ref{lemma:nv} collects implications of the assumptions on $\Theta_n^w$.
\begin{lemma'}\label{lemma:nv}
For all $n\in \mathbb{N}$ and each $\theta^w_n \in \Theta^w_n $, let $\{\left(y_i(1),y_i(0)\right)\}_{i=1}^n$ be the associated potential outcomes and $\{A_i\}_{i=1}^n$ be the associated always-reporter indicators. 
Denote $\A=\{i\in[n]:A_i=1\}$ and $n_A=\sum_{i=1}^n A_i$. Define $v_A^1$, $v_A^0$ and $v_A^{01}$ as in (\ref{eqn:va}) and (\ref{eqn:va01}), and $\overline{y}_{A}(a)=n_A^{-1}\sum_{i\in\A}y_i(a)$ for $a\in\{0,1\}$. Let $k$ be an integer such that $k\in [1,n_A-1]$ and $n_A\geq 2$.

\begin{enumerate}[label=(\roman*)]
    \item The following inequality for the conditional variance holds:\label{eqn:nv1}
\begin{equation}\label{eqn:nv1_eq}
    \AVar_{n,k}\left(\arhj\right)  \geq  \frac{(1-\delta)}{n_A}\left( \frac{n_A-k}{k}v^1_A+\frac{k}{n_A-k}v^0_A\right).
\end{equation}    
\item We have the following inequality: for $a\in \{0,1\},$\label{eqn:nv2}
\begin{equation}
     \frac{\max_{i\in \A} \left| y_i(a)-\overline{y}_{A}(a) \right|}{\sqrt{v_A^a}}   \leq B  n_A^{1/4},
\end{equation}
where $B$ is the constant defined in Assumption\ref{a:theta}-\ref{a:theta3}. 
\end{enumerate}
\end{lemma'}

\begin{proof}
First note,
 \begin{align*}
    v^{01}_{A}&=\frac{1}{n_A-1}\sum_{i\in \A} \left(y_i(1)-y_i(0)-\left(\overline{y}_{A}(1)-\overline{y}_{A}(0)\right)\right)^2  \\
    & =   v_A^1 +v_A^0 -  \frac{2}{n_A-1}\sum_{i\in \A} \left(y_i(1)-\overline{y}_{A}(1)\right)\left(y_i(0)-\overline{y}_{A}(0)\right)\\
     &\leq   v_A^1 +  v_A^0 +  2\delta  \sqrt{v_A^1 v_A^0}\leq v_A^1 + v_A^0+  \frac{\delta \left(n_A-k\right)}{k} v_A^1+  \frac{\delta k}{n_A-k}  v_A^0,
 \end{align*}
where the first inequality is by Assumption \ref{a:theta}-\ref{a:theta2} and the second inequality is by the inequality $2ab\leq a^2+b^2$.
\ref{eqn:nv1} follows by the calculation,
\begin{align*}
   &\frac{1}{k} v^1_{A} +  \frac{1}{n_A-k}v^0_{A}-\frac{1}{n_A} v^{01}_{A}\\
   = & \frac{1}{k} v^1_{A} +  \frac{1}{n_A-k}v^0_{A} - \frac{1}{n_A}\left(v_A^1 + v_A^0+  \frac{\delta \left(n_A-k\right)}{k} v_A^1+  \frac{\delta k}{n_A-k}  v_A^0\right)\\
   = & \frac{1-\delta}{n_A}\left( \frac{n_A-k}{k}v^1_A+  \frac{k}{n_A-k} v^0_A\right).
\end{align*}

\ref{eqn:nv2} follows by the calculation, for each $a\in \{0,1\}$,
\begin{align*}
 & \max_{i\in \A} \vert y_i(a)-\overline{y}_{A}(a)\vert
 =  \max_{i\in \A}  \left(\vert y_i(a)-\overline{y}_{\A}(a)\vert^{4}  \right)^{1/4}\\
 \leq &  \left(n_A\right)^{1/4}\left(\frac{1}{n_A}\sum_{i\in \A} \left( y_i(a)-\overline{y}_{A}(a)\right)^{4}\right)^{1/4}
  \leq B  n_A^{1/4}\sqrt{v_A^a},
\end{align*}
where the last inequality follows by Assumption \ref{a:theta}-\ref{a:theta3}. $v_A^a$ are positive by Assumption \ref{a:theta}-\ref{a:theta2} for $a\in\{0,1\}$ and hence we can divide on both side.
\end{proof}

\begin{lemma'}\label{lemma:CLT}
For all $n\in \mathbb{N}$ and each $\theta_n^w\in \Theta_n^w$, define the vector of always-reporter indicators $A=(A_i)_{i=1}^n$,  $n_A=\sum_{i=1}^n A_i$ and $n_{A}^1=\sum_{i=1}^n D_iA_i$, where $D=\{D_i\}_{i=1}^n\sim\textrm{CR}(n,n_1)$. Define $\tau_n=n_A^{-1}\sum_{i=1}^n A_i\left(y_i(1)-y_i(0)\right)$. 

Let $g:\mathbb{R}\to \mathbb{R}$ be an arbitrary function. Let $Z\sim \textrm{N}(0,1)$ be a standard normal random variable independent of $D$. Define $\arhj$ as in (\ref{eqn:hajek_a}) and $\AVar_{n,k}\left(\arhj\right)$ as in (\ref{eqn:AVnk}).\footnote{When $k=1$ or $k=n_A$, we shall abuse the notation and define $\AVar_{n,k}\left(\arhj\right)=\epsilon$ for some arbitrary $\epsilon>0$. These events happen with probability approaching 0, as we shown in the proof.}

For $n_A\geq 2$, we have the inequality, for every nonnegative real numbers $t$ and $x$,
\begin{align*}
  & \left|\P_{\theta_n^w}\left(  \frac{t\left(\arhj-\tau_n\right)^2}{\AVar_{n,n_{A}^1}\left(\arhj\right)} + g\left(n_{A}^1\right) \geq x\right) - \P_{\theta_n^w}\left(  tZ^2 + g\left(n_{A}^1\right)\geq x\right)\right|\\
  \leq &  \frac{2\left(1-r\right)}{rs^2(n-1)} + \sqrt{\frac{2}{r}}C_{\ref{eqn:53}}(\delta,r)\max_{a\in \{0,1\}} \max_{i\in \A}\frac{\left|y_i(a)-\overline{y}_A(a)\right|}{\sqrt{n_{A}v^a_A}}\\
 \leq  & \frac{2\left(1-r\right)}{rs^2(n-1)} + \sqrt{\frac{2}{r}}C_{\ref{eqn:53}}(\delta,r)\times B n_A^{-\frac{1}{4}},
\end{align*}
where $B$ is the constant defined in Assumption \ref{a:theta}-\ref{a:theta3} , $C_{\ref{eqn:53}}(\delta,r)$ is a constant that depends on $\delta$ and $r$ defined in Assumption \ref{a:theta}-\ref{a:theta2} and Assumption \ref{a:cr}, respectively, and $s$ is the constant defined in Assumption\ref{a:theta}-\ref{a:theta1}.
\end{lemma'}
\begin{proof}
We suppress the dependence on $\theta_n^w$ for simplicity. If $t=0$, the inequality trivially holds. We shall assume $t>0$.
We have the identity
\begin{align*}
&\E\left[  \mathds{1}\left\{\frac{t\left(\arhj-\tau_n\right)^2}{\AVar_{n,n_{A}^1}\left(\arhj\right)} + g\left(n_{A}^1\right) \geq x\right\}\right] - \E\left[\mathds{1}\left\{ tZ^2 + g\left( n_{A}^1\right)\geq x\right\}\right]\\
=&\E\left[  \E\left[\mathds{1}\left\{\frac{t\left(\arhj-\tau_n\right)^2}{\AVar_{n,n_{A}^1}\left(\arhj\right)} + g\left(n_{A}^1\right) \geq x\right\} \Bigg|n_A^1\right]-\E\left[\mathds{1}\left\{ tZ^2 + g\left( n_{A}^1\right)\geq x\right)\Bigg| n_A^1\right]\right]\\
=&\E\left[  \P\left(\frac{t\left(\arhj-\tau_n\right)^2}{\AVar_{n,n_A^1}\left(\arhj\right)} + g\left(n_A^1\right) \geq x\Bigg|n_A^1\right) - \P\left(tZ^2 + g\left(n_A^1\right) \geq x\Bigg|n_A^1\right)\right]\\
= & \E\left[  \P\left(\frac{\left(\arhj-\tau_n\right)^2}{\AVar_{n,n_A^1}\left(\arhj\right)}  \geq \frac{x-g\left(n_A^1\right)}{t}\Bigg|n_A^1\right) -  \P\left(Z^2  \geq \frac{x -g\left(n_A^1\right)}{t}\Bigg|n_A^1\right)\right]
\end{align*}

Recall that $n_A^1=\sum_{i=1}^n D_iA_i$ and $n_A^0=n_A-n_A^1=\sum_{i=1}^n\left(1-D_i\right)A_i$. Define events 
\begin{align}
    &  \mathcal{E}=\left\{n_A^1\leq \frac{n_A n_{1}}{2n}\right\} \cup  \left\{n_A^0\leq \frac{n_A n_{0}}{2n}\right\},\label{eqn:E}\\
    & \mathcal{E}^c=\left\{n_A^1> \frac{n_A n_{1}}{2n}\right\} \cap \left\{ n_A^0> \frac{n_A n_{0}}{2n}\right\}.\label{eqn:Ec}
\end{align}
On the event $\mathcal{E}^c$ and by Assumption \ref{a:cr}, we have,
\begin{equation}\label{eqn:size_lower_bound}
   \min\left\{n_A^1 , n_A^0\right\} \geq \frac{rn_A}{2}, \text{and, } \frac{1}{n_A}\min\left\{n_A^1 , n_A^0\right\} \geq \frac{r}{2}. 
\end{equation}
Hence on the event $\mathcal{E}^c$ we have
\begin{align*}
    \AVar_{n,n_A^1}\left(\arhj\right)  &\geq \min\left\{ \frac{\left(1-\delta\right)n_A^1}{n_A},\frac{\left(1-\delta\right)n_A^0}{n_A}\right\}\left( \frac{1}{n_A^1}v^1_A+\frac{1}{n_A^0}v^0_A\right)\\
    & \geq \frac{\left(1-\delta\right)r}{2}\left( \frac{1}{n_A^1}v^1_A+\frac{1}{n_A^0}v^0_A\right).
\end{align*}
For a fixed $n_A^1$, the premise of Theorem \ref{thm:beb} is satisfied with $c^{-2}=\left(1-\delta\right)r/2$. Conditioning on the event $\mathcal{E}^c$ and for a fixed $n_A^1$, we have the calculation


\begin{align}
     & \left|\P\left(\frac{\left(\arhj-\tau_n\right)^2}{\AVar_{n,n_A^1}\left(\arhj\right)}  \geq \frac{x-g\left(n_A^1\right)}{t}\Bigg|n_A^1\right)-\P\left(Z^2  \geq \frac{x -g\left(n_A^1\right)}{t}\Bigg|n_A^1\right)\right|\nonumber\\
     \leq &  \sup_{q\in\mathbb{R}}\left|\P\left(\frac{\left(\arhj-\tau_n\right)^2}{\AVar_{n,n_A^1}\left(\arhj\right)}  \geq q\Bigg|n_A^1\right)-\P\left(Z^2  \geq q\right)\right|\nonumber\\
       =& \sup_{q\in\mathbb{R}^+}\left|\P\left(\AVar_{n,n_A^1}^{-\frac{1}{2}}\left(\arhj\right)\left|\arhj-\tau_n\right|\geq \sqrt{q}\Bigg|n_A^1\right)-\P\left(|Z|  \geq \sqrt{q}\right)\right|\nonumber\\
     \leq & \sup_{t\in\mathbb{R}}\left|\P\left(\AVar_{n,n_A^1}^{-\frac{1}{2}}\left(\arhj\right)\left(\arhj-\tau_n\right)\leq t\Bigg|n_A^1\right)-\P\left(Z \leq t\right)\right|\nonumber\\
      & + \sup_{t\in\mathbb{R}}\left|\P\left(\AVar_{n,n_A^1}^{-\frac{1}{2}}\left(\arhj\right)\left(\arhj-\tau_n\right)\geq t\Bigg|n_A^1\right)-\P\left(Z  \geq t\right)\right| \nonumber \\
     \leq & C_{\ref{eqn:53}}(\delta,r) \max_{a\in \{0,1\}} \max_{i\in \A}\frac{\left|y_i(a)-\overline{y}_A(a)\right|}{\sqrt{n_{A}^av^a_A}},    \label{eqn:beb}
\end{align}
by Theorem \ref{thm:beb} and Lemma \ref{lemma:nv}-\ref{lemma:ns1}, where the constant $C$ may depend on $\delta$ and $r$.


We have the following calculation:
\begin{align}
& \left|\E\left[  \mathds{1}\left\{\frac{t\left(\arhj-\tau_n\right)^2}{\AVar_{n,n_{A}^1}\left(\arhj\right)} + g\left(n_{A}^1\right) \geq x\right\}\right] - \E\left[\mathds{1}\left\{ tZ^2 + g\left( n_{A}^1\right)\geq x\right\}\right]\right| \nonumber \\
\leq & \P\left(\mathcal{E} \right) +\P\left(\mathcal{E}^c \right)\times \nonumber \\
 &  \E\left[ \sup_{t\in\mathbb{R}}\left.\left|\P\left(\frac{\left(\arhj-\tau_n\right)^2}{\AVar_{n,n_A^1}\left(\arhj\right)}  \geq t\Bigg|n_A^1\right)-\P\left(Z^2  \geq t\Bigg|n_A^1\right)\right| \right|\mathcal{E}^c\right] \nonumber \nonumber \\ 
\leq  & \P\left(\mathcal{E} \right) + C_{\ref{eqn:53}}(\delta,r) \E\left[  \left.\max_{a\in \{0,1\}} \max_{i\in \A}\frac{\left|y_i(a)-\overline{y}_A(a)\right|}{\sqrt{n_{A}^av^a_A}}\right|\mathcal{E}^c\right] \times \P\left(\mathcal{E}^c \right) \label{eqn:52}\\  
\leq &   \frac{2\left(1-r\right)}{rs^2(n-1)} + \sqrt{\frac{2}{r}}C_{\ref{eqn:53}}(\delta,r) \max_{a\in \{0,1\}} \max_{i\in \A}\frac{\left|y_i(a)-\overline{y}_A(a)\right|}{\sqrt{n_{A}v^a_A}}\label{eqn:53},\\
\leq &  \frac{2\left(1-r\right)}{rs^2(n-1)} + \sqrt{\frac{2}{r}}C_{\ref{eqn:53}}(\delta,r)\times B n_A^{-\frac{1}{4}}\label{eqn:54},
\end{align}
where (\ref{eqn:52}) follows from (\ref{eqn:beb}), (\ref{eqn:53}) follows from Lemma \ref{lemma:nonnegligible_share} and (\ref{eqn:size_lower_bound}), and (\ref{eqn:54}) follows from Lemma \ref{lemma:nv}-\ref{eqn:nv2}.
\end{proof}

\subsection{Consistency of the Variance Estimator and Asymptotic Distributions of $\T^0_n$, $\T^1_n$ and $\T^2_n$}\label{section:var_proof}
Recall the variance estimator defined in (\ref{hajek:var}):
\begin{equation*}
    \widehat{\AVar}^{\ha}_n \left( Y,D,A \right)= \frac{\sum_{i=1}^n D_iA_i\left( Y_i-\widehat{\tau}^1_A\right)^2}{\left(\sum_{i=1}^n D_iA_i\right)^2}  +  \frac{\sum_{i=1}^n (1-D_i)A_i\left( Y_i-\widehat{\tau}^0_A\right)^2}{\left(\sum_{i=1}^n \left(1-D_i\right)A_i\right)^2},
\end{equation*}
where $\widehat{\tau}^1_{A}= \sum_{i=1}^n D_iA_iY_i/\sum_{i=1}^n D_iA_i$ and $\widehat{\tau}^0_{A} =\sum_{i=1}^n \left(1-D_i\right)A_iY_i/\sum_{i=1}^n \left(1-D_i\right)A_i$.
Conditioning on the the event that $\sum_{i=1}^n D_iA_i=k$, the variance estimator can be expressed as:
\begin{equation*}
    \widehat{\AVar}^{\ha}_{n,k} \left( Y,D,A \right)= \frac{\sum_{i=1}^n D_iA_i\left( Y_i-\widehat{\tau}^1_{A,k}\right)^2}{k^2}  +  \frac{\sum_{i=1}^n (1-D_i)A_i\left( Y_i-\widehat{\tau}^0_{A,k}\right)^2}{\left(n_A-k\right)^2} 
\end{equation*}
where, $\widehat{\tau}^1_{A,k}= \sum_{i=1}^n D_iA_iY_i/k, \widehat{\tau}^0_{A,k} = \sum_{i=1}^n \left(1-D_i\right)A_iY_i/\left(n_A-k\right)$. For simplicity, we denote,
\begin{equation}
    \widehat{v}^1_{A,k}= \frac{\sum_{i=1}^n D_iA_i\left( Y_i-\widehat{\tau}^1_{A,k}\right)^2}{k},  \widehat{v}^0_{A,k}= \frac{\sum_{i=1}^n \left(1-D_i\right)A_i\left( Y_i-\widehat{\tau}^0_{A,k}\right)^2}{n_A-k},
\end{equation}
and the conditional variance estimator can be written as:
\begin{equation}\label{eqn:conditional_variance}
    \widehat{\AVar}^{\ha}_{n,k} \left( Y,D,A \right)= \frac{1}{k}\widehat{v}^1_{A,k} + \frac{1}{n_A-k }\widehat{v}^0_{A,k}. 
\end{equation}
Define the target variance,
\begin{equation}\label{eqn:target_variance}
    \widetilde{\AVar}_{n,k}\left(\arhj\right)=\frac{1}{k}v^1_{A}+\frac{1}{n_A-k}v^0_{A},
\end{equation}
with $v^1_{A}$ and $v^0_{A}$ defined in (\ref{eqn:va}). Lemma \ref{lemma:var_est_tail} provides a tail inequality for the variance estimator.
\begin{lemma'}\label{lemma:var_est_tail}
For all $n\in\mathbb{N}$ and any $\theta_n^w\in \Theta_n^w$,  let $\{\left(y_i(1),y_i(0)\right)\}_{i=1}^n$ be the associated potential outcomes and $\{A_i\}_{i=1}^n$ be the associated always-reporter indicators. 
Denote $\A=\{i\in[n]:A_i=1\}$ and $n_A=\sum_{i=1}^n A_i$. Define $n_A^1=\sum_{i=1}^n D_iA_i$.

Consider the conditional variance estimator $\widehat{\AVar}^{\ha}_{n,k}$ in  (\ref{eqn:conditional_variance}), the variance target $\widetilde{\AVar}_{n,k}$ (\ref{eqn:target_variance}).
Suppose $n_A\geq 2$. Let $k$ be a positive integer with $k\in [1,n_A-1]$. For every $\epsilon \in (0,2]$,
\begin{align*}
      & \P_{\theta_n^w}\left( \left. \left|\frac{ \widetilde{\AVar}_{n,n_A^1} }{ \widehat{\AVar}^{\ha}_{n,n_A^1}}-1\right|\geq \epsilon \right| n_A^1=k \right)\\
      \leq &     C_{\ref{eqn:ceb}}\left(\epsilon\right)\left( \frac{n_A-k}{k}+\frac{k}{n_A-k}\right) \frac{1}{n_A}\max_{a\in \{0,1\}}\frac{\max_{i\in\A} \left(y_i(a)-\overline{y}_A\left(a\right)\right)^2}{v^a_{A}}\\
      & +  \frac{8(2+\epsilon)}{\epsilon}n_A^{-1} + \frac{2(2+\epsilon)}{\epsilon}n_A^{-1}\left(\frac{n_A-k}{k}+\frac{k}{n_A-k}\right),\\
      \leq &  \left( C_{\ref{eqn:ceb}}\left(\epsilon\right)B^2 + \frac{2(2+\epsilon)}{\epsilon}\right) n_A^{-\frac{1}{2}} \left(\frac{n_A-k}{k}+\frac{k}{n_A-k}\right)+  \frac{8(2+\epsilon)}{\epsilon}n_A^{-1} ,
\end{align*}
where,
\begin{align}\label{eqn:ceb}
   C_{\ref{eqn:ceb}}\left(\epsilon\right)=  \left(\frac{8(2+\epsilon)}{\epsilon}\right)^2 + \left(\frac{2\left(2-\epsilon\right)}{\epsilon}\right)^2.
\end{align}
with the constant $B$ defined in Assumption \ref{a:theta}-\ref{a:theta3}.
\end{lemma'}
\begin{proof}

Recall that by Lemma \ref{lemma:cr}, conditioning on the event $n_A^1=k$, $\left(D_i\right)_{i\in \A}\sim \textrm{CR}(n_A,k)$. We suppress the dependence on $\theta_n^w$ and the conditioning event for simplicity. By Assumption \ref{a:theta2}, $\widetilde{\AVar}_{n,k}$ is positive. On the event where $\widehat{\AVar}^{\ha}_{n,k}=0$, we interpret $\widetilde{\AVar}_{n,k}/\widehat{\AVar}^{\ha}_{n,k}=\infty$.

We have the inequality:
\begin{align}
    & \P\left( \frac{ \widetilde{\AVar}_{n,k} }{ \widehat{\AVar}^{\ha}_{n,k}}\geq 1+\epsilon \right)   =   \P\left( \frac{ \widetilde{\AVar}_{n,k}-\widehat{\AVar}^{\ha}_{n,k} }{ \widehat{\AVar}^{\ha}_{n,k}}\geq \epsilon \right)  \nonumber \\
   =  &  \P\left( \frac{k^{-1}v_{A}^1+(n_A-k)^{-1}v_{A}^0-k^{-1}\widehat{v}_{A,k}^1-(n_A-k)^{-1}\widehat{v}_{A,k}^0  }{ k^{-1}\widehat{v}_{A,k}^1+(n_A-k)^{-1}\widehat{v}_{A,k}^0}\geq \epsilon \right) \nonumber \\
     \leq &  \P\left( \frac{ k^{-1}\left(v_{A}^1-\widehat{v}_{A,k}^1\right)}{ k^{-1}\widehat{v}_{A,k}^1+(n_A-k)^{-1}\widehat{v}_{A,k}^0}\geq \frac{\epsilon}{2} \right) + \P\left( \frac{ (n_A-k)^{-1}\left(v_{A}^0-\widehat{v}_{A,k}^0\right)  }{ k^{-1}\widehat{v}_{A,k}^1+(n_A-k)^{-1}\widehat{v}_{A,k}^0}\geq \frac{\epsilon}{2} \right)\nonumber \\
\leq  &\P\left( \frac{ \left(v_{A}^1-\widehat{v}_{A,k}^1\right)}{ \widehat{v}_{A,k}^1}\geq \frac{\epsilon}{2} \right) +\P\left( \frac{ \left(v_{A}^0-\widehat{v}_{A,k}^0\right)}{ \widehat{v}_{A,k}^0}\geq \frac{\epsilon}{2} \right), \label{eqn:2025102161} 
\end{align}
where we use the fact that $\widehat{v}_{A,k}^1$ and $\widehat{v}_{A,k}^0$ are nonnegative in (\ref{eqn:2025102161}). Similarly, we have,
\begin{align}
    & \P\left( \frac{ \widetilde{\AVar}_{n,k} }{ \widehat{\AVar}^{\ha}_{n,k}}\leq 1-\epsilon \right)   =   \P\left( \frac{ \widetilde{\AVar}_{n,k}-\widehat{\AVar}^{\ha}_{n,k} }{ \widehat{\AVar}^{\ha}_{n,k}}\leq -\epsilon \right)  \nonumber \\
   =  &  \P\left( \frac{k^{-1}v_{A}^1+(n_A-k)^{-1}v_{A}^0-k^{-1}\widehat{v}_{A,k}^1-(n_A-k)^{-1}\widehat{v}_{A,k}^0  }{ k^{-1}\widehat{v}_{A,k}^1+(n_A-k)^{-1}\widehat{v}_{A,k}^0}\leq -\epsilon \right) \nonumber \\
     \leq &  \P\left( \frac{ k^{-1}\left(v_{A}^1-\widehat{v}_{A,k}^1\right)}{ k^{-1}\widehat{v}_{A,k}^1+(n_A-k)^{-1}\widehat{v}_{A,k}^0}\leq -\frac{\epsilon}{2} \right) + \P\left( \frac{ (n_A-k)^{-1}\left(v_{A}^0-\widehat{v}_{A,k}^0\right)  }{ k^{-1}\widehat{v}_{A,k}^1+(n_A-k)^{-1}\widehat{v}_{A,k}^0}\leq -\frac{\epsilon}{2} \right)\nonumber \\
\leq  &\P\left( \frac{ \left(v_{A}^1-\widehat{v}_{A,k}^1\right)}{\widehat{v}_{A,k}^1}\leq -\frac{\epsilon}{2} \right) +\P\left( \frac{ \left(v_{A}^0-\widehat{v}_{A,k}^0\right)}{ \widehat{v}_{A,k}^0}\leq -\frac{\epsilon}{2} \right), \label{eqn:2025102162} 
\end{align}
where we note that $ \widetilde{\AVar}_{n,k}/ \widehat{\AVar}^{\ha}_{n,k}\leq 1-\epsilon$ implicitly implies that $\widehat{\AVar}^{\ha}_{n,k}>0$.
Combining (\ref{eqn:2025102161}) and (\ref{eqn:2025102162}) we have
\begin{align}
   &\P\left( \left|\frac{ \widetilde{\AVar}_{n,k} }{ \widehat{\AVar}^{\ha}_{n,k}}-1\right|\geq \epsilon \right) =   \P\left( \frac{ \widetilde{\AVar}_{n,k} }{ \widehat{\AVar}^{\ha}_{n,k}}\geq 1+\epsilon \right)  +   \P\left( \frac{ \widetilde{\AVar}_{n,k} }{ \widehat{\AVar}^{\ha}_{n,k}}\leq 1-\epsilon \right)\nonumber \\
  \leq  &\P\left( \left|\frac{ v_{A}^1-\widehat{v}_{A,k}^1}{\widehat{v}_{A,k}^1}\right|\geq \frac{\epsilon}{2} \right)+\P\left( \left|\frac{ v_{A}^0-\widehat{v}_{A,k}^0}{\widehat{v}_{A,k}^0}\right|\geq \frac{\epsilon}{2} \right) \label{eqn:20251021632}
\end{align}
Our result follows from bounding the two terms in (\ref{eqn:20251021632}).
We bound the first term in (\ref{eqn:20251021632}). The bound for the second term follows from an analogous argument.
\begin{align}
    & \P\left( \frac{\left|v_{A}^1-\widehat{v}_{A,k}^1\right| }{ \widehat{v}_{A,k}^1}\geq \frac{\epsilon}{2} \right)= \P\left( \frac{v_{A}^1-\widehat{v}_{A,k}^1 }{ \widehat{v}_{A,k}^1}\geq \frac{\epsilon}{2} \right) + \P\left( \frac{v_{A}^1-\widehat{v}_{A,k}^1 }{ \widehat{v}_{A,k}^1}\leq -\frac{\epsilon}{2} \right) \nonumber \\
    = & \P\left( \frac{v_{A}^1}{ \widehat{v}_{A,k}^1}\geq 1+\frac{\epsilon}{2} \right) + \P\left( \frac{v_{A}^1 }{ \widehat{v}_{A,k}^1}\leq 1-\frac{\epsilon}{2} \right). \label{eqn:2025102175}
\end{align}
We have,
\begin{align}
    & \P\left( \frac{v_{A}^1}{ \widehat{v}_{A,k}^1}\geq 1+\frac{\epsilon}{2} \right) =   \P\left( \frac{\widehat{v}_{A,k}^1}{ v_{A}^1}\leq \frac{2}{2+\epsilon} \right) = \P\left( \frac{\widehat{v}_{A,k}^1-v_{A}^1}{ v_{A}^1}\leq -\frac{\epsilon}{2+\epsilon} \right) \nonumber\\
   = & \P\left( \frac{k^{-1}\sum_{i=1}^n D_iA_i\left( Y_i-\ytA \right)^2 - \left( \widehat{\tau}^1_{A,k} - \ytA\right)^2 -v_{A}^1}{ v_{A}^1}\leq -\frac{\epsilon}{2+\epsilon} \right) \nonumber\\
   \leq & \P\left( \frac{k^{-1}\sum_{i=1}^n D_iA_i\left( Y_i-\ytA \right)^2 - v^1_{A} }{ v_{A}^1}\leq -\frac{\epsilon}{2\left(2+\epsilon\right)} \right) \label{eqn:2025102163}\\
    & + \P\left( -\frac{ \left(\widehat{\tau}^1_{A,k} - \ytA\right)^2 }{ v_{A}^1}\leq -\frac{\epsilon}{2\left(2+\epsilon\right)} \right)  \label{eqn:2025102164},
\end{align}
where we use the fact that
\begin{align*}
      \frac{1}{k}\sum_{i=1}^n D_iA_i\left( Y_i-\widehat{\tau}^1_{A,k}\right)^2 =\frac{1}{k}\sum_{i=1}^n D_iA_i\left( Y_i-\ytA \right)^2 - \left( \widehat{\tau}^1_{A,k} - \ytA\right)^2.   
\end{align*}

For (\ref{eqn:2025102164}), we have,
\begin{align}
    & \P\left( -\frac{ \left(\widehat{\tau}^1_{A,k} - \ytA\right)^2 }{ v_{A}^1}\leq -\frac{\epsilon}{2\left(2+\epsilon\right)} \right)  =  \P\left( \frac{ \left(\widehat{\tau}^1_{A,k} - \ytA\right)^2 }{ v_{A}^1}\geq \frac{\epsilon}{2\left(2+\epsilon\right)} \right) \\
   \leq  & \frac{2(2+\epsilon)}{\epsilon} \frac{\E\left[\left(\widehat{\tau}^1_{A,k} - \ytA\right)^2\right]}{v_{A}^1}=\frac{2(2+\epsilon)}{\epsilon} \frac{\textrm{V}\left(\widehat{\tau}^1_{A,k}\right) }{v^1_{A}} =\frac{2\left(2+\epsilon\right)}{\epsilon}\frac{n_A-k}{n_Ak},
\end{align}
where the last equality follows from 
\begin{equation}
\textrm{V}\left(\widehat{\tau}^1_{A,k}\right) = \frac{n_A-k}{n_Ak}\frac{1}{n_A-1}\sum_{i\in \A}\left(y_i(1)-\ytA\right)^2= \frac{n_A-k}{n_Ak}v^1_A.
\end{equation}
Notice we can write:
\begin{align*}
 &\frac{1}{k}\sum_{i=1}^n D_iA_i\left( Y_i-\ytA \right)^2 - v^1_{A}\\
=&
\frac{n_A}{\left(n_A-1\right)k}\sum_{i=1}^n D_iA_i\left( Y_i-\ytA \right)^2 -v^1_A - \frac{1}{\left(n_A-1\right)k}\sum_{i=1}^n D_iA_i\left( Y_i-\ytA \right)^2
\end{align*}
Now, 
\begin{align}
  & \P\left( \frac{k^{-1}\sum_{i=1}^n D_iA_i\left( Y_i-\ytA \right)^2 - v^1_{A} }{ v_{A}^1}\leq -\frac{\epsilon}{2\left(2+\epsilon\right)} \right) \nonumber \\
  \leq & \P\left( \frac{ \frac{n_A}{k}\frac{1}{n_A-1}\sum_{i=1}^n D_iA_i\left( Y_i-\ytA \right)^2 - v^1_{A} }{ v_{A}^1}\leq -\frac{\epsilon}{4\left(2+\epsilon\right)} \right) \label{eqn:2025102173} \\
   & +    \P\left( \frac{ - \frac{1}{\left(n_A-1\right)k}\sum_{i=1}^n D_iA_i\left( Y_i-\ytA \right)^2}{ v_{A}^1}\leq -\frac{\epsilon}{4\left(2+\epsilon\right)} \right)  \label{eqn:2025102174}
\end{align}
A bound for (\ref{eqn:2025102173}) follows from the calculation
\begin{align}
    &\P\left( \frac{ \frac{n_A}{k}\frac{1}{n_A-1}\sum_{i=1}^n D_iA_i\left( Y_i-\ytA \right)^2 - v^1_{A} }{ v_{A}^1}\leq -\frac{\epsilon}{4\left(2+\epsilon\right)} \right)\nonumber \\
    \leq & \left(\frac{4(2+\epsilon)}{\epsilon}\right)^2 \left(\frac{n_A}{n_A-1}\right)^2  \left(v^1_{A}\right)^{-2}\V\left(\frac{1}{k}\sum_{i=1}^n D_iA_i\left(Y_i-\ytA\right)^2 \right) \label{eqn:2025102169}\\
    = & \left(\frac{4(2+\epsilon)}{\epsilon}\right)^2 \left(\frac{n_A}{n_A-1}\right)^2  \left(v^1_{A}\right)^{-2} \frac{n_A-k}{kn_A}\frac{1}{n_A-1}\sum_{i\in \A} \left(y_i(1)-\ytA\right)^4 \nonumber \\
   \leq  & \left(\frac{4(2+\epsilon)}{\epsilon}\right)^2\left(\frac{n_A}{n_A-1}\right)^2 \frac{n_A-k}{kn_A}\frac{\max_{i\in\A} \left(y_i(1)-\ytA\right)^2}{v^1_{A}} \label{eqn:2025102170}\\
    \leq &  \left(\frac{4(2+\epsilon)}{\epsilon}\right)^2 \left(\frac{n_A}{n_A-1}\right)^2 \frac{n_A-k}{kn_A} B^2n_A^{\frac{1}{2}},\label{eqn:2025102171}
\end{align}
where (\ref{eqn:2025102169}) uses the Chebyshev inequality and the fact that 
\begin{equation}\label{eqn:20260108}
    \E\left[\frac{n_A}{k}\frac{1}{n_A-1}\sum_{i=1}^n D_iA_i\left( Y_i-\ytA \right)^2\right]=v_A^1,
\end{equation}
(\ref{eqn:2025102170}) uses the fact that
\begin{equation*}
    \frac{\sum_{i\in \A}\left(y_i(1)-\ytA\right)^4}{n_A-1}\leq     \frac{\sum_{i\in \A}\left(y_i(1)-\ytA\right)^2}{n_A-1}\times \max_{i\in \A}\left(y_i(1)-\ytA\right)^2,
\end{equation*}
and (\ref{eqn:2025102171}) uses Lemma \ref{lemma:nv}-\ref{eqn:nv2}. A bound for (\ref{eqn:2025102174}) follows from the calculation
\begin{align*}
  &\P\left( \frac{ - \frac{1}{k\left(n_A-1\right)}\sum_{i=1}^n D_iA_i\left( Y_i-\ytA \right)^2  }{ v_{A}^1}\leq -\frac{\epsilon}{4\left(2+\epsilon\right)} \right)       \\
= & \P\left( \frac{  \frac{1}{k\left(n_A-1\right)}\sum_{i=1}^n D_iA_i\left( Y_i-\ytA \right)^2  }{ v_{A}^1}\geq \frac{\epsilon}{4\left(2+\epsilon\right)} \right)         \\  
\leq  & \frac{4\left(2+\epsilon\right)}{\epsilon} \frac{\frac{1}{n_A\left(n_A-1\right)}\sum_{i\in \A}\left(y_i(1)-\ytA\right)^2}{v_{A}^1}=  \frac{4\left(2+\epsilon\right)}{n_A\epsilon},
\end{align*}
by the Markov inequality and \ref{eqn:20260108}. Collecting terms we have:
\begin{align*}
    & \P\left( \frac{v_{A}^1}{ \widehat{v}_{A,k}^1}\geq 1+\frac{\epsilon}{2} \right) \\
     \leq &  \left(\frac{4(2+\epsilon)}{\epsilon}\right)^2 \left(\frac{n_A}{n_A-1}\right)^2 \frac{n_A-k}{kn_A} B^2n_A^{\frac{1}{2}} + \frac{4\left(2+\epsilon\right)}{n_A\epsilon} + \frac{2\left(2+\epsilon\right)}{\epsilon}\frac{n_A-k}{n_Ak}.
\end{align*}
For the second term in (\ref{eqn:2025102175}). We use the premise that $\epsilon \leq 2$ and 
\begin{align}
 & \P\left( \frac{v_{A}^1 }{ \widehat{v}_{A,k}^1}\leq 1-\frac{\epsilon}{2} \right)=\P\left( \frac{v_{A}^1 }{ \widehat{v}_{A,k}^1}\leq \frac{2-\epsilon}{2} \right) = \P\left( \frac{\widehat{v}_{A,k}^1 }{ v_{A}^1}\geq \frac{2}{2-\epsilon} \right) \nonumber\\
= & \P\left( \frac{k^{-1}\sum_{i=1}^n D_iA_i\left( Y_i-\ytA \right)^2 - \left( \widehat{\tau}^1_{A,k} - \ytA\right)^2 -v_{A}^1}{ v_{A}^1}\geq \frac{\epsilon}{2-\epsilon} \right) \nonumber\\
\leq & \P\left( \frac{k^{-1}\sum_{i=1}^n D_iA_i\left( Y_i-\ytA \right)^2-v_{A}^1}{ v_{A}^1}\geq \frac{\epsilon}{2-\epsilon} \right) \nonumber\\
\leq &  \P\left( \frac{\frac{n_A}{\left(n_A-1\right)k}\sum_{i=1}^n D_iA_i\left( Y_i-\ytA \right)^2 - v^1_{A}}{ v_{A}^1}\geq \frac{\epsilon}{2-\epsilon} \right)\label{eqn:2025102176} \\
= &  \P\left( \left(\frac{\frac{n_A}{\left(n_A-1\right)k}\sum_{i=1}^n D_iA_i\left( Y_i-\ytA \right)^2 - v^1_{A}}{ v_{A}^1}\right)^2 \geq \left(\frac{\epsilon}{2-\epsilon}\right)^2 \right) \nonumber\\
\leq & \left(\frac{2-\epsilon}{\epsilon}\right)^2  \left(\frac{n_A}{n_A-1}\right)^2 \left(v^1_{A}\right)^{-2}\V\left(\frac{1}{k}\sum_{i=1}^n D_iA_i\left(Y_i-\ytA\right)^2 \right)\\
    \leq &  \left(\frac{2-\epsilon}{\epsilon}\right)^2 \left(\frac{n_A}{n_A-1}\right)^2  \left(v^1_{A}\right)^{-2} \frac{n_A-k}{kn_A}\frac{1}{n_A-1}\sum_{i\in \A}\left(y_i(1)-\ytA\right)^4 \nonumber \\
    \leq & \left(\frac{2-\epsilon}{\epsilon}\right)^2\left(\frac{n_A}{n_A-1}\right)^2 \frac{n_A-k}{kn_A}\frac{\max_{i\in\A} \left(y_i(1)-\ytA\right)^2}{v^1_{A}} \nonumber \\
    \leq &   \left(\frac{2-\epsilon}{\epsilon}\right)^2 \left(\frac{n_A}{n_A-1}\right)^2 \frac{n_A-k}{kn_A} B^2 n_A^{\frac{1}{2}}, \nonumber
\end{align}
where in (\ref{eqn:2025102176}) follows because $\sum_{i=1}^n D_iA_i\left( Y_i-\ytA \right)^2$ is nonnegative and $n_A\left(n_A-1\right)^{-1}\geq 1$ for $n_A\geq 2$ and rest calculations are similar to the bound for (\ref{eqn:2025102173}).
Hence for $\epsilon \leq 2$,
\begin{align*}
&\P\left( \frac{\left|v_{A}^1-\widehat{v}_{A,k}^1\right| }{ \widehat{v}_{A,k}^1}\geq \frac{\epsilon}{2} \right)     \leq  \left(\frac{4(2+\epsilon)}{\epsilon}\right)^2 \left(\frac{n_A}{n_A-1}\right)^2 \frac{n_A-k}{kn_A} B^2 n_A^{\frac{1}{2}} \\
 & + \frac{4\left(2+\epsilon\right)}{n_A\epsilon} + \frac{2\left(2+\epsilon\right)}{\epsilon}\frac{n_A-k}{n_Ak} + \left(\frac{2-\epsilon}{\epsilon}\right)^2 \left(\frac{n_A}{n_A-1}\right)^2 \frac{n_A-k}{kn_A} B^2 n_A^{\frac{1}{2}}.
\end{align*}
For $n_A\geq 2$,  we have $0\leq n_A/(n_A-1)\leq 2$. Hence we can simplify,
\begin{align*}
    & \P\left( \frac{\left|v_{A}^1-\widehat{v}_{A,k}^1\right| }{ \widehat{v}_{A,k}^1}\geq \frac{\epsilon}{2} \right) \\\leq  &C_{\ref{eqn:ceb}}(\epsilon)B^2\frac{n_A-k}{k}n_A^{-\frac{1}{2}}+ \frac{4(2+\epsilon)}{\epsilon}n_A^{-1} + \frac{2(2+\epsilon)}{\epsilon}\frac{n_A-k}{k} n_A^{-1},
\end{align*}
where
\begin{align*}
     C_{\ref{eqn:ceb}}(\epsilon)=   \left(\frac{8(2+\epsilon)}{\epsilon}\right)^2 + \left(\frac{2\left(2-\epsilon\right)}{\epsilon}\right)^2. 
\end{align*}
By symmetry, we have
\begin{align}
&\P\left( \frac{\left|v_{A}^0-\widehat{v}_{A,k}^0\right| }{ \widehat{v}_{A,k}^0}\geq \frac{\epsilon}{2} \right) \nonumber\\
\leq &    C_{\ref{eqn:ceb}}(\epsilon)B^2\frac{k}{n_A-k}n_A^{-\frac{1}{2}}+ \frac{4(2+\epsilon)}{\epsilon}n_A^{-1} + \frac{2(2+\epsilon)}{\epsilon}\frac{k}{n_A-k} n_A^{-1}. \label{eqn:2025110568}
\end{align}
Combining these two bounds gives us the result in the lemma.
\end{proof}

\begin{corollary'}\label{cor:var_est_tail}
Under the setup of Lemma \ref{lemma:var_est_tail}, we have, for $n\geq 2$,
\begin{align}
     & \P_{\theta_n^w}\left(  \left|\frac{\widetilde{\V}_{n,n_A^1}}{\widehat{\AVar}^{\ha}_{n}}-1\right| \geq  \epsilon \right)      =    \E\left[ \P_{\theta_n^w}\left(  \left|\frac{\widetilde{\V}_{n,n_A^1}}{\widehat{\AVar}^{\ha}_{n,n_A^1}}-1\right| \geq  \epsilon \bigg| n_{A}^1\right)  \right] \nonumber \\
    \leq &  C_{\ref{constant:2}}(r,s,\epsilon)n^{-1} +C_{\ref{constant:3}}\left(\epsilon,r\right) \frac{1}{n_A}\max_{a\in \{0,1\}}\frac{\max_{i\in\A} \left(y_i(a)-\overline{y}_A\left(a\right)\right)^2}{v^a_{A}} \label{eqn:2025102979}\\
    \leq &  C_{\ref{constant:2}}(r,s,\epsilon)n^{-1} +C_{\ref{constant:3}}\left(\epsilon,r\right) B^2n_A^{-\frac{1}{2}}\label{eqn:2025102980}
\end{align}
where $C_{\ref{constant:3}}\left(\epsilon,r\right)$ is defined as
\begin{equation}\label{constant:3}
    C_{\ref{constant:3}}(\epsilon,r)=C_{\ref{eqn:ceb}}\left(\epsilon\right)\frac{2\left(2-r\right)}{r}, 
\end{equation}
with $C_{\ref{eqn:ceb}}\left(\epsilon\right)$ defined in (\ref{eqn:ceb}), and  $C_{\ref{constant:2}}(r,s,\epsilon)$ is defined as
\begin{equation}\label{constant:2}
    C_{\ref{constant:2}}\left(r,s,\epsilon\right)=\frac{4\left(1-r\right)}{rs^2}+ \frac{8\left(2+\epsilon\right)}{\epsilon s}+ \frac{2\left(2+\epsilon\right)}{\epsilon}\frac{2\left(2-r\right)}{r s}
\end{equation}
with constants $r$, $s$ and $B$ defined in Assumption-\ref{a:cr}, Assumption\ref{a:theta}-\ref{a:theta1} and  Assumption\ref{a:theta}-\ref{a:theta3}, respectively.
\end{corollary'}
\begin{proof}
We suppress the dependence on $\theta_n^w$ and the conditioning event for simplicity. Define events $\mathcal{E}$ and $\mathcal{E}^c$ as in (\ref{eqn:E}) and (\ref{eqn:Ec}).
Note on the event $\mathcal{E}^c$, we have the inequality,
\begin{equation*}
   n_A^1 > \frac{n_A n_{1}}{2n} \geq  \frac{rn_A}{2} ,n_A^0 > \frac{n_A n_{0}}{2n} \geq  \frac{rn_A}{2}.
\end{equation*}
\begin{equation*}
    n_A^1 = n_A- n_A^0 <  n_A - \frac{rn_A}{2}= \left(1-\frac{r}{2}\right)n_A,
\end{equation*}
\begin{equation*}
  n_A^0 = n_A-  n_A^1 < n_A - \frac{r n_A }{2}= \left(1-\frac{r}{2}\right) n_A. 
\end{equation*}
Hence, conditioning on the event $\mathcal{E}^c$, we have
\begin{equation}\label{eqn:ratio_bound}
 \frac{n_A^1}{n_A^0}\in \left(  \frac{r/2}{1-r/2}, \frac{1-r/2}{r/2}\right),  \frac{n_A^0}{n_A^1}\in \left(  \frac{r/2}{1-r/2}, \frac{1-r/2}{r/2}\right).
\end{equation}
The result follows by
\begin{align*}
  & \E\left[ \P\left(  \left|\frac{\widetilde{\V}_{n,n_A^1}}{\widehat{\AVar}^{\ha}_{n,n_A^1}} -1\right|\geq \epsilon \bigg| n_{A}^1\right)  \right] \\
  \leq &  \P\left(\mathcal{E}\right) +\E\left[ \P\left(  \left|\frac{\widetilde{\V}_{n,n_A^1}}{\widehat{\AVar}^{\ha}_{n,n_A^1}} -1\right|\geq \epsilon \bigg| n_{A}^1\right) \bigg| \mathcal{E}^c\right] \times\P\left(\mathcal{E}^c\right)\\
   \leq  & \P\left(\mathcal{E}\right) +\\
   &    \E\left[ C_{\ref{eqn:ceb}}\left(\epsilon\right)\left(\frac{n_A^0}{n_A^1}+\frac{n_A^1}{n_A^0}\right)\frac{1}{n_A}\max_{a\in \{0,1\}}\frac{\max_{i\in\A} \left(y_i(a)-\overline{y}_A\left(a\right)\right)^2}{v^a_{A}}\bigg| \mathcal{E}^c\right]+\\
      & \E\left[ \frac{8(2+\epsilon)}{\epsilon}n_A^{-1} + \frac{2(2+\epsilon)}{\epsilon}n_A^{-1}\left(\frac{n_A^0}{n_A^1}+\frac{n_A^1}{n_A^0}\right)\bigg| \mathcal{E}^c\right],\\
    \leq & \frac{2(1-r)}{rs^2(n-1)}+  C_{\ref{eqn:ceb}}\left(\epsilon\right)\frac{2\left(2-r\right)}{r} \frac{1}{n_A}\max_{a\in \{0,1\}}\frac{\max_{i\in\A} \left(y_i(a)-\overline{y}_A\left(a\right)\right)^2}{v^a_{A}}\\
      & + \frac{8\left(2+\epsilon\right)}{\epsilon}n_A^{-1} + \frac{2\left(2+\epsilon\right)}{\epsilon}\frac{2\left(2-r\right)}{r}n_A^{-1}\\
   \leq & \frac{2(1-r)}{rs^2(n-1)}+ C_{\ref{eqn:ceb}}\left(\epsilon\right)B^2 \frac{2\left(2-r\right)}{r}n_A^{-\frac{1}{2}}+\frac{8\left(2+\epsilon\right)}{\epsilon}n_A^{-1} + \frac{2\left(2+\epsilon\right)}{\epsilon}\frac{2\left(2-r\right)}{r}n_A^{-1}
\end{align*}
where the second inequality follows from Lemma \ref{lemma:var_est_tail} with the constant $C(\epsilon)$ defined in (\ref{eqn:ceb}) and the third inequality follows from Lemma \ref{lemma:nonnegligible_share} and (\ref{eqn:ratio_bound}). Constants $r$, $s$ and $B$ are defined in Assumption-\ref{a:cr}, Assumption \ref{a:theta}-\ref{a:theta1} and  Assumption \ref{a:theta}-\ref{a:theta2}, respectively.

For each $\theta_n^w\in \Theta_n^w$, we have $n_A^{-1}\leq s^{-1}n^{-1}$. We write, for $n\geq 2$
\begin{align*}
 & \frac{2(1-r)}{rs^2(n-1)}n^{-1}n +  \frac{8\left(2+\epsilon\right)}{\epsilon}s^{-1}n^{-1} + \frac{2\left(2+\epsilon\right)}{\epsilon}\frac{2\left(2-r\right)}{r}s^{-1}n^{-1}\\
\leq   &  \left(\frac{4\left(1-r\right)}{rs^2}+ \frac{8\left(2+\epsilon\right)}{\epsilon}s^{-1}+ \frac{2\left(2+\epsilon\right)}{\epsilon}\frac{2\left(2-r\right)}{r}s^{-1}\right)n^{-1}\\     
=& C_{\ref{constant:2}}(r,s,\epsilon)n^{-1}.
\end{align*}
\end{proof}
Anticipating the proof in Lemma \ref{lemma:E10} and recalling the definition of $\widehat{v}_A^{1,k}$ and $\widehat{v}_A^{0,k}$ in (\ref{eqn:conditional_variance}), we note that a similar calculation can be carried out for $\epsilon\leq 1$
\begin{align}
 & \P\left(  \left|\frac{v_A^1}{\widehat{v}_{A,n_A^1}^{1}}-1 \right|\geq  \epsilon \right)= \E\left[ \P\left(  \left|\frac{v_A^1}{\widehat{v}_{A,n_A^1}^{1}}-1 \right|\geq  \epsilon \bigg| n_{A}^1\right)  \right] \nonumber \\
  \leq &  \P\left(\mathcal{E}\right) + \E\left[ \P\left(  \left|\frac{v_A^1}{\widehat{v}_{A,n_A^1}^{1}}-1 \right|\geq  \frac{2\epsilon}{2} \bigg| n_{A}^1\right) \bigg| \mathcal{E}^c \right] \times\P\left(\mathcal{E}^c\right)\nonumber \\
   \leq  & \P\left(\mathcal{E}\right) +\E\left[ C_{\ref{eqn:ceb}}\left(2\epsilon\right)B^2\frac{n_A^0}{n_A^1}n_A^{-\frac{1}{2}}+ \frac{4(1+\epsilon)}{\epsilon}n_A^{-1} + \frac{2(1+\epsilon)}{\epsilon}\frac{n_A^0}{n_A^1} n_A^{-1}\bigg| \mathcal{E}^c \right] ,\nonumber \\
   \leq & \frac{2\left(1-r\right)}{rs^2\left(n-1\right)}+ C_{\ref{eqn:ceb}}\left(2\epsilon\right)B^2 \frac{2-r}{r}n_A^{-\frac{1}{2}}+\frac{8\left(2+\epsilon\right)}{\epsilon}n_A^{-1} + \frac{2\left(2+\epsilon\right)}{\epsilon}\frac{2-r}{r}n_A^{-1} \nonumber \\
   = & O(n^{-\frac{1}{2}}),    \label{eqn:va1}
\end{align}
where the second inequality is by \eqref{eqn:2025110568} and the third inequality is by Lemma \ref{lemma:nonnegligible_share} and \eqref{eqn:ratio_bound}. A similar calculation holds for $\widehat{v}_{A,n_A^0}^{0}$. 

\begin{lemma'}\label{lemma:beb_bound}
For all $n\in\mathbb{N}$ and any $\theta_n^w\in \Theta_n^w$,  let $\{\left(y_i(1),y_i(0)\right)\}_{i=1}^n$ be the associated potential outcomes and $\{A_i\}_{i=1}^n$ be the associated always-reporter indicators. Denote $D=\{D_i\}_{i=1}^n\sim\textrm{CR}(n,n_1)$, $\A=\{i\in[n]:A_i=1\}$ and $n_A=\sum_{i=1}^n A_i$. Define $\tau_n=n_A^{-1}\sum_{i\in \A} \left(y_i(1)-y_i(0)\right)$.  

Consider the variance estimator $\widehat{\AVar}^{\ha}_{n}$ in  (\ref{hajek:var}) and the Hajek estimator $\arhj$ in (\ref{eqn:hajek_a}). Let $Z$ be a normal variable with mean 0 and unit variance, $Z\sim \N(0,1)$, that is independent from $D$. There exists a positive integer $N$ such that for all $n\geq N$, every $\epsilon\in (0,2]$ and $t\in \mathbb{R}$,
\begin{align*}
& \P_{\theta_n^w}\left(  \frac{\left(\arhj-\tau_n\right)^2}{    \widehat{\AVar}^{\ha}_n} + g\left(n_{A}^1\right) \geq t\right)\\
 \leq  & \P_{\theta_n^w}\left(  Z^2\left(1+\epsilon\right) + g\left(n_{A}^1\right)\geq t \right)\\
 &+ C_{\ref{constant:2}}(r,s,\epsilon)n^{-1} +C_{\ref{constant:3}}\left(\epsilon,r\right) B^2n_A^{-\frac{1}{2}}+ \frac{2\left(1-r\right)}{rs^2(n-1)} + \sqrt{\frac{2}{r}}C_{\ref{eqn:53}}(\delta,r)\times B n_A^{-\frac{1}{4}}
\end{align*}
where the constant $ C_{\ref{constant:3}}(r,\epsilon)$ is defined in (\ref{constant:3}), $ C_{\ref{constant:2}}(r,s,\epsilon)$ is defined in (\ref{constant:2}), $C_{\ref{eqn:53}}(\delta,r)$ is defined in Lemma \ref{lemma:CLT}, and constants $r$, $s$, $\delta$ and $B$ are defined in Assumption \ref{a:cr}, Assumption \ref{a:theta}-\ref{a:theta1}, Assumption \ref{a:theta}-\ref{a:theta2} and Assumption \ref{a:theta}-\ref{a:theta3}, respectively.
\end{lemma'}
\begin{proof}
We suppress the dependence on $\theta_n^w$ and the conditioning event for simplicity. Pick $N=\lceil 2s^{-1}\rceil $. We have $n_A\geq sn \geq 2$ for all $n\geq N$. Recall the definitions of $\widetilde{\V}_{n,n_A^1}$ and $\widehat{\AVar}^{\ha}_{n,n_A^1}$ defined in (\ref{eqn:target_variance}) and (\ref{eqn:conditional_variance}), respectively. We remind readers that $\V_{n,n_A^1}$ is defined in \ref{eqn:AVnk} with k as $n_A^1$.
\begin{align}
    & \P\left(  \frac{\left(\arhj-\tau_n\right)^2}{    \widehat{\AVar}^{\ha}_n} + g\left(n_{A}^1\right) \geq t\right) \nonumber=   \P\left(  \frac{\left(\arhj-\tau_n\right)^2}{  \V_{n,n_A^1}}\frac{\V_{n,n_A^1}}{\widetilde{\V}_{n,n_A^1}}\frac{\widetilde{\V}_{n,n_A^1}}{\widehat{\AVar}^{\ha}_n} + g\left(n_{A}^1\right) \geq t\right)\nonumber\\
    \leq & \P\left(  \frac{\left(\arhj-\tau_n\right)^2}{  \V_{n,n_A^1}}\frac{\widetilde{\V}_{n,n_A^1}}{\widehat{\AVar}^{\ha}_n} + g\left(n_{A}^1\right) \geq t\right) \label{eqn:2025102178}\\
    = &\E\left[ \P\left(  \frac{\left(\arhj-\tau_n\right)^2}{  \V_{n,n_A^1}}\frac{\widetilde{\V}_{n,n_A^1}}{\widehat{\AVar}^{\ha}_{n,n_A^1}} + g\left(n_{A}^1\right) \geq t\bigg| n_{A}^1\right) \right] \nonumber\\
    \leq  &\E\left[ \P\left(  \frac{\left(\arhj-\tau_n\right)^2}{  \V_{n,n_A^1}}\frac{\widetilde{\V}_{n,n_A^1}}{\widehat{\AVar}^{\ha}_{n,n_A^1}} + g\left(n_{A}^1\right) \geq t,\frac{\widetilde{\V}_{n,n_A^1}}{\widehat{\AVar}^{\ha}_{n,n_A^1}} \leq 1+\epsilon\bigg| n_{A}^1\right)  \right] \\ & +\E\left[ \P\left(  \frac{\widetilde{\V}_{n,n_A^1}}{\widehat{\AVar}^{\ha}_{n,n_A^1}} \geq  1+\epsilon \bigg| n_{A}^1\right)  \right] \nonumber  \\ \leq  &\E\left[ \P\left(  \frac{\left(\arhj-\tau_n\right)^2}{  \V_{n,n_A^1}}\left(1+\epsilon\right) + g\left(n_{A}^1\right) \geq t\bigg| n_{A}^1\right)  \right]\nonumber +\E\left[ \P\left(  \frac{\widetilde{\V}_{n,n_A^1}}{\widehat{\AVar}^{\ha}_{n,n_A^1}} \geq  1+\epsilon \bigg| n_{A}^1\right)  \right] \nonumber  \\
     \leq &  \E\left[ \P\left(  Z^2\left(1+\epsilon\right) + g\left(n_{A}^1\right) \geq t\bigg| n_{A}^1\right)  \right]  + \E\left[ \P\left(  \frac{\widetilde{\V}_{n,n_A^1}}{\widehat{\AVar}^{\ha}_{n,n_A^1}} \geq  1+\epsilon \bigg| n_{A}^1\right)  \right] \nonumber\\
      &  +\Bigg( \E\left[ \P\left(  \frac{\left(\arhj-\tau_n\right)^2}{  \V_{n,n_A^1}}\left(1+\epsilon\right) + g\left(n_{A}^1\right) \geq t\bigg| n_{A}^1\right)  \right]\nonumber\\
     & -\E\left[ \P\left( Z^2\left(1+\epsilon\right) + g\left(n_{A}^1\right) \geq t\bigg| n_{A}^1\right)  \right] \Bigg) \nonumber
\end{align}
where for (\ref{eqn:2025102178}) we use the fact that $\V_{n,n_A^1}/\widetilde{\V}_{n,n_A^1}\leq 1$. The results follow by applying Lemma \ref{lemma:CLT} and Corollary \ref{cor:var_est_tail}.
\end{proof}

Note because for all $\theta_n^w\in \Theta_n^w$, we have $n_A^{-1}\leq s^{-1}n^{-1}$
For $n\geq 2$, we can simplify the upper bounds by
\begin{align*}
& C_{\ref{constant:2}}(r,s,\epsilon)n^{-1} +C_{\ref{constant:3}}\left(\epsilon,r\right) B^2n_A^{-\frac{1}{2}}+ \frac{2\left(1-r\right)}{rs^2(n-1)} + \sqrt{\frac{2}{r}}C_{\ref{eqn:53}}(\delta,r)\times B n_A^{-\frac{1}{4}}\\
\leq & C_{\ref{constant:2}}(r,s,\epsilon)n^{-1} +C_{\ref{constant:3}}\left(\epsilon,r\right) B^2n_A^{-\frac{1}{2}}+ \frac{4\left(1-r\right)}{rs^2}n^{-1} + \sqrt{\frac{2}{r}}C_{\ref{eqn:53}}(\delta,r)\times B n_A^{-\frac{1}{4}}\\
\leq  & \left(\frac{6\left(1-r\right)}{rs^2}+ \frac{8\left(2+\epsilon\right)}{\epsilon s}+ \frac{2\left(2+\epsilon\right)}{\epsilon}\frac{2\left(2-r\right)}{rs}\right) n^{-1}  \\
+ &   C_{\ref{constant:3}}\left(\epsilon,r\right) B^2n_A^{-\frac{1}{2}} +\sqrt{\frac{2}{r}}C_{\ref{eqn:53}}(\delta,r)\times B n_A^{-\frac{1}{4}} \\
\leq  & \left(\frac{6\left(1-r\right)}{rs^2}+ \frac{8\left(2+\epsilon\right)}{\epsilon s}+ \frac{2\left(2+\epsilon\right)}{\epsilon}\frac{2\left(2-r\right)}{rs}\right) n^{-1}  \\
+ &  C_{\ref{constant:3}}\left(\epsilon,r\right) B^2s^{-\frac{1}{2}}n^{-\frac{1}{2}} +\sqrt{\frac{2}{r}}C_{\ref{eqn:53}}(\delta,r)\times B s^{-\frac{1}{4}}n^{-\frac{1}{4}}\\
\leq &   C_{\ref{constant:1}}(r,s,\epsilon,B,\delta)n^{-\frac{1}{4}}
\end{align*}
with
\begin{align}
\begin{split}\label{constant:1}
    C_{\ref{constant:1}}(r,s,\epsilon,B,\delta) & = \frac{6\left(1-r\right)}{rs^2}+ \frac{8\left(2+\epsilon\right)}{\epsilon s}+ \frac{2\left(2+\epsilon\right)}{\epsilon}\frac{2\left(2-r\right)}{rs} + \\
     & C_{\ref{constant:3}}\left(\epsilon,r\right) B^2s^{-\frac{1}{2}}+\sqrt{\frac{2}{r}}C_{\ref{eqn:53}}(\delta,r)\times B s^{-\frac{1}{4}}.
\end{split}
\end{align}
We shall use this constant for the remaining calculations.
\begin{corollary'}\label{corollary:g}
Under the same setup as in Lemma \ref{lemma:beb_bound}, suppose 
\begin{equation}\label{eqn:corollarygc}
\sup_{n\in\mathbb{N}}\sup_{\theta^w_n\in\Theta_n^w}\E_{\theta^w_n}\left[\left|g(n_A^1)\right|\right]\leq C < \infty
\end{equation}
for some positive constant $C$.
For all $t\in\mathbb{R}$,
\begin{align*}
\limsup_{n\to\infty}\sup_{\theta^w_n\in\Theta_n^w}\left(\P_{\theta^w_n}\left(  \frac{\left(\arhj-\tau_n\right)^2}{    \widehat{\AVar}^{\ha}_n} + g\left(n_{A}^1\right) \geq t\right) -\P_{\theta_n^w}\left(  Z^2+ g(n_A^1) \geq t\right)\right)\leq 0.
\end{align*}    
\end{corollary'}
\begin{proof}
For $\epsilon\in (0,2]$ and $t\in \mathbb{R}$, the bound in Lemma \ref{lemma:beb_bound} applies for all parameters $\theta_n^w\in \Theta_n^w$ and all $n\in\mathbb{N}$, we shall write, for simplicity, 
\begin{align*}
& \P_{\theta_n^w}\left(  \frac{\left(\arhj-\tau_n\right)^2}{    \widehat{\AVar}^{\ha}_n} + g\left(n_{A}^1\right) \geq t\right) \\
\leq & \P_{\theta_n^w}\left(  Z^2\left(1+\epsilon\right) + g\left(n_{A}^1\right)\geq t \right) + C_{\ref{constant:1}}(r,s,\epsilon,B,\delta)n^{-\frac{1}{4}}.
\end{align*}
Hence  we have, for $\epsilon\in (0,2]$ and $t\in\mathbb{R}$,
\begin{align}
    & \sup_{\theta_n^w\in\Theta_n^w}\left(\P_{\theta_n^w}\left(  \frac{\left(\arhj-\tau_n\right)^2}{    \widehat{\AVar}^{\ha}_n} + g\left(n_{A}^1\right) \geq t\right) -\P_{\theta_n^w}\left(  Z^2+ g(n_A^1) \geq t\right)\right)\nonumber \\
    \leq & \sup_{\theta_n^w\in\Theta_n^w}\left( \P_{\theta_n^w}\left(  Z^2\left(1+\epsilon\right)+ g(n_A^1) \geq t\right)-\P_{\theta_n^w}\left(  Z^2+ g(n_A^1) \geq t\right)\right)\nonumber \\
     & + C_{\ref{constant:1}}(r,s,\epsilon,B,\delta)n^{-\frac{1}{4}} \nonumber \\
    \leq &  \sqrt{\frac{\epsilon\left(|t|+C\right)}{2\pi\left(1+\epsilon\right)}}+ C_{\ref{constant:1}}(r,s,\epsilon,B,\delta)n^{-\frac{1}{4}},   \label{eqn:2025102082}
\end{align}
where for (\ref{eqn:2025102082}) we used the fact that
\begin{equation*}
     \sup_{\theta_n^w\in\Theta_n^w}\left( \P_{\theta_n^w}\left(  Z^2\left(1+\epsilon\right)+ g(n_A^1) \geq t\right)-\P_{\theta_n^w}\left(  Z^2+ g(n_A^1) \geq t\right)\right) \leq \sqrt{\frac{\epsilon\left(|t|+C\right)}{2\pi\left(1+\epsilon\right)}},
\end{equation*}
which is used in Lemma \ref{lemma:squareroot_continuity}. 
Taking $\limsup_{n\to\infty}$ followed by $\lim\sup_{\epsilon\to 0}$ yields the stated result.
\end{proof}

We note that $\T_n^0$, $\T_n^1$ and $\T_n^2$ defined in (\ref{eqn:t0}), (\ref{eqn:t1}) and (\ref{eqn:t2}) has the form
\begin{equation*}
  \frac{\left(\arhj-\tau_n\right)^2}{    \widehat{\AVar}^{\ha}_n} + g_i\left(n_{A}^1\right)
\end{equation*}
with associated the $g_i$, $i\in\{0,1,2\}$, functions:
\begin{align}
    & g_0(n_A^1) =0,g_1(n_A^1) = \V_n^{-1}(A)\left(\frac{n_A^1}{n_1} - \frac{n_A-n_A^1}{n_0}\right)^2 \label{eqn:2025102385}\\
    & g_2(n_A^1) = \V_n^{-1}(A)\left(\left\lfloor\frac{n_A^1}{n_1} - \frac{n_A-n_A^1}{n_0}\right\rfloor_{-}\right)^2 .\label{eqn:2025102386}
\end{align}
Theorem \ref{thm:feasible_clt} characterizes the asymptotic distributions of $\T_n^0$, $\T_n^1$, and $\T_n^2$, evaluated at the true always-reporter table $A$ implied by each $\theta_n^w$.

\begin{theorem'}\label{thm:feasible_clt}
Under the same setup as in Lemma \ref{lemma:beb_bound}, for every $t\in\mathbb{R}$,
\begin{align*}
\limsup_{n\to\infty}\ &\sup_{\theta_n^w\in\Theta_n^w}\left(
  \P_{\theta_n^w}\left(\T_n^i\ge t \right)
-\P_{\theta_n^w}\left( Z^2+g_i(n_A^1)\ge t \right)\right)\le 0.
\end{align*}
for $i\in \{0,1,2\}$, where $\T_n^0$, $\T_n^1$ and $\T_n^2$ are defined in (\ref{eqn:t0}), (\ref{eqn:t1}) and (\ref{eqn:t2}) respectively.
\end{theorem'}
\begin{proof}
It is clear that $\sup_{n\in\mathbb{N}}\sup_{\theta_n\in\Theta_n^w}\E_{\theta_n}\left[\left|g_0(n_A^1)\right|\right]=0$. Since $0\leq g_2(n_A^1)\leq g_1(n_A^1)$, we only need to verify the condition for $g_1(n_A^1)$. Now,
\begin{equation*}
    \E_{\theta_n^w}\left[g_1(n_A^1)\right]= \V_n^{-1}\left(A\right)\E_{\theta_n^w}\left[\left(\frac{n_A^1}{n_1} - \frac{n_A-n_A^1}{n_0}\right)^2\right]=  \V_n^{-1}\left(A\right) \V_n\left(A\right)=1,
\end{equation*}
for all $\theta_n^w\in \Theta_n^w$ and all $n\in\mathbb{N}$. Then the conclusion follows from Corollary \ref{corollary:g}.
\end{proof}
\subsection{Randomization Critical Value}\label{section:rand_cv}
This section analyzes the properties of randomization critical values. Our main result, Theorem \ref{thm:thmA24}, establishes their convergence and thus justifies the asymptotic validity of using the randomization-based critical value.

Section \ref{section:quantile_properties} states some properties of the quantiles of the random variables $Z^2+g(n_A^1)$, where $Z\sim \mathbb{N}(0,1)$. Section \ref{section:quantileusage} uses these results to prove Theorem \ref{thm:thmA24}.
\subsubsection{Preliminary Notations}

We use the shorthand notation $\Theta^w_n =\Theta^w_n (\delta, s, r , B)$, as defined in \eqref{defn:theta_w}. We typically denote an element in $\Theta_n^w$ by $\theta_n^w$. 

We use $\chi^2_{1-\alpha}$ to denote the $(1-\alpha)$-quantile of a chi-square
random variable with one degree of freedom. Note that if
$Z \sim \N(0,1)$, then $Z^2$ follows a chi-square distribution with
one degree of freedom. 


Let \(Y=(Y_i)_{i=1}^n\) denote the observed outcomes and \(A=(A_i)_{i=1}^n\) a reporting table.
Let \(D^*=(D_i^*)_{i=1}^n \sim \mathrm{CR}(n,n_1)\) be an independently generated treatment assignment vector. We define the test-statistic based on the randomization distribution as:
\begin{equation}\label{eqn:zs}
    Z^* = \frac{  \arhjs\left(D^*\right)}{    \sqrt{\widehat{\AVar}^{\ha*}_n\left( D^* \right)}},
\end{equation}
where, 
\begin{equation*}
        \arhjs\left(D^*\right)=\frac{\sum_{i=1}^n D^*_iA_iY_i}{\sum_{i=1}^n D^*_iA_i} -\frac{\sum_{i=1}^n \left(1-D^*_i\right)A_iY_i}{\sum_{i=1}^n \left(1-D^*_i\right)A_i} , 
    \end{equation*}
and
\begin{equation*}
     \widehat{\AVar}^{\ha*}_n\left( D^* \right)= \frac{\sum_{i=1}^n D^*_iA_i\left( Y_i-\widehat{\tau}^{1*}_A\right)^2}{\left(\sum_{i=1}^n D^*_iA_i\right)^2}  +  \frac{\sum_{i=1}^n (1-D^*_i)A_i\left( Y_i-\widehat{\tau}^{0*}_A\right)^2}{\left(\sum_{i=1}^n \left(1-D^*_i\right)A_i\right)^2} ,
\end{equation*}
with,
\begin{align*}
    \widehat{\tau}^{1*}_A=\frac{\sum_{i=1}^n D^*_iA_iY_i}{\sum_{i=1}^n D^*_iA_i}, \text{ and, }
    \widehat{\tau}^{0*}_A=\frac{\sum_{i=1}^n \left(1-D^*_i\right)A_iY_i}{\sum_{i=1}^n \left(1-D^*_i\right)A_i}.
\end{align*}
We also define 
\begin{equation}\label{eqn:nas}
    n_{A}^{1*}=\sum_{i=1}^n D_i^* A_i,n_{A}^{0*}=\sum_{i=1}^n \left(1-D_i^*\right) A_i.
\end{equation}
Given a measurable function \(g\), define the randomization distribution function of \((Z^*)^2+g(n_{A}^{1*})\) based on the observed data by
\begin{equation}\label{eqn:rand_dist}
   t \;\longmapsto\; \P_{\theta_n^w}\!\left(\left(Z^*\right)^2 + g\left(n_{A}^{1*}\right) \le t \,\middle|\, D^{\obs}\right), \qquad t\in\mathbb{R}.
\end{equation}
We write “\(|\,D^{\obs})\)” to emphasize the fact that the distribution is conditioned on the realized assignments. Unless noted otherwise, we define $n_A^1=\sum_{i=1}^n D_iA_i$.

\subsubsection{Properties of the Quantiles}\label{section:quantile_properties}
Lemma \ref{lemma:bounded_quantile} states that the quantiles of $Z^2 + g(n_A^1)$ are uniformly bounded in $n$, under suitable regularity conditions on $g$.

\begin{lemma'}\label{lemma:bounded_quantile}
Let $g:\mathbb{R}\to \mathbb{R} $ be a function such that
\begin{equation*}
\sup_{n\in\mathbb{N}}\sup_{\theta_n^w\in\Theta_n^w}\E_{\theta_n^w}\left[\left|g(n_A^1)\right|\right]\leq C<\infty,
\end{equation*}
for a positive constant $C$. Given an $\alpha\in \left(0,1\right)$, let $q_n^{1-\alpha}$ be the $\left(1-\alpha\right)$-th quantile of the random variable $Z^2+g(n_A^1)$, where $Z\sim \N\left(0,1\right)$ is independent of $n_A^1$. We have
\begin{equation}
    \sup_{n\in\mathbb{N}}\sup_{\theta_n^w\in\Theta_n^w} q_n^{1-\alpha} \leq \chi^2_{1-\alpha/2} + \frac{2C}{\alpha}
\end{equation}
\end{lemma'}
\begin{proof}
We suppress the dependence on $\theta_n^w$. We have
\begin{align*}
    &\P\left( Z^2+g(n_A^1) \geq \chi^2_{1-\alpha/2} + \frac{C}{\alpha/2} \right)\leq  \P\left( Z^2\geq \chi^2_{1-\alpha/2}\right)+\P\left( g\left(n_A^1\right)\geq \frac{C}{\alpha/2}\right)\\
    \leq & \P\left( Z^2\geq \chi^2_{1-\alpha/2}\right)+\P\left( \left|g\left(n_A^1\right)\right|\geq \frac{2C}{\alpha}\right)\leq \frac{\alpha}{2} + \frac{\alpha}{2} \frac{\E\left[ \left|g\left(n_A^1\right)\right|\right]}{C}\\
    \leq & \alpha.
\end{align*}
Hence we must have $q_n^{1-\alpha}\leq  \chi^2_{1-\alpha/2} + 2C/\alpha$ for all $n\in \mathbb{N}$.
\end{proof}
We note that $Z^2$ with $Z\sim \N\left(0,1\right)$ follows a Chi-squared $\chi^2_1$ distribution with one degree of freedom. $Z^2$ has a density function:
\begin{equation*}
    f_{\chi^2_1}(t) = \frac{1}{\sqrt{2\pi}}t^{-\frac{1}{2}}e^{-\frac{t}{2}},t\in \left(0,\infty\right),
\end{equation*}
which is strictly decreasing on its support. 

\begin{lemma'}\label{lemma:positve_quantile}
Consider a vector of always–reporter indicators $A=\left(A_i\right)_{i=1}^n$, and a completely randomized design $D=\left(D_i\right)_{i=1}^n\sim \mathrm{CR}(n,n_{1})$. Let $n_{0}=n-n_{1}$ and $n_{A}^1=\sum_{i=1}^n D_iA_i$.  Let $Z\sim \textrm{N}(0,1)$ be a normal random variable with mean 0 and variance 1 that is independent of $D$ and $g:\mathbb{R}\to \mathbb{R}$ be an arbitrary function. For every $\alpha\in(0,1)$, $Z^2+g(n_A^1)$ has a unique quantile $q_n^{1-\alpha}$. In particular, $\P\left(Z^2+g(n_A^1)\leq q_n^{1-\alpha}\right)=1-\alpha$.
\end{lemma'}
\begin{proof}
Define $n_A=\sum_{i=1}^n A_i$. The random variable $n_A^1$ has finite support $\{0,\ldots,n_A \}$. Therefore, $Z^2+g(n_A^1)$ has the following density
\begin{equation}
    f_{Z^2+g(n_A^1)}\left(t\right)=\sum_{k=0}^{n_A} f_{\chi^2_1}\left(t-g\left(k\right)\right)p_{n_A^1}(k)\mathds{1}\left\{t\geq g\left(k\right)\right\},
\end{equation}
where $p_{n_A^1}$ is the probability mass function of $n_A^1$. It can be easily seen that the density function is strictly positve on its support. Hence the quantile is unique. 
\end{proof}

\begin{lemma'}\label{lemma:non_trivial_bound}
Let $q_n^{1-\alpha}$ be the $(1-\alpha)$-th quantile of $Z^2+g(n_A^1)$, where $Z\sim \N(0,1)$ is independent of $n_A^1$. Assume that there exist positive constants $C_1$, $C_2$ and $\gamma$, such that
\begin{equation}\label{eqn:lower_bound}
   \inf_{n\in\mathbb{N}}\inf_{\theta_n^w\in \Theta_n^w}\P_{\theta_n^w}\left( q_n^{1-\alpha}-C_1< g\left(n_A^1\right) < q_n^{1-\alpha}-C_2\right) \geq \gamma.
\end{equation}
For any $\eta\in(0,C_2]$, the following inequalities hold for all $\theta_n^w\in\Theta_n^w$ and all $n\in\mathbb{N}$:
\begin{equation*}\label{eqn:20251024105}
    \P_{\theta_n^w}\left(Z^2+g(n_A^1)\geq q_n^{1-\alpha}-\eta\right) - \P_{\theta_n^w}\left(Z^2+g(n_A^1)\geq q_n^{1-\alpha}\right) \geq \eta\gamma f_{\chi_1^2}\left(C_1\right)
\end{equation*}
\begin{equation*}\label{eqn:20251024106}
       \P_{\theta_n^w}\left(Z^2+g(n_A^1)\geq q_n^{1-\alpha}\right)- \P_{\theta_n^w}\left(Z^2+g(n_A^1)\geq q_n^{1-\alpha}+\eta\right) \geq \eta\gamma f_{\chi_1^2}\left(C_1+C_2\right).
\end{equation*}
\end{lemma'}
\begin{proof}
We suppress the dependence on $\theta_n^w$.
Given a $\eta\in (0,C_2]$, the event $\{q_n^{1-\alpha}-C_1< g\left(n_A^1\right) < q_n^{1-\alpha}-C_2\}$ is equivalent to the event $  \{\eta\leq C_2 <  q_n^{1-\alpha}-g\left(n_A^1\right) < C_1\}$. Then,
\begin{align*}
  &  \P\left(Z^2+g(n_A^1)\geq q_n^{1-\alpha}-\eta\right) - \P\left(Z^2+g(n_A^1)\geq q_n^{1-\alpha}\right)\\
 \geq  &  \E\left[\P\left( \left. q_n^{1-\alpha}-\eta-g\left(n_A^1\right)  \leq Z^2 < q_n^{1-\alpha}-g(n_A^1)\right|n_A^1\right) \mathds{1}\{C_2<q_n^{1-\alpha}- g\left(n_A^1\right) < C_1\}\right]\\
 = & \E\left[ \int_{q_n^{1-\alpha}-\eta-g\left(n_A^1\right)}^{q_n^{1-\alpha}-g(n_A^1)}f_{\chi_1^2}\left(t\right)dt \mathds{1}\{C_2<  q_n^{1-\alpha}-g(n_A^1) < C_1\}\right]\\
\geq  &  \eta  f_{\chi_1^2}\left(C_1\right)P\left(  C_2< q_n^{1-\alpha}-g(n_A^1)< C_1\right)\geq\eta\gamma f_{\chi_1^2}\left(C_1\right),
\end{align*}
where the last inequality follows from (\ref{eqn:lower_bound}), the fact that on the event $ C_2< q_n^{1-\alpha}-g(n_A^1)< C_1$ we have $q_n^{1-\alpha}-g(n_A^1)-\eta\geq 0$, and the fact the $f_{\chi_1^2}$ is a decreasing function.
Similarly,
\begin{align*}
  &  \P\left(Z^2+g(n_A^1)\geq q_n^{1-\alpha}\right)- \P\left(Z^2+g(n_A^1)\geq q_n^{1-\alpha}+\eta\right) \\
 \geq  &  \E\left[\P\left( \left. q_n^{1-\alpha}-g\left(n_A^1\right)   \leq Z^2 < q_n^{1-\alpha}-g(n_A^1)+\eta\right|n_A^1\right) \mathds{1}\{C_2< q_n^{1-\alpha}-g(n_A^1)< C_1\}\right]\\
 = & \E\left[ \int_{q_n^{1-\alpha}-g\left(n_A^1\right)}^{q_n^{1-\alpha}-g(n_A^1)+\eta}f_{\chi_1^2}\left(t\right)dt \mathds{1}\{C_2< q_n^{1-\alpha}-g(n_A^1) < C_1\}\right]\\
\geq  &  \eta  f_{\chi_1^2}\left(C_1+\eta\right)\P\left(  C_2< q_n^{1-\alpha}-g(n_A^1)< C_1\right)\\
\geq & \eta  f_{\chi_1^2}\left(C_1+C_2\right)\P\left(  C_2< q_n^{1-\alpha}-g(n_A^1)< C_1\right)\geq \eta\gamma f_{\chi_1^2}\left(C_1+C_2\right).
\end{align*}
\end{proof}

We verify equation (\ref{eqn:lower_bound}) for the $g_0,g_1$ and $g_2$ functions defined in (\ref{eqn:2025102385}) and (\ref{eqn:2025102386}).
\begin{lemma'}\label{lemma:lower bound_quantile}
Let $g:\mathbb{R}\to [0,\infty)$ be a non-negative function. For an $\alpha\in(0,1)$, let $q_n^{1-\alpha}$ be the $\left(1-\alpha\right)$-th quantile of  $Z^2+g(n_A^1)$, where $Z\sim \N(0,1)$ is independent of $n_A^1$. We have 
\begin{equation*}
    q_n^{1-\alpha} \geq \chi^2_{1-\alpha},
\end{equation*}
for all $\theta_n^w\in\Theta_n^w$ and $n\in\mathbb{N}$.
\end{lemma'}
\begin{proof}
Note that for all $\theta_n^w\in\Theta_n^w$ and $n\in\mathbb{N}$
\begin{align*}
     &\P_{\theta_n^w}\left(Z^2+g(n_A^1) \leq \chi^2_{1-\alpha}\right) \leq  \P_{\theta_n^w}\left(Z^2 \leq \chi^2_{1-\alpha}\right)= 1-\alpha.  
\end{align*}
The result then follows immediately.
\end{proof}

The 75-percentile of a Chi-squared random variable with one degree of freedom is 1.32. Hence $\chi^2_{1-\alpha}\geq 1.32$ for $\alpha \in (0,0.25]$.
\begin{lemma'}
For $\alpha\in (0,0.25]$, the $g_0,g_1$ and $g_2$ functions defined in (\ref{eqn:2025102385}) and (\ref{eqn:2025102386}) satisfy (\ref{eqn:lower_bound}) with constants
\begin{equation*}
    C_1=\chi^2_{1-\alpha/2}+\frac{2}{\alpha}+1, C_2=0.1, \gamma=\frac{0.22}{1.22}.
\end{equation*}
\end{lemma'}
\begin{proof}

We use $q_{i,n}^{1-\alpha}$ to denote the quantile associated with $Z^2+g_i(n_A^1)$ for $i\in\{0,1,2\}$, where $Z\sim \N(0,1)$ is independent of $n_A^1$. 
Following the calculations in the proof of Theorem \ref{thm:feasible_clt}, we have
\begin{equation*}
\sup_{n\in\mathbb{N}}\sup_{\theta_n^w\in\Theta_n^w}\E_{\theta_n^w}\left[\left|g_i\left(n_A^1\right)\right|\right]\leq 1, i\in\{0,1,2\}.
\end{equation*}
Hence by Lemma \ref{lemma:bounded_quantile}, $\sup_{n\in\mathbb{N}}\sup_{\theta_n^w\in\Theta_n^w} q_{i,n}^{1-\alpha} \leq \chi^2_{1-\alpha/2}+2/\alpha$. Let $C_1=\chi^2_{1-\alpha/2}+2/\alpha+1$ and $C_2=0.1$. Notice $q_{i,n}^{1-\alpha}-C_1<0$ for all $i\in\{0,1,2\}$ and $n\in\mathbb{N}$.\\
For $g_0$, we have $q_{0,n}^{1-\alpha}=\chi^2_{1-\alpha}$,
\begin{equation*}
    \P_{\theta_n^w}\left(q_{0,n}^{1-\alpha}-C_1 < g_0 < q_{0,n}^{1-\alpha}-C_2\right) \geq    \P_{\theta_n^w}\left( 0 \leq  g_0 < \chi^2_{1-\alpha}-0.1\right)=1,
\end{equation*}
for all $\theta_n^w\in \Theta_n^w$ and all $n\in\mathbb{N}$.
For $g_1$, we have
    \begin{align}
        & \P\left(  q_{1,n}^{1-\alpha}-C_1< g_1 < q_{1,n}^{1-\alpha}-C_2\right) \geq  \P\left(  0\leq  g_1 < \chi^2_{1-\alpha}-C_2\right) \label{eqn:2025102888}\\
      = & 1- \P\left(g_1 \geq  \chi^2_{1-\alpha}-C_2\right) \nonumber  \\
      =  &1- \P\left( \left(\frac{n_A^1}{n_1}-\frac{n_A^0}{n_0}\right)^2 \geq  \left(\chi^2_{1-\alpha}-C_2\right) \V_n\left(A\right)\right)\nonumber \\
      \geq  & 1- \frac{\V_n\left(A\right)}{\V_n\left(A\right)\left(\chi^2_{1-\alpha}-C_2\right)} = 1-\frac{1}{\chi^2_{1-\alpha}-0.1} \nonumber \\
      \geq &  1- \frac{1}{1.32-0.1} = 1-\frac{1}{1.22}=\frac{0.22}{1.22},\label{eqn:2025102889}
    \end{align}
where for (\ref{eqn:2025102888}) we use Lemma \ref{lemma:lower bound_quantile} and for (\ref{eqn:2025102889}) we use the fact that $\chi^2_{1-\alpha}\geq 1.32$ for $ \alpha\in (0, 0.25]$. A similar argument holds for $g_2$.

\end{proof}

\subsubsection{Asymptotically Valid Inference with Randomization-Based Critical Values}\label{section:quantileusage}

We state a high-level theorem establishing asymptotically valid inference using a estimator of the quantile.

\begin{theorem'}\label{theorem:E5}
Let $g:\mathbb{R}\to \mathbb{R} $ be a function such that
\begin{equation*}
\sup_{n\in\mathbb{N}}\sup_{\theta_n^w\in\Theta_n^w}\E_{\theta_n^w}\left[\left|g(n_A^1)\right|\right]\leq C<\infty,
\end{equation*}
for a positive constant $C$. For an $\alpha\in(0,1)$, let $q_n^{1-\alpha}$ be the $\left(1-\alpha\right)$-th quantile of the random variable $Z^2+g(n_A^1)$, where $Z\sim \N(0,1)$ is independent of $n_A^1$.  Let $\widehat{q}_n$ be a quantile estimator such that for each $\eta>0$ we have 
\begin{equation*}
\limsup_{n\to\infty}\sup_{\theta_n^w\in\Theta_n^w}\P_{\theta_n^w}\left(\left|\widehat{q}_n-q_n^{1-\alpha}\right|\geq \eta\right)=0,
\end{equation*}
Given a $\theta_n^w\in \Theta_n^w$,  let $\{\left(y_i(1),y_i(0)\right)\}_{i=1}^n$ be the associated potential outcomes and $\{A_i\}_{i=1}^n$ be the associated always-reporter indicators. Define $\A=\{i\in[n]:A_i=1\}$ and $n_A=\sum_{i=1}^n A_i$, $\tau_n=n_A^{-1}\sum_{i\in \A} \left(y_i(1)-y_i(0)\right)$. Define the variance estimator $\widehat{\AVar}^{\ha}_{n}$ as in  (\ref{hajek:var}) and the Hajek estimator $\arhj$ as in (\ref{eqn:hajek_a}).
Then,
\begin{equation}
    \limsup_{n\to\infty}\sup_{\theta_n^w\in\Theta_n^w}\P_{\theta^w_n}\left(  \frac{\left(\arhj-\tau_n\right)^2}{    \widehat{\AVar}^{\ha}_n} + g\left(n_{A}^1\right) \geq \widehat{q}_n\right)\leq \alpha.
\end{equation}
\end{theorem'}

\begin{proof}
We suppress the dependence on $\theta_n^w$. By Corollary \ref{corollary:g} and for any $\epsilon \in (0,2]$ and $\eta>0$, we have the calculations,
\begin{align}
    & \P\left(  \frac{\left(\arhj-\tau_n\right)^2}{    \widehat{\AVar}^{\ha}_n} + g\left(n_{A}^1\right) \geq \widehat{q}_n\right) \nonumber\\
    =& \P\left(  \frac{\left(\arhj-\tau_n\right)^2}{    \widehat{\AVar}^{\ha}_n} + g\left(n_{A}^1\right) \geq \widehat{q}_n, \left|\widehat{q}_n-q_n^{1-\alpha}\right|\leq \eta \right) \nonumber \\
    & +\P\left(  \frac{\left(\arhj-\tau_n\right)^2}{    \widehat{\AVar}^{\ha}_n} + g\left(n_{A}^1\right) \geq \widehat{q}_n, \left|\widehat{q}_n-q_n^{1-\alpha}\right|> \eta \right) \nonumber \\
    \leq & \P\left(  \frac{\left(\arhj-\tau_n\right)^2}{    \widehat{\AVar}^{\ha}_n} + g\left(n_{A}^1\right) \geq  q_n^{1-\alpha}-\eta\right)+ \P\left(\left|\widehat{q}_n-q_n^{1-\alpha}\right|\geq \eta\right) \nonumber \\
    \leq & \P\left(  Z^2+ g(n_A^1) \geq q_n^{1-\alpha}-\eta\right)+ C_{\ref{constant:1}}(r,s,\epsilon,B,\delta)n^{-\frac{1}{4}}+   \label{eqn:2025102891}\\  
     & \sqrt{\frac{\epsilon \left(\left|q_n^{1-\alpha}-\eta\right|+C\right)}{2\pi\left(1+\epsilon\right)}} +\P\left(\left|\widehat{q}_n-q_n^{1-\alpha}\right|\geq \eta\right) \label{eqn:2025102892}\\
     \leq &  \P\left(  Z^2+ g(n_A^1) \geq q_n^{1-\alpha}-\eta\right)+  C_{\ref{constant:1}}(r,s,\epsilon,B,\delta)n^{-\frac{1}{4}}\nonumber \\ & +\sqrt{\frac{\epsilon }{2\pi\left(1+\epsilon\right)}} \sqrt{
     \chi^2_{1-\alpha/2} +  \frac{2C}{\alpha}+C+\eta}+\P\left(\left|\widehat{q}_n-q_n^{1-\alpha}\right|\geq \eta\right) \label{eqn:2025102893}\\
     = & \P\left(  Z^2+ g(n_A^1) \geq q_n^{1-\alpha}\right)+  C_{\ref{constant:1}}(r,s,\epsilon,B,\delta)n^{-\frac{1}{4}}\nonumber\\ 
     & +\sqrt{\frac{\epsilon }{2\pi\left(1+\epsilon\right)}} \sqrt{
     \chi^2_{1-\alpha/2} +  \frac{2C}{\alpha}+C+\eta}+\P\left(\left|\widehat{q}_n-q_n^{1-\alpha}\right|\geq \eta\right)\nonumber\\
     &+ \P\left(  Z^2+ g(n_A^1) \geq q_n^{1-\alpha}-\eta\right)-\P\left(  Z^2+ g(n_A^1) \geq q_n^{1-\alpha}\right)\nonumber\\
     \leq & \alpha + C_{\ref{constant:1}}(r,s,\epsilon,B,\delta)n^{-\frac{1}{4}} +\sqrt{\frac{\epsilon }{2\pi\left(1+\epsilon\right)}} \sqrt{
     \chi^2_{1-\alpha/2} + \frac{2C}{\alpha}+C+\eta}\nonumber \label{eqn:2025102894}\\
     &+\P\left(\left|\widehat{q}_n-q_n^{1-\alpha}\right|\geq \eta\right)+  \sqrt{\frac{2}{\pi}}\sqrt{\eta} ,
\end{align}
where we use the estimate in (\ref{eqn:2025102082}) for (\ref{eqn:2025102891}) and (\ref{eqn:2025102892}), Lemma \ref{lemma:bounded_quantile} for (\ref{eqn:2025102893}), and \eqref{eqn:square_cont} for (\ref{eqn:2025102894}), which follows from the calculation
\begin{align*}
    & \P\left(  Z^2+ g(n_A^1) \geq q_n^{1-\alpha}-\eta\right) - \P\left(  Z^2+ g(n_A^1) \geq q_n^{1-\alpha}\right)\\
   =  &\E\left[  \P\left(  \left. q_n^{1-\alpha}-g(n_A^1)-\eta\leq Z^2 \leq q_n^{1-\alpha}-g(n_A^1) \right| n_A^1  \right)\right]\\
   = &\E\left[  \P\left(  \left.q_n^{1-\alpha}-g(n_A^1)-\eta\leq Z^2 \leq q_n^{1-\alpha}-g(n_A^1) \right| n_A^1  \right)1\{q_n^{1-\alpha}-g(n_A^1)-\eta\leq 0 \}\right]\\
      &+ \E\left[  \P\left(  \left.q_n^{1-\alpha}-g(n_A^1)-\eta\leq Z^2 \leq q_n^{1-\alpha}-g(n_A^1) \right| n_A^1  \right)1\{q_n^{1-\alpha}-g(n_A^1)-\eta> 0 \}\right]\\
    \leq &\E\left[  \P\left(  \left.0\leq Z^2 \leq \eta \right| n_A^1  \right)1\{q_n^{1-\alpha}-g(n_A^1)-\eta\leq 0 \}\right]\\
      &+ \E\left[  \P\left(  \left.q_n^{1-\alpha}-g(n_A^1)-\eta\leq Z^2 \leq q_n^{1-\alpha}-g(n_A^1) \right| n_A^1  \right)1\{q_n^{1-\alpha}-g(n_A^1)-\eta> 0 \}\right]\\ 
     \leq  & \sqrt{\frac{2}{\pi}}\sqrt{\eta},
\end{align*}
Taking $\limsup_{n\to\infty}$ on both sides yields,
\begin{align*}
    & \limsup_{n\to\infty} \sup_{\theta_n^w\in \Theta_n^w} \P_{\theta_n^w}\left(  \frac{\left(\arhj-\tau_n\right)^2}{    \widehat{\AVar}^{\ha}_n} + g\left(n_{A}^1\right) \geq  \widehat{q}_n\right) \\
    \leq &  \alpha +\sqrt{\frac{\epsilon }{2\pi\left(1+\epsilon\right)}} \sqrt{
     \chi^2_{1-\alpha/2} + \frac{2C}{\alpha}+C+\eta}+  \sqrt{\frac{2}{\pi}}\sqrt{\eta}.
\end{align*}
We note that the LHS of the inequality above does not depend on $\epsilon$ or $\eta$. Taking $\lim_{\epsilon\to 0}$ and $\lim_{\eta\to 0}$ gives the desired results.
\end{proof}

Hence all we need to show is that, for all $\eta>0$,
\begin{equation*}
\limsup_{n\to\infty}\sup_{\theta_n^w\in\Theta_n^w}\P_{\theta_n^w}\left(\left|\widehat{q}_n-q_n^{1-\alpha}\right|\geq \eta\right)=0.
\end{equation*}

\begin{lemma'}\label{lemma:E6}
Let $g$ be a non-negative function such that
\begin{equation*}
\sup_{n\in\mathbb{N}}\sup_{\theta_n^w\in\Theta_n^w}\E_{\theta_n^w}\left[\left|g(n_A^1)\right|\right]\leq C<\infty,
\end{equation*} 
for a positive constant $C$ and satisfies the condition (\ref{eqn:lower_bound}) in Lemma \ref{lemma:non_trivial_bound} with positive constants $\gamma$, $C_1$ and $C_2$. Let $q_n^{1-\alpha}$ be the $\left(1-\alpha\right)$-th quantile of  $Z^2+g(n_A^1)$, where $Z\sim \N(0,1)$ is independent of $n_A^1$.

Let $\{\widehat{q}^{1-\alpha}_n\}_{n}^{\infty}$ be the $1-\alpha$ quantile of the randomization distribution
\begin{equation*}
  \widehat{q}^{1-\alpha}_n=\inf\left\{ t\in \mathbb{R}:  \P_{\theta_n^w}\left(\left.\left(Z^*\right)^2+g\left(n_A^{1*}\right)\leq t \right| D^\obs\right)\geq 1-\alpha\right\},
\end{equation*}
where $Z^*$ and $n_A^{1*}$ are defined in (\ref{eqn:zs}) and (\ref{eqn:nas}), and the randomization distribution is defined in \eqref{eqn:rand_dist}. \\
\noindent For a given $\eta>0$, define the interval,
\begin{equation}
    I_{\eta}=\left[\chi^2_{1-\alpha}-\eta,\chi^2_{1-\alpha/2}+ \frac{2C}{\alpha}+\eta\right].
\end{equation}
On the event,
\begin{equation}
    \left|\widehat{q}^{1-\alpha}_n-q_n^{1-\alpha}\right|>\eta,
\end{equation}
we must have the event $\mathcal{E}_n$:
\begin{align*}
      &\sup_{t\in I_{\eta}}\left| \P_{\theta_n^w}\left(Z^2+g(n_A^1)\leq t \right) -  \P_{\theta_n^w}\left(\left.\left(Z^*\right)^2+g\left(n_A^{1*}\right)\leq t \right| D^\obs\right)\right|\\
 \geq &  \min\{ \eta\gamma f_{\chi_1^2}\left(C_1\right), \eta\gamma f_{\chi_1^2}\left(C_1+C_2\right)\}  
\end{align*}
In particular, if 
\begin{equation*}
\limsup_{n\to\infty}\sup_{\theta_n^w\in\Theta_n^w}\P_{\theta_n^w}\left( \left|\widehat{q}^{1-\alpha}_n-q_n^{1-\alpha}\right|>\eta \right)>\epsilon,
\end{equation*}
then
\begin{equation*}
\limsup_{n\to\infty}\sup_{\theta_n^w\in\Theta_n^w}\P_{\theta_n^w}\left(\mathcal{E}_n \right)>\epsilon.
\end{equation*}
\end{lemma'}

\begin{proof}
We suppress the dependence on $\theta_n^w$ for simplicity. On the event $ \left|\widehat{q}^{1-\alpha}_n-q_n^{1-\alpha}\right|>\eta$, we have either $  \widehat{q}^{1-\alpha}_n-q_n^{1-\alpha}>\eta$ or $\widehat{q}^{1-\alpha}_n-q_n^{1-\alpha}<-\eta$.

For the case $  \widehat{q}^{1-\alpha}_n-q_n^{1-\alpha}>\eta$, we have, $\widehat{q}^{1-\alpha}_n>q_n^{1-\alpha}+\eta>q_n^{1-\alpha}$. By the definition of $\widehat{q}^{1-\alpha}_n$, we have
\begin{equation*}
\P\left( \left.\left(Z^*\right)^2 + g\left(n_A^{1*}\right) \leq  q_n^{1-\alpha}+\eta\right| D^\obs\right) \leq 1-\alpha
\end{equation*}
Lemma \ref{lemma:non_trivial_bound} gives
\begin{align*}
        &\P\left(Z^2+g(n_A^1)\leq q_n^{1-\alpha}+\eta\right) \geq  \P\left(Z^2+g(n_A^1)\leq q_n^{1-\alpha}\right)+\eta\gamma f_{\chi_1^2}\left(C_1+C_2\right)\\
      =  & 1-\alpha+\eta\gamma f_{\chi_1^2}\left(C_1+C_2\right),
\end{align*}
where the second inequality is by Lemma \ref{lemma:positve_quantile}.

Hence, 
\begin{equation}\label{eqn:eq1}
\begin{split}
            & \P\left(Z^2+g(n_A^1)\leq q_n^{1-\alpha}+\eta\right) - \P\left( \left.\left(Z^*\right)^2 + g\left(n_A^{1*}\right) \leq  q_n^{1-\alpha}+\eta\right| D^\obs\right) \\
        \geq &\eta\gamma f_{\chi_1^2}\left(C_1+C_2\right)  
\end{split}
\end{equation}
Alternatively, for the case $\widehat{q}^{1-\alpha}_n-q_n^{1-\alpha}<-\eta$, we have $\widehat{q}^{1-\alpha}_n<q_n^{1-\alpha}-\eta<q_n^{1-\alpha}$. By the definition of $ \widehat{q}^{1-\alpha}_n$, we have
\begin{equation*}
    \P\left(\left.Z^* + g\left(n_A^{1*}\right) \leq  q_n^{1-\alpha}-\eta\right|D^\obs\right) \geq 1-\alpha
\end{equation*}
Lemma \ref{lemma:non_trivial_bound} gives
\begin{align*}
    &\P\left(Z^2+g(n_A^1)\leq q_n^{1-\alpha}-\eta\right) \leq   \P\left(Z^2+g(n_A^1)\leq q_n^{1-\alpha}\right)  - \eta\gamma f_{\chi_1^2}\left(C_1\right)\\
    =  &  \left(1-\alpha\right) - \eta\gamma f_{\chi_1^2}\left(C_1\right) 
\end{align*}
Hence, 
\begin{equation}\label{eqn:eq2}
\begin{split}
 & \P\left(Z^2+g(n_A^1)\leq q_n^{1-\alpha}-\eta\right) -  \P\left( \left.Z^* + g\left(n_A^{1*}\right) \leq  q_n^{1-\alpha}-\eta\right|D^\obs\right) \\
 \leq & - \eta\gamma f_{\chi_1^2}\left(C_1\right).    
\end{split}
\end{equation}
Notice by Lemma \ref{lemma:bounded_quantile} and Lemma \ref{lemma:lower bound_quantile}, $q_n^{1-\alpha}-\eta \in I_{\eta}$ and $q_n^{1-\alpha}+\eta \in I_{\eta}$. Combining equations (\ref{eqn:eq1}) and (\ref{eqn:eq2}) yields the stated results. 

\end{proof}
Given a vector of always-reporter indicators $A=\left(A_i\right)_{i=1}^n$. Define $\A=\{i\in[n]:A_i=1\}$ and $n_A=\sum_{i=1}^n A_i$. Given realized outcomes, we define
\begin{equation}\label{eqn:ybarA}
    \overline{Y}_A=\frac{1}{n_A}\sum_{i\in \A} Y_i,   v_A^*=\frac{1}{n_A-1}\sum_{i\in \A} \left(Y_i -\overline{Y}_A\right)^2.
\end{equation}
As (\ref{eqn:target_variance}), we define the target variance conditioning on the realized outcomes, with $v^{1*}_{A} = v^{0*}_{A}= v_A^*$,
\begin{equation*}
    \AVar_{n,k}^* = \frac{1}{k}v^{1*}_{A}+\frac{1}{n_A-k}v^{0*}_{A},
\end{equation*}
for $k\in [1,n_A-1]$. We note that if $\AVar^*_{n,k}=0$ for some $k$, then all $Y_i,i\in \A$ are identical. Hence $\AVar^*_{n,k}=0$ for all $k\in [1,n_A-1]$, and $\arhjs$ and $ \widehat{\AVar}^{\ha*}_n$ are 0 conditionally almost surely. We shall define $0/0\equiv 0$ when such event happens.  

\begin{theorem'}\label{thm:E7}
Let $g$ be a non-negative function such that
\begin{equation*}
\sup_{n\in\mathbb{N}}\sup_{\theta_n^w\in\Theta_n^w}\E_{\theta_n^w}\left[\left|g(n_A^1)\right|\right]\leq C<\infty,
\end{equation*}
for a positive constant $C$. Given a $\theta_n^w\in \Theta_n^w$,  let $\{A_i\}_{i=1}^n$ be the associated always-reporter indicators and define $\A=\{i\in[n]:A_i=1\}$. Suppose $n_A\geq 2$. For each $t\geq 0$ and $\epsilon\in(0,1)$, we have,
\begin{align*}
& \left|\left(\P_{\theta_n^w}\left( \left.\left(Z^*\right)^2 + g\left(n_A^{1*}\right) \leq  t\right| D^{\obs}\right) \right)- \P_{\theta_n^w}\left(Z^2+g(n_A^1)\leq t\right) \right|\\
 \leq &  \frac{2(1-r)}{rs^2(n-1)} + \sqrt{\frac{2}{r}}C_{\ref{eqn:53}}(\delta,r) \frac{1}{\sqrt{n_A}}\frac{\max_{i\in\A} \left|Y_i-\overline{Y}_A\right|}{\sqrt{v_A^*}} \\
      & +  C_{\ref{constant:2}}(r,s,\epsilon)n^{-1} +C_{\ref{constant:3}}\left(\epsilon,r\right) \frac{1}{n_A}\frac{\max_{i\in\A} \left|Y_i-\overline{Y}_A\right|^2}{v_A^*}  \\
    &+\sqrt{\frac{2\epsilon\left(t+C\right)}{\pi\left(1-\epsilon\right)}} +\sqrt{\frac{2\epsilon\left(t+C\right)}{\pi\left(1+\epsilon\right)}} ,    
\end{align*}
where $C_{\ref{eqn:53}}(\delta,r)$ is defined in \eqref{eqn:53}, $C_{\ref{constant:3}}\left(\epsilon,r\right)$ is defined in \eqref{constant:3}, $ C_{\ref{constant:2}}(r,s,\epsilon)$ is defined in \eqref{constant:2}, $\{Y_i\}_{i\in \A}$ are observed outcomes, and $\Bar{Y}_A$ and $v_A^*$ are defined in \eqref{eqn:ybarA}. Constants $r$, $s$, $\delta$ and $B$ are defined in Assumption \ref{a:cr}, Assumption \ref{a:theta}-\ref{a:theta1}, Assumption \ref{a:theta}-\ref{a:theta2} and Assumption \ref{a:theta}-\ref{a:theta3}, respectively.
\end{theorem'}

\begin{proof}
Recall the definition of $Z^*$ from \eqref{eqn:zs}. We write $\arhjs\left( D^* \right)$ as $\arhjs$, and $\widehat{\AVar}^{\ha*}_n\left( D^* \right)$ as $\widehat{\AVar}^{\ha*}_n$. We suppress the dependence on $\theta_n^w$.
We have the following inequality:
\begin{align}
    &\P\left( \left.\frac{\left(\arhjs\right)^2}{ \widehat{\AVar}^{\ha*}_n}+g\left(n_A^{1*}\right)\leq t \right| D^{\obs}\right) \nonumber\\
  =  & \P\left( \left.\frac{\left(\arhjs\right)^2}{\AVar_{n,n_A^{1*}}^* }\frac{\AVar_{n,n_A^{1*}}^*}{\widehat{\AVar}^{\ha*}_n}+g\left(n_A^{1*}\right)\leq t \right| D^{\obs}\right) \nonumber \\
  = & \P\left( \left.\frac{\left(\arhjs\right)^2}{\AVar_{n,n_A^{1*}}^* }\frac{\AVar_{n,n_A^{1*}}^*}{\widehat{\AVar}^{\ha*}_n}+g\left(n_A^{1*}\right)\leq t ,\frac{\AVar_{n,n_A^{1*}}^*}{\widehat{\AVar}^{\ha*}_n} \geq 1-\epsilon\right| D^{\obs}\right) \nonumber \\
  & + \P\left( \left.\frac{\left(\arhjs\right)^2}{\AVar_{n,n_A^{1*}}^* }\frac{\AVar_{n,n_A^{1*}}^*}{\widehat{\AVar}^{\ha*}_n}+g\left(n_A^{1*}\right)\leq t ,\frac{\AVar_{n,n_A^{1*}}^*}{\widehat{\AVar}^{\ha*}_n} < 1-\epsilon\right| D^{\obs}\right) \nonumber \\
   \leq & \P\left( \left.\frac{\left(\arhjs\right)^2}{\AVar_{n,n_A^{1*}}^* }\left(1-\epsilon\right)+g\left(n_A^{1*}\right)\leq t \right| D^{\obs}\right) + \P\left(\left.\frac{\AVar_{n,n_A^{1*}}^*}{\widehat{\AVar}^{\ha*}_n} < 1-\epsilon \right| D^{\obs} \right)\nonumber
\end{align}
Similarly,
\begin{align*}
    & \P\left( \left.\frac{\left(\arhjs\right)^2}{ \widehat{\AVar}^{\ha*}_n}+g\left(n_A^{1*}\right)> t \right| D^{\obs}\right) \nonumber\\
  =  & \P\left( \left.\frac{\left(\arhjs\right)^2}{\AVar_{n,n_A^{1*}}^* }\frac{\AVar_{n,n_A^{1*}}^*}{\widehat{\AVar}^{\ha*}_n}+g\left(n_A^{1*}\right)> t \right| D^{\obs}\right) \nonumber \\    
 = & \P\left( \left.\frac{\left(\arhjs\right)^2}{\AVar_{n,n_A^{1*}}^* }\frac{\AVar_{n,n_A^{1*}}^*}{\widehat{\AVar}^{\ha*}_n}+g\left(n_A^{1*}\right)> t , \frac{\AVar_{n,n_A^{1*}}^*}{\widehat{\AVar}^{\ha*}_n}\leq 1+\epsilon\right| D^{\obs}\right) \nonumber \\  
  & + \P\left( \left.\frac{\left(\arhjs\right)^2}{\AVar_{n,n_A^{1*}}^* }\frac{\AVar_{n,n_A^{1*}}^*}{\widehat{\AVar}^{\ha*}_n}+g\left(n_A^{1*}\right)> t , \frac{\AVar_{n,n_A^{1*}}^*}{\widehat{\AVar}^{\ha*}_n}> 1+\epsilon\right| D^{\obs}\right)\\
 \leq  &\P\left( \left.\frac{\left(\arhjs\right)^2}{\AVar_{n,n_A^{1*}}^* }\left(1+\epsilon\right)+g\left(n_A^{1*}\right)> t\right| D^{\obs}\right) \nonumber + \P\left( \left.\frac{\AVar_{n,n_A^{1*}}^*}{\widehat{\AVar}^{\ha*}_n}> 1+\epsilon\right| D^{\obs}\right)
\end{align*}
Hence,
\begin{align*}
     & \P\left( \left.\frac{\left(\arhjs\right)^2}{ \widehat{\AVar}^{\ha*}_n}+g\left(n_A^{1*}\right)\leq t \right| D^{\obs}\right) \\
     \geq &\P\left( \left.\frac{\left(\arhjs\right)^2}{\AVar_{n,n_A^{1*}}^* }\left(1+\epsilon\right)+g\left(n_A^{1*}\right)\leq  t\right| D^{\obs}\right) -  \P\left( \left.\frac{\AVar_{n,n_A^{1*}}^*}{\widehat{\AVar}^{\ha*}_n}> 1+\epsilon\right| D^{\obs}\right)    
\end{align*}

We have the inequalities:
\begin{align*}
     & \P\left( \left.\frac{\left(\arhjs\right)^2}{ \widehat{\AVar}^{\ha*}_n}+g\left(n_A^{1*}\right)\leq  t \right| D^{\obs}\right) -   \P\left( Z^2+g\left(n_A^{1}\right)\leq  t \right)\\
     \geq & \P\left( \left.\frac{\left(\arhjs\right)^2}{ \widehat{\AVar}^{\ha*}_n}\left(1+\epsilon\right)+g\left(n_A^{1*}\right)\leq  t \right| D^{\obs}\right)-   \P\left( Z^2\left(1+\epsilon\right)+g\left(n_A^1\right)\leq  t \right)\\
     & +   \P\left( Z^2\left(1+\epsilon\right)+g\left(n_A^{1}\right)\leq  t \right)- \P\left( Z^2+g\left(n_A^1\right)\leq  t \right) \\
     & -  \P\left( \left.\frac{\AVar_{n,n_A^{1*}}^*}{\widehat{\AVar}^{\ha*}_n}> 1+\epsilon\right| D^{\obs}\right),    
\end{align*}
and,
\begin{align*}
    & \P\left( \left.\frac{\left(\arhjs\right)^2}{ \widehat{\AVar}^{\ha*}_n}+g\left(n_A^{1*}\right)\leq t \right| D^{\obs}\right)- \P\left( Z^2+g\left(n_A^{1}\right)\leq  t \right)\\
    \leq &     \P\left( \left.\frac{\left(\arhjs\right)^2}{\AVar_{n,n_A^{1*}}^* }\left(1-\epsilon\right)+g\left(n_A^{1*}\right)\leq t \right| D^{\obs}\right)- \P\left( Z^2\left(1-\epsilon\right)+g\left(n_A^{1}\right)\leq  t \right)\\
     & + \P\left( Z^2\left(1-\epsilon\right)+g\left(n_A^{1}\right)\leq  t \right)- \P\left( Z^2+g\left(n_A^{1}\right)\leq  t \right)\\
     & + \P\left(\left.\frac{\AVar_{n,n_A^{1*}}^*}{\widehat{\AVar}^{\ha*}_n} < 1-\epsilon \right| D^{\obs} \right)
\end{align*}
Note we have $x_1\leq x \leq x_2$ implies $|x|\leq \max\{ |x_1|,|x_2|\}$. Hence
\begin{align}
    & \left|\P\left( \left.\frac{\left(\arhjs\right)^2}{ \widehat{\AVar}^{\ha*}_n}+g\left(n_A^{1*}\right)\leq t \right| D^{\obs}\right)- \P\left( Z^2+g\left(n_A^{1}\right)\leq  t \right)\right|\nonumber \\
   \leq  & \sup_{s\geq 0,t}\left| \P\left( \left.\frac{s\left(\arhjs\right)^2}{\AVar_{n,n_A^{1*}}^* }+g\left(n_A^{1*}\right)\leq t \right| D^{\obs}\right)- \P\left( sZ^2+g\left(n_A^1\right)\leq  t \right)\right| \nonumber\\
   & +  \left|\P\left( Z^2\left(1-\epsilon\right)+g\left(n_A^1\right)\leq  t \right)- \P\left( Z^2+g\left(n_A^1\right)\leq  t \right)\right|\nonumber\\
   & + \left| \P\left( Z^2\left(1+\epsilon\right)+g\left(n_A^1\right)\leq  t \right)- \P\left( Z^2+g\left(n_A^1\right)\leq  t \right)\right|\nonumber\\
   & +\P\left( \left|\left.\frac{\AVar_{n,n_A^{1*}}^*}{\widehat{\AVar}^{\ha*}_n}-1\right |>\epsilon \right| D^{\obs} \right)\nonumber\\
   \leq &  \frac{2\left(1-r\right)}{rs^2(n-1)} + \sqrt{\frac{2}{r}}C_{\ref{eqn:53}}(\delta,r) \frac{1}{\sqrt{n_A}}\frac{\max_{i\in\A} \left|Y_i-\overline{Y}_A\right|}{\sqrt{v_A^*}} \label{eqn:20261029120} \\
      & +  C_{\ref{constant:2}}(r,s,\epsilon)n^{-1} +C_{\ref{constant:3}}\left(\epsilon,r\right) \frac{1}{n_A}\frac{\max_{i\in\A} \left|Y_i-\overline{Y}_A\right|^2}{v_A^*} \label{eqn:20261029121} \\
    & +\sqrt{\frac{2\epsilon\left(t+C\right)}{\pi\left(1-\epsilon\right)}} +\sqrt{\frac{2\epsilon\left(t+C\right)}{\pi\left(1+\epsilon\right)}} \label{eqn:20261029122},
\end{align}
(\ref{eqn:20261029120}) follows from a same calculation as in (\ref{eqn:53}), (\ref{eqn:20261029121}) follows from a same calculation as in (\ref{eqn:2025102979}), and (\ref{eqn:20261029122}) follows from Lemma \ref{lemma:squareroot_continuity} and the fact that $t \geq 0$.
\end{proof}

We collect the several algebraic identities in the lemma below.
\begin{lemma'}\label{lemma:E8}
Let $\A$ be the set of always-reporters and $D=\left(D_i\right)_{i=1}^n$ be the observed assignments. Define $\overline{Y}_A$ as in (\ref{eqn:ybarA}) and $\widehat{\tau}_A^1$ and $  \widehat{\tau}_A^0$ as in (\ref{def:avg}). Denote $n_A^1=\sum_{i=1}^n D_iA_i$ and $n_A^0=\sum_{i=1}^n \left(1-D_i\right)A_i$. We have the following algebraic identities:
\begin{enumerate}[label=(\roman*)]
    \item $  \widehat{\tau}_A^1- \overline{Y}_A = n_A^{-1}n_A^0\left(\widehat{\tau}_A^1-\widehat{\tau}_A^0\right)$, $\widehat{\tau}_A^0- \overline{Y}_A = -n_A^{-1}n_A^1\left(\widehat{\tau}_A^1-\widehat{\tau}_A^0\right)$.
    \item $ \left|Y_i- \overline{Y}_A \right|\leq \left| y_i(1)-\overline{y}_A(1)\right| + \left|\overline{y}_A(1) - \widehat{\tau}_A^1\right| + n_A^{-1}n_A^0\left|\widehat{\tau}_A^1-\widehat{\tau}_A^0\right|$ for all $i\in\A$ if $D_i=1$.
    \item $ \left|Y_i- \overline{Y}_A \right|\leq \left| y_i(0)-\overline{y}_A(0)\right| + \left|\overline{y}_A(0) - \widehat{\tau}_A^0\right| + n_A^{-1}n_A^1\left|\widehat{\tau}_A^1-\widehat{\tau}_A^0\right|$ for all $i\in\A$ if $D_i=0$.
    \item $ \sum_{i\in\A}\left(Y_i- \overline{Y}_A \right)^2 = \sum_{a\in\{0,1\}}\sum_{i\in\A,D_i=a}\left(y_i(a)-\widehat{\tau}_A^a \right)^2 +  n_A^{-1}n_A^1n_A^0 \left( \widehat{\tau}_A^1 -\widehat{\tau}_A^0 \right)^2.$
\end{enumerate}
\end{lemma'}
\begin{proof}
By definition we have
\begin{equation*}
    \overline{Y}_A =  \frac{n_A^1}{n_A}\widehat{\tau}_A^1+\frac{n_A^0}{n_A}\widehat{\tau}_A^0.  
\end{equation*}
Hence, 
\begin{align*}
    \widehat{\tau}_A^1- \overline{Y}_A = \frac{n_A^0}{n_A}\left(\widehat{\tau}_A^1-\widehat{\tau}_A^0\right),    \widehat{\tau}_A^0- \overline{Y}_A = -\frac{n_A^1}{n_A}\left(\widehat{\tau}_A^1-\widehat{\tau}_A^0\right).
\end{align*}
This proves (i). Suppose $D_i=1$,
\begin{align*}
    & \left|Y_i- \overline{Y}_A \right| =  \left|y_i(1)- \left(  \frac{n_A^1}{n_A}\widehat{\tau}_A^1+\frac{n_A^0}{n_A}\widehat{\tau}_A^0\right) \right| = \left|y_i(1)-\widehat{\tau}_A^1+ \frac{n_A^0}{n_A} \left( \widehat{\tau}_A^1-\widehat{\tau}_A^0\right) \right|\\
   \leq & \left| y_i(1)-\ytA\right| + \left|\ytA - \widehat{\tau}_A^1\right| + \frac{n_A^0}{n_A}\left|\widehat{\tau}_A^1-\widehat{\tau}_A^0\right|.
\end{align*}
This proves (ii) and, by symmetry, (iii). For (iv), we have,
\begin{align*}
    \sum_{i\in\A}\left(Y_i- \overline{Y}_A \right)^2
    =& \sum_{a\in\{0,1\}}\sum_{i\in\A,D_i=a} \left(Y_i - \widehat{\tau}_A^a + \widehat{\tau}_A^a - \overline{Y}_A \right)^2 \\
    =& \sum_{a\in\{0,1\}}\sum_{i\in\A,D_i=a}\left(y_i(a)-\widehat{\tau}_A^a \right)^2 + \sum_{a\in\{0,1\}}n_A^a \left( \widehat{\tau}_A^a -\overline{Y}_A\right)^2\\
    =&  \sum_{a\in\{0,1\}}\sum_{i\in\A,D_i=a}\left(y_i(a)-\widehat{\tau}_A^a \right)^2 +  \frac{n_A^1n_A^0}{n_A} \left( \widehat{\tau}_A^1 -\widehat{\tau}_A^0 \right)^2,
\end{align*}
where the third equality uses (i).
\end{proof}

\begin{lemma'}\label{lemma:E9}
 Given a $\theta_n^w\in \Theta_n^w$,  let $\{A_i\}_{i=1}^n$ be the associated always-reporter indicators and define $\A=\{i\in[n]:A_i=1\}$ and $n_A=\sum_{i=1}^n A_i$. Given the observed outcomes $Y=\left(Y_i\right)_{i=1}^{n}$ and define $\overline{Y}_A$ and $v^*_A$ as in (\ref{eqn:ybarA}).
For each $\eta>0$
\begin{equation}
    \limsup_{n\to\infty}\sup_{\theta_n^w\in \Theta_n^w} \P_{\theta_n^w}\left( \frac{1}{n_A}\frac{\max_{i\in \A}\left(Y_i-\overline{Y}_A\right)^2}{v^*_A}\geq \eta \right)\to 0
\end{equation}
\end{lemma'}

\begin{proof}
Conditional on $D_i=1$ and using the inequality $\left(a+b+c\right)^2 \leq 3a^2+3b^2+3c^2$, we have the inequality
\begin{align*}
    &\frac{ \left|Y_i- \overline{Y}_A \right|^2}{v_A^*}\\
    \leq & \frac{3 \left( y_i(1)-\overline{y}_A(1)\right)^2 + 3\left(\overline{y}_A(1) - \widehat{\tau}_A^1\right)^2 + 3n_A^{-2}\left(n_A^0\right)^2\left(\widehat{\tau}_A^1-\widehat{\tau}_A^0\right)^2}{  \left(n_A-1\right)^{-1} \sum_{i\in\A}\left(Y_i- \overline{Y}_A \right)^2} \\
    \leq & \frac{3 \left( y_i(1)-\overline{y}_A(1)\right)^2 + 3\left(\overline{y}_A(1) - \widehat{\tau}_A^1\right)^2 + 3n_A^{-2}\left(n_A^0\right)^2\left(\widehat{\tau}_A^1-\widehat{\tau}_A^0\right)^2}{  n_A^{-1} \sum_{i\in\A}\left(Y_i- \overline{Y}_A \right)^2}\\
    = & \frac{3 \left( y_i(1)-\overline{y}_A(1)\right)^2 + 3\left(\overline{y}_A(1) - \widehat{\tau}_A^1\right)^2 +3n_A^{-2}\left(n_A^0\right)^2\left(\widehat{\tau}_A^1-\widehat{\tau}_A^0\right)^2}{  n_A^{-1} \left(\sum_{a\in\{0,1\}}\sum_{i\in\A,D_i=a}\left(y_i(a)-\widehat{\tau}_A^a \right)^2 +  n_A^{-1}n_A^1n_A^0 \left( \widehat{\tau}_A^1 -\widehat{\tau}_A^0 \right)^2\right)}\\
   \leq  & \frac{3 \left( y_i(1)-\overline{y}_A(1)\right)^2 }{n_A^{-1}\sum_{i\in\A,D_i=1}\left(y_i(1)-\widehat{\tau}_A^1 \right)^2} + \frac{3\left(\overline{y}_A(1) - \widehat{\tau}_A^1\right)^2 }{n_A^{-1}\sum_{i\in\A,D_i=1}\left(y_i(1)-\widehat{\tau}_A^1 \right)^2}+\frac{3n^0_A}{n_A^1 n_A}.
\end{align*}
$$\frac{3 n_{A}^{0}}{n_{A} n_{A}^{1}}$$ \\
A similar identity holds when $D_i=0$. Hence we have,
\begin{align*}
    & \frac{\max_{i\in \A}\left(Y_i-\overline{Y}_A\right)^2}{v^*_A}\\
    \leq  & \frac{3 \max_{i\in \A}\left( y_i(1)-\overline{y}_A(1)\right)^2 }{n_A^{-1}\sum_{i\in\A,D_i=1}\left(y_i(1)-\widehat{\tau}_A^1 \right)^2}+\frac{3 \max_{i\in \A}\left( y_i(0)-\overline{y}_A(0)\right)^2 }{n_A^{-1}\sum_{i\in\A,D_i=0}\left(y_i(0)-\widehat{\tau}_A^0 \right)^2} \\
     & + \frac{3\left(\overline{y}_A(1) - \widehat{\tau}_A^1\right)^2 }{n_A^{-1}\sum_{i\in\A,D_i=1}\left(y_i(1)-\widehat{\tau}_A^1 \right)^2}+\frac{3\left(\overline{y}_A(0) - \widehat{\tau}_A^0\right)^2 }{n_A^{-1}\sum_{i\in\A,D_i=0}\left(y_i(0)-\widehat{\tau}_A^0 \right)^2}\\
     & +\frac{3n_A^1}{n_A n_A^0}+\frac{3n_A^0}{n_A n_A^1}.
\end{align*}
By a union bound, we have, for each $\theta_n^w\in\Theta_n^w$,
\begin{align*}
& \P_{\theta_n^w}\left( \frac{1}{n_A}\frac{\max_{i\in \A}\left(Y_i-\overline{Y}_A\right)^2}{v^*_A}\geq \eta\right)\\
\leq & \sum_{a\in\{0,1\}}\P_{\theta_n^w}\left( \frac{1}{n_A} \frac{3 \max_{i\in \A}\left( y_i(a)-\overline{y}_A(a)\right)^2 }{n_A^{-1}\sum_{i\in\A,D_i=a}\left(y_i(a)-\widehat{\tau}_A^a \right)^2}\geq \frac{\eta}
{6}\right)\\
 & +  \sum_{a\in\{0,1\}}\P_{\theta_n^w}\left(\frac{1}{n_A}\frac{3\left(\overline{y}_A(a) - \widehat{\tau}_A^a\right)^2 }{n_A^{-1}\sum_{i\in\A,D_i=a}\left(y_i(a)-\widehat{\tau}_A^a \right)^2}\geq \frac{\eta}{6}\right)\\
 & + \sum_{a\in \{0,1\}} \P_{\theta_n^w}\left(\frac{1}{n_A}\frac{3n^{1-a}_A}{n_A^an_A}\geq \frac{\eta}{6}\right)
\end{align*}
By Lemma \ref{lemma:E10} below,
\begin{equation*}
    \limsup_{n\to \infty}\sup_{\theta_n^w\in\Theta_n^w}\P_{\theta_n^w}\left( \frac{1}{n_A}\frac{\max_{i\in \A}\left(Y_i-\overline{Y}_A\right)^2}{v^*_A}\geq \eta\right)=0
\end{equation*}
\end{proof}

\begin{lemma'}\label{lemma:E10}
Given the setup in Lemma \ref{lemma:E8} and Lemma \ref{lemma:E9}, for every $\eta> 0$ and $a\in\{0,1\}$, following statements hold
\begin{equation}\label{eqn:20251030134}
    \limsup_{n\to \infty}\sup_{\theta_n^w\in\Theta_n^w} \P_{\theta_n^w}\left( \frac{1}{n_A} \frac{3 \max_{i\in \A}\left( y_i(a)-\overline{y}_A(a)\right)^2 }{n_A^{-1}\sum_{i\in\A,D_i=a}\left(y_i(a)-\widehat{\tau}_A^a \right)^2}\geq \eta\right)\to 0,
\end{equation}
\begin{equation}\label{eqn:20251030135}
    \limsup_{n\to \infty}\sup_{\theta_n^w\in\Theta_n^w}\P_{\theta_n^w}\left(\frac{1}{n_A}\frac{3\left(\overline{y}_A(a) - \widehat{\tau}_A^a\right)^2 }{n_A^{-1}\sum_{i\in\A,D_i=a}\left(y_i(a)-\widehat{\tau}_A^a \right)^2}\geq \eta\right) \to 0,
\end{equation}
and
\begin{equation}\label{eqn:20251030136}
      \limsup_{n\to \infty}\sup_{\theta_n^w\in\Theta_n^w}\P_{\theta_n^w}\left(\frac{n^{1-a}_A}{n_A^an^2_A} \geq \eta\right) \to 0.
\end{equation}
\end{lemma'}

\begin{proof}
We prove the case for $a=1$. The case for $a=0$ can be proved analogously. Rearranging the left-hand side in (\ref{eqn:20251030134}), we have,
\begin{align*}
     & \frac{1}{n_A}\frac{3 \max_{i\in \A}\left( y_i(1)-\overline{y}_A(1)\right)^2 }{n_A^{-1}\sum_{i\in\A,D_i=1}\left(y_i(1)-\widehat{\tau}_A^1 \right)^2}\\
    =& \frac{1}{n_A}\frac{3 \max_{i\in \A}\left( y_i(1)-\overline{y}_A(1)\right)^2 }{v_A^1}\times \frac{v^1_A}{n_A^{-1}\sum_{i\in\A,D_i=1}\left(y_i(1)-\widehat{\tau}_A^1 \right)^2}\\
    \leq  & \frac{3B^2}{\sqrt{n_A}} \times \frac{n_A}{n_A^1} \times \frac{v^1_A}{ \left(n_A^1\right)^{-1} \sum_{i\in\A,D_i=1}\left(y_i(1)-\widehat{\tau}_A^1 \right)^2}, 
\end{align*}
where the last inequality is by Lemma \ref{lemma:nv}.
We prove
\begin{equation*}
    \limsup_{n\to\infty}\sup_{\theta_n^w\in\Theta_n^w}\P_{\theta_n^w}\left(\frac{B^2}{\sqrt{n_A}} \times \frac{n_A}{n_A^1} \times \frac{v^1_A}{ \left(n_A^1\right)^{-1} \sum_{i\in\A,D_i=1}\left(y_i(1)-\widehat{\tau}_A^1 \right)^2} \geq \eta\right )\to 0.
\end{equation*}
For $\epsilon$ sufficiently small and applying a union bound, we prove
\begin{equation}\label{eqn:20251030137}
     \limsup_{n\to\infty}\sup_{\theta_n^w\in\Theta_n^w}\P_{\theta_n^w}\left( \frac{B^2}{\sqrt{n_A}} \times \frac{n_A}{n_A^1} \geq \frac{\eta}{1+\epsilon}\right) \to 0.
\end{equation}
\begin{equation}\label{eqn:20251030138}
      \limsup_{n\to\infty}\sup_{\theta_n^w\in\Theta_n^w}\P_{\theta_n^w}\left( \frac{v^1_A}{ \left(n_A^1\right)^{-1} \sum_{i\in\A,D_i=1}\left(y_i(1)-\widehat{\tau}_A^1 \right)^2}  \geq 1+\epsilon\right) \to 0   
\end{equation}

For (\ref{eqn:20251030137}), because both sides in the event are positive, we note
\begin{align}
     & \frac{B^2}{\sqrt{n_A}} \times \frac{n_A}{n_A^1} \geq \frac{\eta}{1+\epsilon} \Leftrightarrow \frac{n_A^1}{n_A}\frac{\sqrt{n_A}}{B^2} \leq \frac{1+\epsilon}{\eta} \\
\Leftrightarrow    & \frac{n_A^1}{n_A}\leq \frac{1+\epsilon}{\eta}\frac{B^2}{\sqrt{n_A}} \Rightarrow \frac{n_A^1}{n_A}\leq \frac{1+\epsilon}{\eta\sqrt{s}}\frac{B^2}{\sqrt{n}},
\end{align}
where the last implication is by Assumption \ref{a:theta}-\ref{a:theta1}. Notice $\E\left[n^a_A\right]=n^{-1}n_1n_A$ and $n_A^{-1}\E\left[n^a_A\right]=n^{-1}n_1=\pi\in [r,1-r]$ by  Assumption \ref{a:cr}. For $n$ large enough we have,
\begin{equation*}
\frac{1+\epsilon}{\eta\sqrt{s}}\frac{B^2}{\sqrt{n}} - \pi \leq \frac{1+\epsilon}{\eta\sqrt{s}}\frac{B^2}{\sqrt{n}} - r  < 0.
\end{equation*}
Hence we have,
\begin{align}
    &\P_{\theta_n^w}\left(  \frac{B^2}{\sqrt{n_A}} \times \frac{n_A}{n_A^1} \geq \frac{\eta}{1+\epsilon} \right)\leq     \P_{\theta_n^w}\left( \frac{n_A^1}{n_A}\leq \frac{1+\epsilon}{\eta\sqrt{s}}\frac{B^2}{\sqrt{n}} \right) \nonumber \\
   =  &\P_{\theta_n^w}\left( \frac{n_A^1}{n_A}-\pi \leq \frac{1+\epsilon}{\eta\sqrt{s}}\frac{B^2}{\sqrt{n}} -\pi  \right) \nonumber  \\
   =&  \P_{\theta_n^w}\left( \left(\frac{n_A^1}{n_A}-\pi\right)^2 \geq \left(\frac{1+\epsilon}{\eta\sqrt{s}}\frac{B^2}{\sqrt{n}} -\pi \right)^2 \right) \nonumber  \\
   \leq &\E\left[ \left(\frac{n_A^1}{n_A}-\pi\right)^2\right]/\left(\frac{1+\epsilon}{\eta\sqrt{s}}\frac{B^2}{\sqrt{n}}-\pi \right)^2 \nonumber \\
   \leq &  \frac{\left(1-r\right)^3}{4rs^2}\frac{1}{n-1} /\left(\frac{1+\epsilon}{\eta\sqrt{s}}\frac{B^2}{\sqrt{n}}-\pi \right)^2 = O(n^{-1}) \nonumber,
\end{align}

where the last inequality follows from Lemma \ref{lemma:var_bound_A} and
\begin{align*}
     &\V\left(\frac{n^1_A}{n_A}\right) = \left(\frac{n_1}{n_A}\right)^2\V\left(\frac{n^1_A}{n_1}\right) = \left(\frac{n_1}{n_A}\right)^2\frac{n_0}{n_1n}\frac{1}{n-1}\sum_{i=1}^n\left(A_i-\overline{A}\right)^2 \\
    \leq & \left(\frac{1-r}{s}\right)^2 \frac{1-r}{r} \frac{1}{n-1}\frac{1}{4} = \frac{\left(1-r\right)^3}{4rs^2}\frac{1}{n-1}. 
\end{align*}

 Define events $\mathcal{E}$ and $\mathcal{E}^c$ as in (\ref{eqn:E}) and (\ref{eqn:Ec}). Recall the property in \eqref{eqn:ratio_bound} on $\mathcal{E}^c$. (\ref{eqn:20251030136}) follows from a calculation:
\begin{align*}
    \P_{\theta_n^w}\left(\frac{n^{1-a}_A}{n_A^an^2_A} \geq \eta\right)\leq  \P_{\theta_n^w}\left(\mathcal{E}\right)+  \P_{\theta_n^w}\left(\frac{1-r/2}{r/2}\frac{1}{n^2_A} \geq \eta\right)\to 0
\end{align*}
We hence omit.
For $\epsilon$ small enough,  (\ref{eqn:20251030138}) follows from the calculation in (\ref{eqn:va1}).

We now show (\ref{eqn:20251030135}). The event can be written as
\begin{align}
    & \frac{1}{n_A}\frac{3\left(\overline{y}_A(1) - \widehat{\tau}_A^1\right)^2 }{v^1_A} \frac{v^1_A} { \left(n_A^1\right)^{-1}\sum_{i\in\A,D_i=1}\left(y_i(1)-\widehat{\tau}_A^1 \right)^2}\frac{n_A}{n_A^1}\\
   \geq  & \frac{\eta}{\left(1+\epsilon\right)^2} \left(1+\epsilon\right)^2
\end{align}

Again by a union bound, we can shown as in (\ref{eqn:20251030137})
\begin{align}
    \limsup_{n\to\infty} \sup_{\theta_n^w\in \Theta_n^w}\P_{\theta_n^w}\left( \frac{1}{\sqrt{n_A}}\frac{n_A}{n_A^1} \geq \frac{\eta}{ \left(1+\epsilon\right)^2}\right)= 0,
\end{align}
and as in (\ref{eqn:20251030138})
\begin{equation*}
      \limsup_{n\to\infty}\sup_{\theta_n^w\in\Theta_n^w}\P_{\theta_n^w}\left( \frac{v^1_A}{ \left(n_A^1\right)^{-1} \sum_{i\in\A,D_i=1}\left(y_i(1)-\widehat{\tau}_A^1 \right)^2}  \geq 1+\epsilon\right) = 0.   
\end{equation*}
We only need to show 
\begin{equation}
      \limsup_{n\to\infty}\sup_{\theta_n^w\in\Theta_n^w}\P_{\theta_n^w}\left( \frac{1}{\sqrt{n_A}} \frac{3\left(\widehat{\tau}_A^1-\overline{y}_A(1)\right)^2}{v_A^1}\geq 1+\epsilon\right)=0
\end{equation}
Recall the definition of the event $\mathcal{E}$ from (\ref{eqn:E}) and  $\mathcal{E}^c$ from (\ref{eqn:Ec}) 
\begin{align*}
    &\E \left[\P\left( \left.\frac{1}{\sqrt{n_A}}\frac{3\left(\overline{y}_A(1) - \widehat{\tau}_A^1\right)^2 }{v_A^1} \geq 1+\epsilon\right|n_A^1\right)\right]    \\
   \leq& \P\left(\mathcal{E}\right) +  \E \left[ \left.\P\left( \left.\frac{1}{\sqrt{n_A}}\frac{3\left(\widehat{\tau}_A^1-\overline{y}_A(1) \right)^2 }{v_A^1} \geq 1+\epsilon\right|n_A^1\right)\right|\mathcal{E}^c\right]\times \P\left(\mathcal{E}^c\right)\\
   \leq & \P\left(\mathcal{E}\right) +  \E \left[ \left.\frac{3}{1+\epsilon}\frac{n_A^0}{n_A^1n_A\sqrt{n_A}}\right|\mathcal{E}^c\right]\times \P\left(\mathcal{E}^c\right)\\
   \leq & \frac{2(1-r)}{rs^2(n-1)} + \frac{3}{1+\epsilon}\frac{2-r}{r} s^{-\frac{3}{2}}n^{-\frac{3}{2}}=O(n^{-1}).
\end{align*}
\end{proof}

\begin{theorem'}\label{thm:E11}
Let $g:\mathbb{R}\to \mathbb{R} $ be a function such that
\begin{equation*}
\sup_{n\in\mathbb{N}}\sup_{\theta_n^w\in\Theta_n^w}\E_{\theta_n^w}\left[\left|g(n_A^1)\right|\right]\leq C<\infty,
\end{equation*}
for a positive constant $C$. 

Given a $\theta_n^w\in \Theta_n^w$,  let $\{A_i\}_{i=1}^n$ be the associated always-reporter indicators and denote a completely randomized design $D=\left(D_i\right)_{i=1}^n\sim \mathrm{CR}(n,n_{1})$. Define $n_A^1=\sum_{i=1}^n D_iA_i$. Let $Z\sim \textrm{N}(0,1)$ be a normal random variable with mean 0 and variance 1 that is independent of $D$. Recall the definition of $Z^*$ from \eqref{eqn:zs} and $n_A^1* $ from \eqref{eqn:nas}.
For a given $\eta>0$, define the interval,
\begin{equation*}
    I_{\eta}=\left[\chi^2_{1-\alpha}-\eta,\chi^2_{1-\alpha/2}+ \frac{2C}{\alpha}+\eta\right],
\end{equation*}    
We have, for each $\mu>0$, define the event $\mathcal{E}_n$:
\begin{align*}
      \mathcal{E}_n: \sup_{t\in I_{\eta}}\left| \P_{\theta_n^w}\left(Z^2+g(n_A^1)\leq t \right) -  \P_{\theta_n^w}\left(\left.\left(Z^*\right)^2+g\left(n_A^{1*}\right)\leq t \right| D^\obs\right)\right|\geq \mu.
\end{align*}
We have
\begin{align*}
    \limsup_{n\to\infty}\sup_{\theta_n^w\in \Theta_n^w}\P_{\theta_n^w}\left( \mathcal{E}_n\right)=0
\end{align*}

\begin{proof}
We note that because the underlying probability space is discrete $\mathcal{E}_n$ is measurable. Define $t_{\textrm{max}}=\left|\chi^2_{1-\alpha/2}+ \frac{2C}{\alpha}+\eta\right|$.
By Theorem \ref{thm:E7}, for sufficiently large $n$ and small $\epsilon>0$, we have
\begin{align*}
&\frac{2(1-r)}{rs^2(n-1)} +  C_{\ref{constant:2}}(r,s,\epsilon)n^{-1} + \left(\frac{1}{\sqrt{2\pi}}\sqrt{\frac{\epsilon}{1-\epsilon}}  +  \frac{1}{\sqrt{2\pi}}\sqrt{\frac{\epsilon}{1+\epsilon}} \right)\sqrt{t_{\textrm{max}}+C}< \frac{\mu}{2}.        
\end{align*}
Then the result follows from Lemma \ref{lemma:E9}.

\end{proof}
\end{theorem'}

\begin{theorem'}\label{thm:thmA24}
Let $g$ be a non-negative function such that
\begin{equation*}
\sup_{n\in\mathbb{N}}\sup_{\theta_n^w\in\Theta_n^w}\E_{\theta^w_n}\left[\left|g(n_A^1)\right|\right]\leq C<\infty,
\end{equation*} 
and satisfies the condition \eqref{eqn:lower_bound} in Lemma \ref{lemma:non_trivial_bound} with positive constants $\gamma$, $C_1$ and $C_2$. Given $\alpha\in (0,1)$, let $q_n^{1-\alpha}$ be the $\left(1-\alpha\right)$-th quantile of  $Z^2+g(n_A^1)$, where $Z\sim \N(0,1)$ is independent of $n_A^1$. 

Let $\{\widehat{q}^{1-\alpha}_n\}_{n}^{\infty}$ be the $1-\alpha$ quantile of the randomization distribution
\begin{equation*}
  \widehat{q}^{1-\alpha}_n=\inf\left\{ t\in \mathbb{R}:  \P_{\theta_n^w}\left(\left.\left(Z^*\right)^2+g\left(n_A^{1*}\right)\leq t \right| D^\obs\right)\geq 1-\alpha\right\},
\end{equation*}
where $Z^*$ and $n_A^{1*}$ are defined in (\ref{eqn:zs}) and (\ref{eqn:nas}). For each $\eta>0$, we have 
\begin{equation*}
\limsup_{n\to\infty}\sup_{\theta_n^w\in\Theta_n^w}\P_{\theta_n^w}\left( \left|\widehat{q}^{1-\alpha}_n-q_n^{1-\alpha}\right|>\eta \right)=0.
\end{equation*}
Moreover,
\begin{equation}
    \limsup_{n\to\infty}\sup_{\theta_n^w\in\Theta_n^w}\P_{\theta^w_n}\left(  \frac{\left(\arhj-\tau_n\right)^2}{    \widehat{\AVar}^{\ha}_n} + g\left(n_{A}^1\right) \geq \widehat{q}_n^{1-\alpha}\right)\leq \alpha
\end{equation}   
\begin{proof}
This is proved by Theorem \ref{thm:E11}, Lemma \ref{lemma:E6} (through a proof by contradiction), and Theorem \ref{theorem:E5}.
\end{proof}
\end{theorem'}
\subsection{Auxiliary Lemmas}\label{section:auxiliary_lemma}
We note that $Z^2$ with $Z\sim \N\left(0,1\right)$ follows a Chi-squared $\chi^2_1$ distribution with one degree of freedom. $Z^2$ has a density function:
\begin{equation*}
    f_{\chi^2_1}(t) = \frac{1}{\sqrt{2\pi}}t^{-\frac{1}{2}}e^{-\frac{t}{2}},
\end{equation*}
for $t\in \left(0,\infty\right)$. Note that $f_{\chi^2_1}(t)$ is a decreasing function of $t\in \left(0,\infty\right)$. We have the following inequality:
\begin{align}\label{eqn:square_cont}
\begin{split}
    &\sup_{x\geq 0,s\in [0,\delta]} \int_{x}^{x+s}f_{\chi^2_1}(t)dt = \int_{0}^{\delta}f_{\chi^2_1}(t)dt\leq \frac{1}{\sqrt{2\pi}}\int_0^{\delta}t^{-\frac{1}{2}}dt =\sqrt{\frac{2}{\pi}}\sqrt{\delta}. 
\end{split}
\end{align}

\begin{lemma'}\label{lemma:squareroot_continuity}
Let  $Z^2$ with $Z\sim \N\left(0,1\right)$ be a random variable independent of $D=\left(D_i\right)_{i=1}^n\sim \textrm{CR}(n,n_1)$, and $g:\mathbb{R}\to\mathbb{R}$ be a function such that
\begin{equation}\label{eqn:lemma_constant_C}
\sup_{n\in\mathbb{N}}\sup_{\theta^w_n\in\Theta_n^w}\E_{\theta^w_n}\left[\left|g\left(n_A^1\right)\right|\right]\leq C < \infty
\end{equation}
for some positive constant $C$. For every $\epsilon\in (0,1)$ and $t\in\mathbb{R}$, we have the inequalities
\begin{equation*}
   \left| \P_{\theta_n^w}\left(  Z^2\left(1+\epsilon\right) + g\left(n_{A}^1\right)\geq t \right) -    \P_{\theta_n^w}\left(  Z^2 + g\left(n_{A}^1\right)\geq t \right)\right| \leq  \sqrt{\frac{2\epsilon\left(|t|+C\right)}{\pi\left(1+\epsilon\right)}} ,
\end{equation*}
\begin{equation*}
    \left|\P_{\theta_n^w}\left(  Z^2 + g\left(n_{A}^1\right)\geq t \right) -\P_{\theta_n^w}\left(  Z^2\left(1-\epsilon\right) + g\left(n_{A}^1\right)\geq t \right)\right|\leq   \sqrt{\frac{2\epsilon\left(|t|+C\right)}{\pi\left(1-\epsilon\right)}} ,
\end{equation*}
for all $\theta_n^w\in \Theta_n^w$ and all $n\in\mathbb{N}$. In addition, we have,

\end{lemma'}


\begin{proof}
We first show the result for $\left|\P_{\theta_n^w}\left(  Z^2 + g\left(n_{A}^1\right)\geq t \right) -\P_{\theta_n^w}\left(  Z^2\left(1-\epsilon\right) + g\left(n_{A}^1\right)\geq t \right)\right|$. For simplicity, we suppress the dependence on $\theta_n^w$. For every $\epsilon\in (0,1)$, we have
\begin{align}
    0 &\leq \P\left(  Z^2\left(1+\epsilon\right) + g\left(n_{A}^1\right)\geq t \right) -    \P\left(  Z^2 + g\left(n_{A}^1\right)\geq t \right)  \nonumber\\
   =  &  \P\left( \frac{t-g\left(n_A^1\right)}{1+\epsilon}  \leq Z^2 < t-g\left(n_A^1\right)\right) \nonumber\\
   = &  \E\left[ \P\left(\left.\frac{t-g\left(n_A^1\right)}{1+\epsilon}  \leq Z^2 < t-g\left(n_A^1\right)\right|n_A^1\right)\right] \nonumber \\
   = &  \E\left[ \P\left(\left.\frac{t-g\left(n_A^1\right)}{1+\epsilon}  \leq Z^2 < t-g\left(n_A^1\right)\right|n_A^1\right)\mathds{1}\left\{t-g\left(n_A^1\right)\geq 0\right\}\right] \nonumber \\
   \leq & \sqrt{\frac{2}{\pi}}\E\left[\sqrt{\frac{\epsilon}{1+\epsilon}}\sqrt{t-g\left(n_A^1\right)}\mathds{1}\left\{t-g\left(n_A^1\right)\geq 0\right\}\right] \label{eqn:2025102882} \\
   \leq &  \sqrt{\frac{2}{\pi}}\sqrt{\frac{\epsilon}{1+\epsilon}}\sqrt{\E\left[ \left|t-g\left(n_A^1\right)\right|\right]}  \label{eqn:2025102883} \\
   \leq & \sqrt{\frac{2}{\pi}}\sqrt{\frac{\epsilon}{1+\epsilon}} \sqrt{|t| + \E\left[\left|g\left(n_A^1\right)\right|\right]}\leq  \sqrt{\frac{2}{\pi}}\sqrt{\frac{\epsilon}{1+\epsilon}} \sqrt{|t|+C},\nonumber
\end{align}
where the positive constant $C$ is defined in (\ref{eqn:lemma_constant_C}), (\ref{eqn:2025102882}) follows from (\ref{eqn:square_cont}), and (\ref{eqn:2025102883}) follows from $\E[X]\leq \sqrt{\E[X^2]}$ for any square-integrable random variable $X$. The result for $\left|\P_{\theta_n^w}\left(  Z^2 + g\left(n_{A}^1\right)\geq t \right) -\P_{\theta_n^w}\left(  Z^2\left(1-\epsilon\right) + g\left(n_{A}^1\right)\geq t \right)\right|$ can be shown similarly.
\end{proof}

\subsection{Proof of Theorem \ref{thm:main} and Theorem \ref{thm:asy}}\label{section:prooftheoremmain}

We now prove Theorem \ref{thm:main}. 
\begin{proof}
We write the underlying true table as $A^0$.
Under the sharp-null hypothesis,
\begin{align}
     \Prob_{\theta_n^s}\left( p^{\textrm{worst}} \leq \alpha\right) \leq &   \Prob_{\theta_n^s}\left( p\left(A^0\right)\leq \alpha-\beta\right) +  \Prob_{\theta_n^s}\left(A^0\not \in \mathbb{A}\left(D,R\right)\right)    \\
   \leq  & \alpha-\beta + \beta = \alpha.
\end{align}

Similarly, under the weak-null hypothesis,
\begin{align*}
     \Prob_{\theta_n^w}\left( p^{\textrm{worst}} \leq \alpha\right) \leq &   \Prob_{\theta_n^w}\left( p\left(A^0\right)\leq \alpha-\beta\right) +  \Prob_{\theta_n^w}\left( A^0\not \in \mathbb{A}\left(D,R\right)\right)    \\
   \leq  & \Prob_{\theta_n^w}\left( \mathcal{T}^0_n + g_i\left(n_{A}^1\right) \geq \widehat{q}^{i,\alpha-\beta}_n\right)+\beta, 
\end{align*}
for $i\in \{0,1,2\}$, where the function $g_i(n_A^1)$ is defined in \eqref{eqn:2025102385} or \eqref{eqn:2025102386} and $\widehat{q}^{i,1-\alpha+\beta}_n$ is the randomization-based critical value of the corresponding function $g_i$. $\mathcal{T}^0_n$ is defined in \eqref{eqn:t0}.

We have shown that $g_i(n_A^1)$ satisfies the premise of Theorem \ref{thm:thmA24} in Theorem \ref{thm:feasible_clt}.
Hence by Theorem \ref{thm:thmA24},
\begin{align*}
    & \limsup_{n\to\infty}\sup_{\theta_n^w\in\Theta_n^w}   \Prob_{\theta_n^w}\left( p^{\textrm{worst}} \leq \alpha\right)\\
   \leq  &  \limsup_{n\to\infty}\sup_{\theta_n^w\in\Theta_n^w}  \Prob_{\theta_n^w}\left(  \mathcal{T}_n^0 + g_i\left(n_{A}^i\right) \geq \widehat{q}^{i,1-\alpha+\beta}_n\right)+\beta
   \leq  \alpha-\beta+\beta=\alpha.
\end{align*}

\end{proof}

We now prove Theorem \ref{thm:asy}.
\begin{proof}
Let $i\in \{0,1,2\}$. We write the underlying true table as $A^0$ and define $k_0=\sum_{i=1}^n A_i^0$. Algorithm \ref{alg:asy} will reject the weak-null hypothesis if $ \mathcal{T}^{\max,k}_i\geq q^k_{i,1-\alpha+\beta} $ for all $k$ in $\textrm{Cardinality-A}$. Then we have,
    \begin{align*}
        &   \Prob_{\theta_n^w}\left( \mathcal{T}^{\max,k}_i \geq q^k_{i,1-\alpha+\beta},\forall k\in \textrm{Cardinality-A}\right)\\
        \leq &  \Prob_{\theta_n^w}\left( \mathcal{T}^{\max,k_0}_i \geq q^{k_0}_{i,1-\alpha+\beta} \right)+ \Prob_{\theta_n^w}\left(A^0\not \in \mathbb{A}\left(D,R\right)\right)   \\
         \leq &  \Prob_{\theta_n^w}\left(  \mathcal{T}_n^i+ g_i\left(n_{A}^1\right) \geq q^{k_0}_{i,1-\alpha+\beta}\right) + \beta\\
         \leq & \Prob_{\theta_n^w}\left( Z^2+g_i(n_A^1) \geq q^{k_0}_{i,1-\alpha+\beta}\right)+ \beta\\
          & + \left|\underbrace{ \Prob_{\theta_n^w}\left(  \mathcal{T}_n^i+ g_i\left(n_{A}^1\right) \geq q^{k_0}_{i,1-\alpha+\beta}\right)- \Prob_{\theta_n^w}\left( Z^2+g_i(n_A^1) \geq q^{k_0}_{i,1-\alpha+\beta}\right) }_{(*)}\right|.
    \end{align*}
By Lemma \ref{lemma:bounded_quantile}, Theorem \ref{thm:feasible_clt} and Lemma \ref{lemma:lower bound_quantile}, the set of all possible quantiles for $ q^k_{i,1-\alpha+\beta}$ lies in a bounded set. Using an argument similar as the justification from \eqref{eqn:2025102892} to \eqref{eqn:2025102893}, we can show that,
\begin{equation*}
    \limsup_{n\to\infty}\sup_{\theta_n^w\in\Theta_n^w}\left|(*)\right|=0.
\end{equation*}
Hence we have,
\begin{equation*}
   \limsup_{n\to\infty}\sup_{\theta_n^w\in\Theta_n^w}\Prob_{\theta_n^w}\left( \mathcal{T}^{\max,k}_i \geq q^k_{i,1-\alpha+\beta},\forall k\in \textrm{Cardinality-A}\right)  \leq \alpha.
\end{equation*}
\end{proof}
\section{Proof for Results in Section \ref{section:implementation}}
\subsection{Proof of Lemma \ref{lemma:v}}
\begin{proof}
Let $A^*$ be the optimizing always-reporter vector associated with problem \eqref{eqn:worst_case_p_value}.
By definition, there must exist an interval $[t_{i-1},t_i]$ such that $\T_n(Y,D^\obs,A^*)\in [t_{i-1},t_i]$. Let $A^{i*}$ be the optimizing vector corresponding to the subproblem of \eqref{eqn:inner_problem} associated with $[t_{i-1},t_i]$. We have the inequality:
\begin{align*}
&\E_D\left[\mathds{1}\left(\T_n(Y,D,A^{i*})\geq t_{i-1}\right)\right] \geq \E_D\left[\mathds{1}\left(\T_n(Y,D,A^{*})\geq t_{i-1}\right)\right]\\     
& \geq \E_D\left[\mathds{1}\left(\T_n(Y,D,A^{*})\geq \T_n(Y,D^\obs,A^{*})\right)\right]=p^{\textrm{worst}},
\end{align*}
where $D\sim \textrm{CR}(n,n_1)$. For every interval $[t_{i-1},t_i]$ and its associated optimized value, we have the inequality:
\begin{align*}
v_i=& \E_D\left[\mathds{1}\left(\T_n(Y,D,A^{i*})\geq t_{i-1}\right)\right] \\
= & \E_D\left[\mathds{1}\left(\T_n(Y,D,A^{i*})\geq t_{i}\right)\right] + \E_D\left[\mathds{1}\left(\T_n(Y,D,A^{i*})\in [t_{i-1},t_i)\right)\right]\\
\leq &\E_D\left[\mathds{1}\left(\T_n(Y,D,A^{i*})\geq \T_n(Y,D^\obs,A^{i*})\right)\right] + \E_D\left[\mathds{1}\left(\T_n(Y,D,A^{i*})\in [t_{i-1},t_i)\right)\right]\\
\leq & p^{\textrm{worst}} + \E_D\left[\mathds{1}\left(\T_n(Y,D,A^{i*})\in [t_{i-1},t_i)\right)\right].
\end{align*}

Combining the inequalities yields the desired results.
\end{proof}

\section{Computational Subroutines}\label{section:algorithms}
Section \ref{section:master_algorithm} includes two master algorithms that computer the worst-case p-values for different outcome types:
\begin{enumerate}
    \item \texttt{A\_master\_continuous}: for continuous outcomes.
    \item \texttt{A\_master\_finite\_support}: for outcomes with a small support.  
\end{enumerate}
Section \ref{section:algo_heur} includes subroutines that are used in \texttt{A\_master\_continuous}:
\begin{enumerate}
    \item \texttt{B\_Calculate\_Heuristic\_p\_Value}: given a number of always-reporters, the subroutine finds a heuristic solution for the always-reporter vector that maximizes the p-value.
\item \texttt{B\_Create\_Lower\_Bound:} an MIQCP-based bisection method for constructing a lower bound for the test statistic
\begin{equation*}
    \widetilde{\mathcal{T}}^0_n(Y,D^A,x) + g_i(n_{A,1}),
\end{equation*}
subject to $x \in x(k,D^{\obs},R)$, for a given $k$, an assignment vector $D^A$, and a statistic type $i\in\{0,1,2\}$. Here, $\widetilde{\mathcal{T}}^0_n(Y,D^A,x)$ is defined in \eqref{eqn:tildeTn0}, and the functions $g_i(n_{A,1})$ are defined in \eqref{eqn:2025102385} and \eqref{eqn:2025102386}.

\item \texttt{B\_Create\_Upper\_Bound:} an MIQCP-based bisection method for constructing an upper bound for the same constrained problem considered in \texttt{B\_Create\_Lower\_Bound}.

    \item \texttt{B\_Create\_Lower\_Bound\_2:} a MIQCP method for constructing a lower bound of the statistic
    \begin{equation*}
        \widehat{\mu}_n^2(Y,D^A,x^A) -c\times \widehat{\sigma}^{2,\ha}_n\!\left(Y,D^A,x^A\right)
    \end{equation*}
    subject to $x\in x(k,D^\obs,R)$ for a given $k$, an assignment vector $D^A$ and a given scalar $c$, where the functions $\mu^2_n$ and $\widehat{\sigma}^{2,\ha}_n\!\left(Y,D^A,x^A\right)$ are defined below \eqref{eqn:tildeTn0}. 
     \item \texttt{B\_Calculate\_p\_Value\_Upper\_Bound}: an algorithm that solves the subproblems described in Theorem~\ref{thm:reduction}.
\end{enumerate}
Section \ref{section:alg_asy} includes algorithms for asymptotic inferences:
\begin{enumerate}
    \item \texttt{A\_master\_asymptotics}: a master algorithm that solves the asymptotic inference problem.
    \item  \texttt{B\_Asymptotic\_Inferences\_Inner}: a subroutine that is used in the master algorithm.
\end{enumerate}
Section \ref{section:boundstd} includes
\begin{enumerate}
    \item \texttt{C\_dim2\_bounds}: calculate analytical upper and lower bounds for the squared difference-in-means estimator.
    \item \texttt{C\_v1\_min} and \texttt{C\_v0\_min}: calculate analytical lower bounds for the treated-group and control-group variance estimators.
    \item \texttt{C\_v1\_max} and \texttt{C\_v0\_max}: calculate analytical upper bounds for the treated-group and control-group variance estimators.    
\end{enumerate}

We adopt the indexing convention from Section~\ref{section:IP}. Recall that the first \(r_0\) units are always-reporters assigned to the control group, regardless of whether we index by \(i\) or by \(a\). Also recall the set of matching variables associated with a fixed number of always-reporters \(k \in [\,r_0,\; r_0+\sum_{i=1}^n D_iR_i\,]\), as defined in \eqref{eqn:assignment_variables}:
\begin{equation*}
\begin{split}
    &x(k,D^\obs,R)=\\
    &\left\{ \{x_{ai}\}_{a\in [k],i\in [n]}: \begin{array}{ll}
            x_{ai}\in \{0,1\} & \forall a\in [k],i\in [n],\\
        x_{ai}=1, &\forall a=i, a\leq r_0\\
        \sum_{i}x_{ai}=1,& \forall a\in [k],\\
         \sum_{a}x_{ai}\leq 1, &\forall i\in [n],\\
        x_{ai}\in \{0,1\}, &\forall a\in [k], D_i^\obs=1, R_i=1,\\
        x_{ai}=0,& \forall a\in [k], D_i^\obs=1, R_i=0.
    \end{array}\right\}.    
\end{split}
\end{equation*}

Recall our notation for the observed dataset $\mathcal{D}=\left(Y,D^\obs,R\right)$ as defined in \eqref{eqn:data}. We define the subroutines $\texttt{zeros}$($n_1$,$n_2$) and $\texttt{ones}$($n_1$,$n_2$), which create a $n_1$-by-$n_2$ matrix of zeros and ones, respectively. 

For brevity, we assume that our programs have access to basic information about the experimental design and statistical inferential parameters, including the sample size $n$, the number of treated units $n_1$, the number of control units $n_0$, the number of always-reporters in the control group $r_0$, the number of simulated assignments $n_{mc}$, the significance level $\alpha$, and the statistic type $i$ for testing the balance of the number of always-reporters.

Finally, recall the definition of $\widetilde{p}(x)$ and $\widetilde{v}(x,t_{i-1},t_i)$ from \eqref{eqn:px} and \eqref{eqn:vxtt}. For a given number of always-reporters and due to the underlying symmetry of the assignment distribution, the optimal values of the problems 
\begin{equation}
    \max_{x\in x(k,D^\obs,R)}\widetilde{p}(x) \text{ and } \max_{x\in x(k,D^\obs,R)}\widetilde{v}(x,t_{i-1},t_i)
\end{equation}
are equivalent to the optimal values of the problems
\begin{equation*}
       \max_{x\in x(k,D^\obs,R)}\widetilde{p}(x)  \text{ subject to } \tilde{Y}_a\geq \tilde{Y}_{a+1},\forall a\in [r_0+1,n_A-1]
\end{equation*}
\begin{equation*}
       \max_{x\in x(k,D^\obs,R)}\widetilde{v}(x,t_{i-1},t_i)  \text{ subject to } \tilde{Y}_a\geq \tilde{Y}_{a+1},\forall a\in [r_0+1,n_A-1],
\end{equation*}
where $\tilde{Y_a}=\sum_{a,i}x_{ai}Y_i$, $\forall a\in [n_A]$ and note that the outcomes for the first $r_0$ units are fixed under our convention. We include these constraints in the optimization problems below because it reduces the symmetry of the undelrying integer programming problems, possibly making it faster to solve. 

\subsection{Master Algorithms}\label{section:master_algorithm}
\begin{algorithm}[H]
  \caption{\texttt{A\_master\_continuous:} test null hypotheses at a prespecified significance level.} 
  \label{alg:12152025_c}
  \begin{algorithmic}[1]
\Require Dataset $\mathcal{D}$ as in \eqref{eqn:data}, a prespecified significance level $\alpha\in (0,1)$, and statistic type $i\in \{0,1,2\}$, number of randomization draws $n_{mc}$, pruning significance level $\beta\in [0,\alpha)$.
\Statex \textbf{Step I: Pretesting}
\State Set $Rej \gets 1$.
\State Construct and prune the compatible always-reporter set $\mathbb{A}(D^{\obs},R)$.
\State Set $n_A^{\max}=\sum_{i=1}^n R_i$ and $n_A^{\min}=r_0$.
\If{$\beta>0$}
\Statex \Comment{pruned $\mathbb{A}(D^{\obs},R)$ at significance level $\beta$}
\State Set $\alpha\gets\alpha-\beta$.
\State Set $n_A^{\max} \gets \max_{A\in \mathbb{A}(D^{\obs},R)} \sum_{i=1}^n A_i$.
\State Set $n_A^{\min} \gets \min_{A\in \mathbb{A}(D^{\obs},R)} \sum_{i=1}^n A_i$.
\EndIf
\Statex \textbf{Step II: the Heuristic Procedure}
\For{$k = n_A^{\min}$ \textbf{to}  $n_A^{\max}$}
  \State $(pval,L_k,\mathcal{D}^{mc}_k)\gets \texttt{B\_Calculate\_Heuristic\_p\_Value}(\mathcal{D},k,n_{mc},i)$.
  \State Store $L_k$ and $\mathcal{D}^{{mc}}_k$. \Comment{Store the lower bound and the simulated assignments}
  \If{$p\_val \geq \alpha$}
    \State Set $Rej \gets 0$; \textbf{break}. (Worst-case $p$-value $\geq \alpha$, fail to reject)
  \EndIf
\EndFor
\Statex  \textbf{Step III: upper bound the p-value }
    \If{$Rej$=1} 
    \State Set p\_upper\_bound=$\texttt{zeros}$($n_A^{\max}-n_A^{\min}+1$,1). 
    \Statex \Comment{Initialize an array to store results}
    \For{$k=n_A^{\min}$ \textbf{to} $n_A^{\max}$}
    \State p\_upper\_bound(i)= \texttt{B\_Calculate\_p\_Value\_Upper\_Bound($\mathcal{D}$,k,$\mathcal{D}^{{mc}}_k$,$L_k$)}.
    \EndFor
    \If{$\max\left(p\_upper\_bound\right)<\alpha$}
    \State $Rej=1$.
    \Else
    \State $Rej=0$.
    \EndIf
    \EndIf 
    \State \Return $Rej$
    \end{algorithmic}
\end{algorithm}

Recall the definitions of $\mathcal{T}_n^0$, $\mathcal{T}_n^1$, and $\mathcal{T}_n^2$ in \eqref{eqn:t0}, \eqref{eqn:t1}, and \eqref{eqn:t2}. For outcomes with cardinality $K$, also recall the definition of the count vector $c(A)$ (defined above \eqref{eqn:count_vector_element}) for a given always-reporter vector $A$.

\begin{algorithm}[H]
  \caption{\texttt{A\_master\_small\_support:} test null hypotheses at a prespecified significance level $\alpha\in(0,1)$.}
  \label{alg:12152025_2}
  \begin{algorithmic}
    \Require Dataset $\mathcal{D}$ as in \eqref{eqn:data}; significance level $\alpha\in(0,1)$; statistic type $i\in\{0,1,2\}$.
    \State Initialize $Rej \leftarrow 1$.
    \State Initialize $\texttt{worst\_p\_value} \leftarrow 0$.
    \State Construct $\mathbb{C}\!\left(D^{\obs},R\right)$ as in \eqref{eqn:conut_vector}.
    \For{each $c\in \mathbb{C}\!\left(D^{\obs},R\right)$}
      \State Select any $A\in \mathbb{A}\!\left(D^{\obs},R\right)$ such that $c(A)=c$.
      \State Compute $p^{mc}(A)$ using the statistic $\mathcal{T}_n^i$.
      \If{$p^{mc}(A)\geq \alpha$}
        \State $\texttt{worst\_p\_value} \leftarrow p^{mc}(A)$.
        \State $Rej \leftarrow 0$; \textbf{break}. \Comment{Worst-case $p$-value $\geq \alpha$: fail to reject.}
      \EndIf
    \EndFor
    \Return $Rej$.
  \end{algorithmic}
\end{algorithm}

To implement the step ``select $A\in \mathbb{A}\!\left(D^{\obs},R\right)$ such that $c(A)=c$'' efficiently, one can group units by outcome category and then assign always-reporter indicators to match the target count vector $c$ (i.e., by selecting the appropriate number of units within each outcome category).

\subsection{Subroutines for \texttt{A\_master\_continuous} }\label{section:algo_heur}
We define the following functions. Given a number of always-reporters $n_A$, a vector of assignments $D^A=\left\{D_a\right\}_{a\in [n_A]}$ and outcomes $Y^A=\left\{Y_a\right\}_{a\in [n_A]}$.
 \begin{align}
    \mathrm{DIM}^2(D^A,Y^A) &= \left( \frac{\sum_{a}D_aY_a}{\sum_{a}D_a} -\frac{\sum_{a}\left(1-D_a\right)Y_a}{\sum_{a}\left(1-D_a\right)} \right)^2, \label{eqn:dim2}\\
     \mathrm{VAR}(D^A,Y^A) & =   \frac{\sum_{a}D_aY_a^2}{\left(\sum_{a}D_a\right)^2} - \frac{1}{\sum_{a}D_a}\left(\frac{\sum_{a}D_aY_a}{\sum_{a}D_a}\right)^2 \label{eqn:VAR}\\  
     & +\frac{\sum_{a}\left(1-D_a\right)Y_a^2}{\left(\sum_{a}\left(1-D_a\right)\right)^2} - \frac{1}{\sum_{a}\left(1-D_a\right)}\left(\frac{\sum_{a}\left(1-D_a\right)Y_a}{\sum_{a}\left(1-D_a\right)}\right)^2\\    
   \textrm{V}(n_A)&= \frac{n^2}{n_1n_0(n-1)} \frac{n_A}{n} \left(1-\frac{n_A}{n}\right),\\
       \textrm{N}_A^0(D^A)& =0,\\
       \textrm{N}_A^1(D^A)& =\frac{\left( n_1^{-1}\sum_a D_a - n_0^{-1}\sum_{a} \left(1-D_a\right)\right)^2}{ \textrm{V}(n_A)},\\
     \textrm{N}_A^2(D^A)& =\frac{\left(\lfloor n_1^{-1}\sum_a D_a - n_0^{-1}\sum_{a} \left(1-D_a\right)\rfloor_{-}\right)^2}{ \textrm{V}(n_A)}.
\end{align}

For a given number of always-reporters $k$, recall the definition of the distribution of $\{D^A_a\}_{a\in [k]}\sim \mathcal{L}(n,n_1,k)$ as defined in Section \ref{section:IP}. 

The function below provides a heuristic solution to the worst-case $p$-value problem for a given number of always-reporters. We note that $\textrm{NUM\_L}$ and $\textrm{NUM\_U}$ are analytical lower and upper bounds for the variance estimators, while $\textrm{DIM2\_L}$ and $\textrm{DIM2\_U}$ are analytical lower and upper bounds for the squared difference-in-means estimator. The corresponding construction routines are provided in Section~\ref{section:boundstd}. The analytical bounds are generally looser than those obtained via computational approaches.

\begin{algorithm}[H]
  \caption{\texttt{B\_Calculate\_Heuristic\_p\_Value:} calculate lower bounds for the worst-case p-value.} 
  \label{alg:12152025_1}
  \begin{algorithmic}[1]
    \Require Dataset $\mathcal{D}$, number of always-reporters $n_A$, (optional) $tol=10^{-4}$.
    \State Create assignment vector $D^A_\obs=[\texttt{ones}(n_A,1),\texttt{zeros}(r_0,1) ]$.
     \State  Set  $N_A\gets\mathrm{N}^i_A(D^A_\obs)$.
    \State Set $\textrm{DIM2\_L},\textrm{DIM2\_U}\gets\texttt{C\_dim2\_bounds}(D^A_\obs,\mathcal{D})$.
    \State Set $\textrm{NUM\_L}\gets\texttt{C\_v1\_min}(D^A_\obs,\mathcal{D})+\texttt{C\_v0\_min}(D^A_\obs,\mathcal{D})$.
    \State Set $\textrm{NUM\_U}\gets\texttt{C\_v1\_max}(D^A_\obs,\mathcal{D})+\texttt{C\_v0\_max}(D^A_\obs,\mathcal{D})$.
    \State Set $L\gets \textrm{DIM2\_L}/\textrm{NUM\_U}+N_A$.
    \State Set $U\gets \textrm{DIM2\_U}/\textrm{NUM\_L}+N_A$.   
    \State $L,Y^A\gets \texttt{B\_Create\_Lower\_Bound}(D^A_\obs,n_A,\mathcal{D},L,U,tol).$
    \State Set
    \begin{equation}
        pval=\frac{1}{n_{mc}}\sum_{s=1}^{n_{mc}}1\left\{\mathcal{T}^\obs \left(D^A_{\obs},Y^A\right) \geq \mathcal{T}^\obs \left(D^A_{s},Y^A\right) \right\},
    \end{equation}
    where $D^A_s\sim \mathcal{L}(n,n_1,n_A),\forall s\in [n_{mc}]$,
    \begin{equation*}
        \mathcal{T}^\obs(D^A_{\obs},Y^A)= \frac{ \mathrm{DIM}^2(D^A_{\obs},Y^A) }{ \mathrm{VAR}(D^A_{\obs},Y^A) } +  \mathrm{N}_A^i(D^A_\obs),
    \end{equation*}
    and,
    \begin{equation*}
        \mathcal{T}^\obs(D^A_{s},Y^A)= \frac{ \mathrm{DIM}^2(D^A_{s},Y^A) }{ \mathrm{VAR}(D^A_{s},Y^A) } +  \mathrm{N}_A^i(D^A_s), \quad\forall s\in [n_{mc}].
    \end{equation*}    
    \State Store the \(n_{\mathrm{mc}}\) Monte Carlo draws \(\{D_s^{A}\}_{s=1}^{n_{\mathrm{mc}}}\) as \(\mathcal{D}^{\mathrm{mc}}_{n_A}\).
    \State \Return \textrm{pval}, $L$, $\mathcal{D}^{mc}_{n_A}$
    \end{algorithmic}
\end{algorithm}
The following algorithm uses a bisection method to compute a lower bound for a given test statistic. The initial feasibility problem identifies a feasible set of outcomes whose test statistic exceeds this lower bound; by construction, this feasibility problem is always feasible. Recall that $i\in\{0,1,2\}$ denotes the type of statistics that is used to test the balance of the number of always-reporters.
\begin{algorithm}[H]
  \caption{\texttt{B\_Create\_Lower\_Bound:} construct a lower bound for the test statistic \eqref{eqn:t0}, \eqref{eqn:t1} and \eqref{eqn:t2}.}
  \label{alg:6}
  \begin{algorithmic}[1]
    \Require assignment vector $D^A$, number of always-reporters $n_A$, dataset $\mathcal{D}$ as in \eqref{eqn:data}, lower bound $L$, upper bound $U$, tolerance $tol$.
    \State Calculate $N_A \gets \mathrm{N}^i_A(D^A)$.
    \State Solve the following feasibility problem: find $x=\left(x_{ai}\right)_{a,i}$ and $Y^A=\left(Y_a\right)_{a}$ that solve the problem,
    \Statex\Comment{If the bisection below does not update $L$, return compatiable outcomes.}
       \[
        \max_{x,Y^A} 0
      \]
      subject to,
      \[
        Y_a = \sum_{i\in[n]} x_{ai}Y_i, \quad \forall a\in[n_A],
      \]
      \[
        Y_a \ge Y_{a+1}, \quad \forall a\in[r_0+1,\,n_A-1],
      \]
      \[
        \mathrm{DIM}^2(D^A,Y^A)
        + \bigl(N_A^i-L\bigr)\,\mathrm{VAR}(D^A,Y^A)\geq 0,
      \]
      \[x\in x(n_A,D^{\obs},R).\]
    \State Store the optimizer $Y^{A*}\gets\left(Y_a^*\right)_{a}$.
    \While{$U-L\ge tol$}
      \State Set $M \gets (L+U)/2$.
      \State Solve the following feasibility problem:
      \[
        \max_{x,Y^A} 0
      \]
      \State subject to,
      \[
        Y_a = \sum_{i\in[n]} x_{ai}Y_i, \quad \forall a\in[n_A],
      \]
      \[
        Y_a \ge Y_{a+1}, \quad \forall a\in[r_0+1,\,n_A-1],
      \]
      \[
        \mathrm{DIM}^2(D^A,Y^A)
        + \bigl(N_A^i-M\bigr)\,\mathrm{VAR}(D^A,Y^A)
        \le 0,
      \]
      \[
      x\in x(n_A,D^{\obs},R).
      \]
      \If{the problem is feasible}
        \State Set $U \gets M$. \Comment{The lower bound must be smaller than $U$.}
        \State Store the optimizer $Y^{A*}\gets\left(Y_a^*\right)_{a}$.
      \Else
        \State Set $L \gets M$. \Comment{The lower bound must be larger than $U$.}
      \EndIf
    \EndWhile
    \State \Return $L$, $\{Y_a^*\}_{a\in[n_A]}$
  \end{algorithmic}
\end{algorithm}

\begin{algorithm}[H]
  \caption{\texttt{B\_Create\_Lower\_Bound\_2:} construct a lower bound for statistics of the form $\mathrm{DIM}^2(D^A,Y^A)+c\mathrm{VAR}^2(D^A,Y^A)$ given $Y$ and $D^{A}$ and $c\in \mathbb{R}$}
  \label{alg:6}
  \begin{algorithmic}
    \Require assignment vector $D^A$, dataset $D$ as in \eqref{eqn:data}, scalar $c$, (optional) lower bound $L$. 
    \State Solve the following feasibility problem: find $x=\left(x_{ai}\right)_{a,i}$, $Y^A=\left(Y_a\right)_a$ and $t$ that solves the problem:
    \Statex \Comment{find the smallest $t$ that's compatible with an always-reporter table}
      \[
        \min_{x,Y^A,t} t
      \]
      \State subject to
      \[
        Y_a = \sum_{i\in[n]} x_{ai}Y_i, \quad \forall a\in[n_A], 
      \]
      \[
        Y_a \ge Y_{a+1}, \quad \forall a\in[r_0+1,\,n_A-1],
      \]
      \[x\in x(n_A,D^\obs,R),\]
      \[
        \mathrm{DIM}^2(D^A,Y^A)
        + c\mathrm{VAR}(D^A,Y^A)\leq t,
      \]
      \begin{align*}
         t\geq L.  \tag*{\text{// only if $L$ is provided.}}
      \end{align*}
    \State Return the optimal value $t^*$;
  \end{algorithmic}
\end{algorithm}
The following algorithm computes a tighter upper bound for the test statistic, provided that the input $t_{\textrm{max}}$ is a valid upper bound.

\begin{algorithm}[H]
  \caption{\texttt{B\_Create\_Upper\_Bound:} constructing an upper bound for the test statistic \eqref{eqn:t0}, \eqref{eqn:t1} and \eqref{eqn:t2}. }
  \label{alg:7}
  \begin{algorithmic}
    \Require assignment vector $D^{A}$, dataset $\mathcal{D}$ as in \eqref{eqn:data}, upper bound $U$.
    \State Set $N_A \gets \mathrm{N}^i_A(D^A)$.
    \State Set $t_{\max}=U$.
    \State Set $val \gets -\infty$.
    \While{$val=-\infty$}
      \State Solve the following feasibility problem: find $x=\left(x_{ai}\right)_{ai}$ and $Y^A=\left(Y_a\right)_{a}$ that solves the problem
      \[
        \max_{x,Y^A} 0
      \]
      \State subject to
      \[
        Y_a = \sum_{i\in[n]} x_{ai}Y_i, \quad \forall a\in[n_A],
      \]
      \[
        Y_a \ge Y_{a+1}, \quad \forall a\in[r_0+1,\,n_A-1],
      \]
      \[
        \mathrm{DIM}^2(D^A,Y^A)
        + \bigl(\mathrm{N}^i_A-t_{\max}\bigr)\,\mathrm{VAR}(D^a,Y^a)
        \ge 0.
      \]
      \If{the problem is feasible}
        \State Set $val \gets 0$.
        \State Update $t_{\max} \gets 2t_{\max}$ \Comment{$t_{\max}$ is feasible; enlarge the candidate bound}
      \Else
        \State Set $val \gets -\infty$.
        \State Update $t_{\max} \gets t_{\max}/2$ \Comment{$t_{\max}$ is infeasible; shrink the candidate bound}
      \EndIf
    \EndWhile
    \State \Return $\min\{t_{\max},U\}$
  \end{algorithmic}
\end{algorithm}
The algorithm below solves the subproblems described in Theorem~\ref{thm:reduction}. We include a preprocessing step (Block~1) to compute tighter upper bounds for the simulated statistics. In practice, this step is useful because it yields a quick upper bound on the p-value.

\begin{algorithm}[H]
  \caption{\texttt{B\_Calculate\_p\_Value\_Upper\_Bound:} constructing an upper bound for the p-value for a given number of always-reporters $n_A$. }
  \label{alg:7}
  \begin{algorithmic}
    \Require dataset $\mathcal{D}$, number of always-reporters $n_A$, generated simulated assignments $\mathcal{D}^{mc}_{n_A}$, lower bound for the observed test statistics $L_{n_A}$
\Statex \textbf{Block 1: Create tighter upper bounds for the simulated statistics}
   \State Create $UB=\texttt{zeros}(nsim,1)$ \Comment{To hold upper bound values}
    \For{s=1 \textbf{to } $n_{mc}$} 
    \State Set $D^A=D^{mc}_{n_A}(s)$ \Comment{Extract the $i$th assignment vector}
    \State $N_A\gets\mathrm{N}^i_A(D^A)$
    \State Set $\textrm{DIM2\_L},\textrm{DIM2\_U}\gets\texttt{C\_dim2\_bounds}(D^A,\mathcal{D})$.
    \State Set $\textrm{NUM\_L}\gets\texttt{C\_v1\_min}(D^A,\mathcal{D})+\texttt{C\_v0\_min}(D^A,\mathcal{D})$.
    \State Set $\textrm{NUM\_U}\gets\texttt{C\_v1\_max}(D^A,\mathcal{D})+\texttt{C\_v0\_max}(D^A,\mathcal{D})$.
    \State Set $L\gets \textrm{DIM2\_L}/\textrm{NUM\_U}+N_A$.
    \State Set $U\gets \textrm{DIM2\_U}/\textrm{NUM\_L}+N_A$.
    \State $UB(s)= \texttt{B\_Create\_Upper\_Bound}(D^A,\mathcal{D}, U)$
    \EndFor
\Statex \textbf{Block 2: Preprocess based on the upper bounds}
\State Calculate $pval=\texttt{mean}(UB >= L_{n_A} )$. 
\If{$pval<\alpha$}
    \Statex \Comment{Many upper bounds of the simulated statistics are smaller than the lower bound of the observed statistic }
    \State \Return $pval$
\Else
    \State Calculate the $(1-\alpha)\times 100$ percentile of $UB$, $t_{max}$. 
    \State Create a grid $t=L_{n_A}:0.01:t_{max}$. Index $t=\{t_k\}_{k=0}^K$
    \For{k=1 \textbf{to} K }
    \State Calculate $N_{A,s}=\mathrm{N}^i_A\left(D^{mc}_{n_A}(s)\right)$ for all $s\in [n_{mc}]$.
    \State Calculate $N_{A,\obs}=\mathrm{N}^i_A(D^A_\obs)$.
    \State Calculate $L_s=\texttt{B\_Create\_Lower\_Bound\_2}(\mathcal{D},D^{mc}_{n_A}(s),N_{A,s}-t_{i-1})$ for all $s\in[n_{mc}]$.
    \State Solves the optimization problem
    \begin{equation*}
            \max\sum_{s=1}^{n_{mc}} I_s,
    \end{equation*}
      \State subject to
      \[
        Y_a = \sum_{i\in[n]} x_{ai}Y_i, \quad \forall a\in[n_A],
      \]
      \[
        Y_a \ge Y_{a+1}, \quad \forall a\in[r_0+1,\,n_A-1],
      \]
\Statex \Comment{We write $D^{mc}_{n_A}(s)$ as $D^A_s$ below for simplicity.}
\begin{align*}
  & \mathrm{DIM}^2(D^A_s,Y^A)
  +\Bigl(\mathrm{N}^i_{A,s}-t_{i-1}\Bigr)\mathrm{VAR}(D^A_s,Y^A)
  \ \geq\ L_s\,(1-I_s),  \forall s\in [n_{mc}],\\
  &  \mathrm{DIM}^2(D^A_\obs,Y^A)
  +\Bigl(\mathrm{N}^i_{A,\obs}-t_{i}\Bigr)
  \mathrm{VAR}(D^A_\obs,Y^A) \leq\ 0,\\
  & x\in x(k,D^\obs,R), \qquad I_s\in \{0,1\}, \qquad \forall s\in [n_{mc}].
\end{align*}
    \State Store $v(i)=\sum_{s=1}^{n_{mc}}I^*_s$.
    \EndFor
    \State \Return $\max_{i\in [K]}v(i)$.
\EndIf
  \end{algorithmic}
\end{algorithm}

\subsection{Algorithms for Asymptotic Inferences}\label{section:alg_asy}
Recall that $i\in\{0,1,2\}$ denotes the type of statistics that is used to test the balance of the number of always-reporters.
\begin{algorithm}[H]
  \caption{\texttt{A\_master\_asymptotics:} test null hypotheses at a prespecified significance level $\alpha\in (0,1)$.} 
  \label{alg:12152025_c}
  \begin{algorithmic}[1]
\Require Dataset $\mathcal{D}$ as in \eqref{eqn:data}, a prespecified significance level $\alpha\in (0,1)$, and statistic type $i\in \{0,1,2\}$, number of randomization draws $n_{mc}$, pruning significance level $\beta\in [0,\alpha)$.
\Statex \textbf{Step I: Precomputation}
\State Set $Rej \gets 1$.
\State Construct and prune the compatible always-reporter set $\mathbb{A}(D^{\obs},R)$.
\State Set $n_A^{\max}=\sum_{i=1}^n R_i$ and $n_A^{\min}=r_0$.
\If{$\beta>0$} 
\State Set $\alpha\gets\alpha-\beta$.
\State Set $n_A^{\max} \gets \max_{A\in \mathbb{A}(D^{\obs},R)} \sum_{i=1}^n A_i$.
\State Set $n_A^{\min} \gets \min_{A\in \mathbb{A}(D^{\obs},R)} \sum_{i=1}^n A_i$.
\EndIf
\Statex \textbf{Step II: Asymptotic Inferences}
\For{$k = n_A^{\min}$ \textbf{to}  $n_A^{\max}$}
  \State $Rej\gets \texttt{B\_Asymptotic\_Inferences\_Inner}(\mathcal{D},k)$.
  \If{$Rej=0$}
    \State \textbf{break}.
  \EndIf
\EndFor 
    \State \Return $Rej$
    \end{algorithmic}
\end{algorithm}
\begin{algorithm}[H]
  \caption{\texttt{B\_Asymptotic\_Inferences\_Inner:} asymptotic inference given a number of always-reporters} 
  \label{alg:12152025_1}
  \begin{algorithmic}[1]
    \Require Dataset $\mathcal{D}$, number of always-reporters $n_A$, statistic type $i\in \{0,1,2\}$, signifiancne level $\alpha$.
    \State Create assignment vector $D^A_\obs=[\texttt{ones}(n_A,1),\texttt{zeros}(r_0,1) ]$.
     \State  Set  $N_A\gets\mathrm{N}^i_A(D^A_\obs)$.
    \State Set $\textrm{DIM2\_L},\textrm{DIM2\_U}\gets\texttt{C\_dim2\_bounds}(D^A_\obs,\mathcal{D})$.
    \State Set $\textrm{NUM\_L}\gets\texttt{C\_v1\_min}(D^A_\obs,\mathcal{D})+\texttt{C\_v0\_min}(D^A_\obs,\mathcal{D})$.
    \State Set $\textrm{NUM\_U}\gets\texttt{C\_v1\_max}(D^A_\obs,\mathcal{D})+\texttt{C\_v0\_max}(D^A_\obs,\mathcal{D})$.
    \State Set $L\gets \textrm{DIM2\_L}/\textrm{NUM\_U}+N_A$.
    \State Set $U\gets \textrm{DIM2\_U}/\textrm{NUM\_L}+N_A$.   
    \State $L,Y^A\gets \texttt{B\_Create\_Lower\_Bound}(D^A_\obs,\mathcal{D},L,U,tol).$
    \State Calculate $\mathcal{T}^*=\mathcal{T}^\obs\left(D^A_{\obs},Y^A\right)$, where,
    \begin{equation*}
        \mathcal{T}^\obs(D^A_{\obs},Y^A)= \frac{ \mathrm{DIM}^2(D^A_{\obs},Y^A) }{ \mathrm{VAR}(D^A_{\obs},Y^A) } +  \mathrm{N}_A^i(D^A_\obs).
    \end{equation*} 
    \State Calculate the $1-\alpha$ quantile, $q^{1-\alpha}$, of random variables $\textrm{Z}+g_i(n_A^1)$, where $n_A^1=\sum_{i=1}^nD_iA_i$, $\sum_{i=1}^n A_i=n_A$ and $D\sim \textrm{CR}(n,n_1)$.
    \If{$\mathcal{T}^*> q^{1-\alpha}$}
    \State Set $Rej=1$.
    \Else
    \State Set $Rej=0$.
    \EndIf
    \State \Return $Rej$.
    \end{algorithmic}
\end{algorithm}
\subsection{Analytical bounds for the studentized statistic}\label{section:boundstd}

Given the dataset $\mathcal{D}=(Y,D^\obs,R)$, denote the set of potential always-reporters in the observed treated group as $\mathbb{A}^1(D,R)=\{i\in[n]:D^{\obs}_i=1,R_i=1\}$. Denote the cardinality of this set as $n_{A,p}=\left|\mathbb{A}^1(D^\obs,R)\right|$. Recall our convention that first $r_0$ units are always-reporters assigned to the control group.

\subsubsection{Lower Bound and Upper Bound for    $\mathrm{DIM}^2$ of \eqref{eqn:dim2} }\label{section:lbubns}

Given outcome vector $Y=\left(Y_i\right)_{i=1}^n\in \mathbb{R}^n$ and the set of potential always-reporters $\mathbb{A}^1(D^\obs,R)$, we use $\widetilde{Y}_{(k)}$, $k\in [n_{A,p}]$ to denote the $k$th smallest outcome in the set $\{Y_i\}_{i\in \mathbb{A}^1(D,R)}$, with $\widetilde{Y}_{(1)}$ being the smallest. 

For a fixed number of always-reporters, a standard argument gives the following upper bound and lower bound for the squared difference-in-means estimator $\mathrm{DIM}^2(\cdot,\cdot)$ as defined in \eqref{eqn:dim2}. 

\begin{lemma'}
Given a positive integer $n_A$ and the dataset $\mathcal{D}=(Y,D^\obs,R)$, recall the set of assignment variables $ x\left(n_A,D^\obs,R\right)$ from \eqref{eqn:assignment_variables}. Denote the set of all compatible outcome vectors as
\begin{align}
& \mathbb{Y}(n_A,D^\obs,R,Y) \label{eqn:Yna}\\ 
= &\left\{ \left(Y_a\right)_{a\in [n_A]}:  \exists \{x_{ai}\}_{a,i}\in x\left(n_A,D^\obs,R\right), Y_a=\sum_{i=1}^n x_{ai}Y_i, \forall a\in[n_A]\right\}.\nonumber     
\end{align}    
 
Given an assignment vector $D^A=\left(D_a\right)_{a=1}^{n_A}$, we have
\begin{equation*}
  \min_{ Y^A\in \mathbb{Y}(n_A,D^\obs,R,Y)  }\mathrm{DIM}^2(D^A,Y^A)  \geq L(D^A,\mathcal{D})
\end{equation*}
and,
\begin{equation*}
  \max_{Y^A\in \mathbb{Y}(n_A,D^\obs,R,Y) }\mathrm{DIM}^2(D^A,Y^A)   \leq U(D^A,\mathcal{D}).
\end{equation*}
where $L(D^A,\mathcal{D})$ and $U(D^A,\mathcal{D})$ are the lower bound and upper bound returned from  $\texttt{C\_dim2\_bounds}(D^A,\mathcal{D})$ as defined in Algorithm \ref{alg:C_dim2}.
\end{lemma'}
\begin{algorithm}[H]
  \caption{ \texttt{C\_dim2\_bounds}: Calculate a lower bound and an upper bound for the squared difference-in-means estimator }
 \label{alg:C_dim2}
  \begin{algorithmic}
    \Require $D^{A}=(D_a)_{a\in[n_A]}$, Observed dataset $\mathcal{D}=\left(Y,D^\obs,R\right)$
    \State Create $\mathbb{A}^1(D^\obs,R)$ and set $n_{A,p} =
    \left|\mathbb{A}^1(D^\obs,R)\right| $.
    \State Set $n_{A,1}=\sum_{a=1}^{n_A}D_a$, $r_{0,1}=\sum_{a=1}^{r_0}D_a$.
    \State Set $n_{A,0}=\sum_{a=1}^{n_A}\left(1-D_a\right)$, $r_{0,0}=\sum_{a=1}^{r_0}\left(1-D_a\right)$.
    \State Calculate $L$ and $U$ according to 
    {\small
    \begin{align*}
      U =\frac{1}{n_{A,1}}\left(\sum_{a=1}^{r_0} D_aY_a+\sum_{k=1}^{n_{A,1}-r_{0,1}}\widetilde{Y}_{(n_{A,p}-k)} \right)
        -  \frac{1}{n_{A,0}}\left(\sum_{a=1}^{r_0} \left(1-D_a\right)Y_a+ \sum_{k=1}^{n_{A,0}-r_{0,0}}\widetilde{Y}_{(k)}\right),       
    \end{align*}}
and 
{\small
\begin{equation*}
    L =    \frac{1}{n_{A,1}}\left(\sum_{a=1}^{r_0} D_aY_a+\sum_{k=1}^{n_{A,1}-r_{0,1}}\widetilde{Y}_{(k)} \right)-  \frac{1}{n_{A,0}}\left(\sum_{a=1}^{r_0} \left(1-D_a\right)Y_a+ \sum_{k=1}^{n_{A,0}-r_{0,0}}\widetilde{Y}_{(n_{A,p}-k)}\right).
\end{equation*}}
\If{$L\leq U\leq 0$ or $U\geq L\geq 0$}
    \Return $\min\{L^2,U^2\},\max\{L^2,U^2\} $
\ElsIf{$L\leq 0\leq U$}
    \Return $0,\max\{L^2,U^2\} $
\EndIf
  \end{algorithmic}
\end{algorithm}
In words, $ n_{A,1}$ and $n_{A,0}$  are the numbers of always-reporters assigned to the treated and control groups, respectively. $r_{0,1}$ and $r_{0,0}$ are the numbers of known always-reporters that are assigned to the treated and control groups with the assignment vector $\left(D_a\right)_{a=1}^{n_A}$. The upper bound and the lower bound follow from a standard calculation.

\subsubsection{Lower Bound and Upper Bound for $\mathrm{VAR}$ of \eqref{eqn:VAR}}\label{section:lbubds}

For a fixed number of always-reporters and by Lemma \ref{lemma:a2} and Lemma \ref{lemma:a3} below, we have the following upper bound and lower bound for the variance estimator $\mathrm{VAR}(\cdot,\cdot)$ as defined in \eqref{eqn:VAR}. We note that $\mathrm{VAR}$ is a standard two sample variance estimator and it can be written as the sum of the variance estimator for the treated group and the variance estimator for the control group.

\begin{lemma'}\label{lemma:C2_20251215}
Given a positive integer $n_A$ and the dataset $\mathcal{D}=(Y,D^\obs,R)$, recall the set of assignment variables $ x\left(n_A,D^\obs,R\right)$ from \eqref{eqn:assignment_variables}. Recall the set  $\mathbb{Y}(n_A,D^\obs,R,Y)$ of compatible outcome vectors $ x\left(n_A,D^\obs,R\right)$ as defined in \eqref{eqn:Yna}.

Given an assignment vector $D^A=\left(D_a\right)_{a=1}^{n_A}$, we have
\begin{equation*}
  \min_{ Y^A\in \mathbb{Y}(n_A,D^\obs,R,Y)  }\mathrm{VAR}(D^A,Y^A)  \geq v^1_{\min}+v^0_{\min},
\end{equation*}
and,
\begin{equation*}
  \max_{Y^A\in \mathbb{Y}(n_A,D^\obs,R,Y) }\mathrm{VAR}(D^A,Y^A)  \leq v^1_{\max}+v^0_{\max}.
\end{equation*}
where $v^1_{\max}$, $v^0_{\max}$, $v^1_{\min}$ and $v^0_{\min}$ are the outputs of functions $\texttt{C\_v1\_max}(D^A,\mathcal{D})$, $\texttt{C\_v0\_max}(D^A,\mathcal{D})$, $\texttt{C\_v1\_min}(D^A,\mathcal{D})$, and 
$\texttt{C\_v0\_min}(D^A,\mathcal{D})$ respectively.
\end{lemma'}
\begin{proof}
    This is a direct corollary of Lemma \ref{lemma:a2} and Lemma \ref{lemma:a3} proved below.
\end{proof}

\subsubsection{Proof of Lemma \ref{lemma:C2_20251215}}
Let $S=\{s_1,...,s_{k_1}\} \subset \mathbb{R}$ be a set of $k_1$ points and $V=\{v_1,...,v_{k_2}\}\subset   \mathbb{R}$ be a set of $k_2$ points. For each $X=\{x_1,...,x_{|X|}\}$ such that  $S\subset X\subset S\cup V$, we define the variance of $X$ as
\begin{equation}
    \texttt{var}\left(X\right)=\frac{1}{|X|}\sum_{i=1}^{|X|}\left(x_i-\overline{X}\right)^2, \text{ with } \overline{X}=\frac{1}{|X|}\sum_{i=1}^{|X|}x_i.
\end{equation}
We are interested in the maximum and minimum values of $\texttt{var}\left(X\right)$ among the sets with the same cardinality which contain $S$. We call the sets $S$ and $V$ \textit{increasingly-ordered} if $s_1\leq s_2 \leq ...\leq s_{k_1}$ and $v_1\leq v_2 \leq ...\leq v_{k_2}$.

\begin{lemma'}\label{lemma:a2}
Given increasingly-ordered sets $S=\{s_i\}_{i=1}^{k_1}\subset \mathbb{R}$, $V=\{v_i\}_{i=1}^{k_2}\subset \mathbb{R}$, and a positive integer $k_1+t$, with $t\geq 0$, define the set
\begin{equation*}
    \mathbb{X}^t=\{X: S\subset X\subset S\cup V, |X|=k_1+t\}.
\end{equation*}
Denote the set of consecutive subsets of $V$ with size $t$ as
\begin{equation*}
    \mathbb{V}^{t}_c=\left\{ \{v_{i+1},..,v_{i+t}\}: i\in [0,k_2-t],i\in\mathbb{Z}\right\}.
\end{equation*}
Then we have:
\begin{equation*}
    \min_{X\in  \mathbb{X}^t }\texttt{var}\left(X\right)= \min_{X=S\cup V_c, V_c\in  \mathbb{V}_c^{t}}\texttt{var}\left(X\right).
\end{equation*}
\end{lemma'}
\begin{proof}
Let $ \mathbb{V}^t$ be all subsets of $V$ with size $t$.
Let $V_{nc}\in \mathbb{V}^t\backslash \mathbb{V}^{t}_c$  be a non-consecutive set. List its element as $\{\widetilde{v}^1,...,\widetilde{v}^{t}\}$ in the order induced by $V$.

There are two cases.
If 
$\bigl(V\setminus V_{nc}\bigr)\cap [\widetilde v^{1},\,\widetilde v^{\,t}]
\subseteq \{\widetilde v^{1},\,\widetilde v^{\,t}\}
$,
then \(V_{nc}\) is non-consecutive only because of ties and the labeling order of observations. In this case, the variance of \(V_{nc}\) coincides with that of a consecutive subset.

For the other case, we have a $v\in\bigl(V\setminus V_{nc}\bigr)\cap [\widetilde v^{1},\,\widetilde v^{t}]$ such that $v\in (\widetilde{v}^1,\widetilde{v}^{t})$. We show that such a set cannot be a minimizing set, because we can always construct another set that has a strictly smaller variance than that of $V_{nc}$.

To see this, we note by definition $v= \lambda \widetilde{v}^1+ (1-\lambda) \widetilde{v}^{t}$ with $\lambda\in(0,1)$. Define $m=t+k_1$. The variance of the set $\left(S\cup V_{nc}\cup v\right)\backslash \widetilde{v}^1$ can be expressed as:
\begin{align*}
    & \texttt{var}\left( \left(S\cup V_{nc}\cup v\right)\backslash \widetilde{v}^1\right)=  \frac{1}{2m^2}\sum_{a,b\in S\cup V_{nc}\backslash \{\widetilde{v}^1,\widetilde{v}^t,v\}}(a-b)^2\\
    +& \frac{1}{m^2}\left( \sum_{a\in S\cup V_{nc}\backslash \{\widetilde{v}^1, \widetilde{v}^{t}\} }(v-a)^2+  (v-\widetilde{v}^{t})^2+ \sum_{a\in S\cup V_{nc}\backslash \widetilde{v}^1}(\widetilde{v}^{t}-a)^2\right),
\end{align*}
and the variance of the set $\left(S\cup V_{nc}\cup v\right)\backslash \widetilde{v}^t$ can be expressed as:
\begin{align*}
    & \texttt{var}\left(\left(S\cup V_{nc}\cup v\right)\backslash \widetilde{v}^t\right)=  \frac{1}{2m^2}\sum_{a,b\in S\cup V_{nc}\backslash \{\widetilde{v}^1,\widetilde{v}^t,v\}}(a-b)^2\\
    +& \frac{1}{m^2}\left( \sum_{a\in S\cup V_{nc}\backslash \{\widetilde{v}^1, \widetilde{v}^{t}\} }(v-a)^2+  (v-\widetilde{v}^{1})^2+ \sum_{a\in S\cup V_{nc}\backslash \widetilde{v}^t}(\widetilde{v}^{1}-a)^2\right).
\end{align*}
Then, 
\begin{align*}
     & \lambda \texttt{var}\left( \left(S\cup V_{nc}\cup v\right)\backslash \widetilde{v}^1\right) + \left(1-\lambda\right) \texttt{var}\left( \left(S\cup V_{nc}\cup v\right)\backslash \widetilde{v}^t\right)\\
    = & \frac{1}{2m^2}\sum_{a,b\in S\cup V_{nc}\backslash \{\widetilde{v}^1,\widetilde{v}^t,v\}}(a-b)^2+ \frac{1}{m^2}\sum_{a\in S\cup V_{nc}\backslash \{\widetilde{v}^1, \widetilde{v}^{t}\} }(v-a)^2\\
      & + \frac{1}{m^2} \left( \lambda(v-\widetilde{v}^{t})^2+  \left(1-\lambda\right)  (v-\widetilde{v}^1)^2\right)\\
      & + \frac{\lambda}{m^2} \sum_{a\in S\cup V_{nc}\backslash \widetilde{v}^1}(\widetilde{v}^{t}-a)^2  +\frac{1-\lambda}{m^2}\sum_{a\in S\cup V_{nc}\backslash \widetilde{v}^t}(\widetilde{v}^{1}-a)^2 \\
    \leq  &  \frac{1}{2m^2}\sum_{a,b\in S\cup V_{nc}\backslash \{\widetilde{v}^1,\widetilde{v}^t,v\}}(a-b)^2 \\ 
    & + \frac{\lambda}{m^2}\sum_{a\in S\cup V_{nc}\backslash \{\widetilde{v}^1, \widetilde{v}^{t}\} }\left(\widetilde{v}^1-a\right)^2+\frac{1-\lambda}{m^2}\sum_{a\in S\cup V_{nc}\backslash \{\widetilde{v}^1, \widetilde{v}^{t}\} }\left(\widetilde{v}^t-a\right)^2\\
    & + \frac{1}{m^2} \left(\lambda^3+(1-\lambda)^3\right)\left( \widetilde{v}^1-\widetilde{v}^{t}\right)^2\\
    & +  \frac{\lambda}{m^2} \sum_{a\in S\cup V_{nc}\backslash \widetilde{v}^1}(\widetilde{v}^{t}-a)^2  +\frac{1-\lambda}{m^2}\sum_{a\in S\cup V_{nc}\backslash \widetilde{v}^t}(\widetilde{v}^{1}-a)^2 \\
    = &    \frac{1}{2m^2}\sum_{a,b\in S\cup V_{nc}\backslash \{\widetilde{v}^1,\widetilde{v}^t,v\}}(a-b)^2  +\frac{1}{m^2} \left(\lambda^3+(1-\lambda)^3\right)\left( \widetilde{v}^1-\widetilde{v}^{t}\right)^2 \\
     &+  \frac{1}{m^2}\left(\sum_{a\in S\cup V_{nc}\backslash \widetilde{v}^1}(\widetilde{v}^{t}-a)^2  + \sum_{a\in S\cup V_{nc}\backslash \widetilde{v}^t}(\widetilde{v}^{1}-a)^2 \right)\\
     < &    \frac{1}{2m^2}\sum_{a,b\in S\cup V_{nc}\backslash \{\widetilde{v}^1,\widetilde{v}^t,v\}}(a-b)^2  +\frac{1}{m^2} \left( \widetilde{v}^1-\widetilde{v}^{t}\right)^2 \\
     &+  \frac{1}{m^2}\left(\sum_{a\in S\cup V_{nc}\backslash \widetilde{v}^1}(\widetilde{v}^{t}-a)^2  + \sum_{a\in S\cup V_{nc}\backslash \widetilde{v}^t}(\widetilde{v}^{1}-a)^2 \right)\\
     = &\texttt{var}\left(S\cup V_{nc}\right),
\end{align*}
where the first inequality is by convexity and the last inequality is by $\lambda^3+(1-\lambda^3)<1$ for $\lambda \in (0,1)$ because $1=(\lambda+1-\lambda)^3= \lambda^3 + \left(1-\lambda\right)^3 + 3\lambda(1-\lambda)^2 + 3(1-\lambda)^2$. 
Hence either $\texttt{var}\left(S\cup V_{nc}\right)>  \texttt{var}\left( \left(S\cup V_{nc}\cup v\right)\backslash \widetilde{v}^1\right)$ or $\texttt{var}\left(S\cup V_{nc}\right)>  \texttt{var}\left( \left(S\cup V_{nc}\cup v\right)\backslash \widetilde{v}^t\right)$, and $S\cup V_{nc}$ cannot be the optimizing set.

We have shown that for a non-consecutive set, it either realizes a variance equal to that of a consecutive set or cannot be the optimizing set. This proves our claim.

\end{proof}

\begin{lemma'}\label{lemma:a3}
Given increasingly-ordered sets $S=\{s_i\}_{i=1}^{k_1}\subset \mathbb{R}$, $V=\{v_i\}_{i=1}^{k_2}\subset \mathbb{R}$, and a positive integer $k_1+t$, with $0\leq t\leq k_2$, define the set
\begin{equation*}
    \mathbb{X}^t=\{X: S\subset X\subset S\cup V, |X|=k_1+t\}.
\end{equation*}
Denote the set of shell sets of $V$  with size $t$ as
\begin{equation*}
     \mathbb{V}^{t}_s=\{ \{v_1,...,v_s\}\cup \{v_{k_2-(t-s-1)},...,v_{k_2}\}:s\in [0,t]\},
\end{equation*}
where $\{v_1,...,v_s\}=\emptyset$ if $s=0$ and $\{v_{k_2-(t-s-1)},...,v_{k_2}\}=\emptyset$ if $s=t$.  
Then we have:
\begin{equation*}
    \max_{X\in  \mathbb{X}^t }\texttt{var}\left(X\right)= \max_{X=S\cup V_s, V_s\in  \mathbb{V}_s^{t}}\texttt{var}\left(X\right).
\end{equation*}
\end{lemma'}
\begin{proof}

For the case where $t=k_2$, the statement is trivial. We consider the case where $t<k_2$. Let $ \mathbb{V}^t$ be all subsets of $V$ with size $t$.
For each $\widetilde{V}\in \mathbb{V}^t $, we define its left-shell size and right-shell size as:
\begin{equation}
    \textrm{Left-Shell-Size}\left(\widetilde{V}\right)=\max \left\{k: \{v_{1},v_{2},...,v_{k}\}\subset \widetilde{V}\right\},
\end{equation}
\begin{equation}
    \textrm{Right-Shell-Size}\left(\widetilde{V}\right)=\max \left\{k: \{v_{k_2-k+1},...v_{k_2-1},v_{k_2}\}\subset \widetilde{V}\right\}.
\end{equation}
We note that $\widetilde{V}\in \mathbb{V}^t$ is a shell set if and only if 
\begin{equation}
    \textrm{Left-Shell-Size}\left(\widetilde{V}\right)+  \textrm{Right-Shell-Size}\left(\widetilde{V}\right)=t.
\end{equation}
We show that we can always transform a non-shell set to a shell set through consecutive steps and weakly increase the variance.

Suppose the set $V_{ns}$ that is not a shell set. 
Then there must exists a $v_s\in V_{ns}$, and $v_{s_l},v_{s_r}\not\in V_{ns}$ such that $s_l<s<s_r$.

Define
\begin{equation*}
    l=\min\{i\in [k_2]: v_i\leq v, v_i\not \in V_{ns}\},    r=\max\{i\in [k_2]:  v_i\geq v, v_i\not \in V_{ns}\}.
\end{equation*}
Moreover, $v_s=\lambda v_l + (1-\lambda)v_r$ for some $\lambda\in [0,1]$. Let $m=k+t_1$.
The variance of the set $V_{ns}\cup v_l\backslash v_s$ can be expressed as
\begin{align*}
 \texttt{var}\left(V_{ns}\cup v_l\backslash v_s\right) & =\frac{1}{2m^2}\sum_{a,b\in V_{ns}\backslash v_s}(a-b)^2+ \frac{1}{m^2} \sum_{a\in V_{ns}\backslash v_s}(v_l-a)^2,   
\end{align*}
and the variance of the set $V_{ns}\cup v_r\backslash v_s$ can be expressed as
\begin{align*}
 \texttt{var}\left(V_{ns}\cup v_r\backslash v_s\right) & =\frac{1}{2m^2}\sum_{a,b\in V_{ns}\backslash v_s}(a-b)^2+ \frac{1}{m^2} \sum_{a\in V_{ns}\backslash v_s}(v_r-a)^2,   
\end{align*}
By convexity, we have
\begin{equation*}
     \texttt{var}\left(V_{ns}\right) \leq \lambda  \texttt{var}\left(V_{ns}\cup v_l\backslash v_s \right)  + (1-\lambda)\texttt{var}\left(V_{ns}\cup v_r\backslash v_s\right).
\end{equation*}
WLOG, if $ \texttt{var}\left(V_{ns}\right)< \texttt{var}\left(V_{ns}\backslash v \cup v_r\right)$, replacing $v$ with $v_r$ will strictly increase the variance and increase the shell size of the new set. If $ \texttt{var}\left(V_{ns}\right)= \texttt{var}\left(V_{ns} \cup v_r\backslash v_s\right)=\texttt{var}\left(V_{ns} \cup v_l\backslash v_s\right)$, replacing $v_s$ with either $v_r$ and $v_l$ will not decrease the variance and increase the shell size of the new set.

Repeat the same procedure multiple times. Since there are at most $t$ points in the set, the procedure will stop after at most $t$ steps and output a shell set. Since each step we do not decrease the variance, the new shell set has a variance at least as large as the that of the set $V_{ns}$. This proves the claim.

\end{proof}

\subsubsection{Algorithms according to Lemma \ref{lemma:a2} and Lemma \ref{lemma:a3} }

Given the observed dataset $\mathcal{D}=\left(Y,D^\obs,R\right)$, we denote the set of potential always-reporters in the observed treated group as $\mathbb{A}^1(D,R)=\{i\in[n]:D^{\obs}_i=1,R_i=1\}$ and denote $n_{A,p}=\left|\mathbb{A}^1(D,R)\right|$.

Recall the notation that we use $\widetilde{Y}_{(k)}$, $k\in [n_{A,p}]$ to denote the $k$th smallest outcome in the set $\{Y_i\}_{i\in \mathbb{A}^1(D,R)}$, with $\widetilde{Y}_{(1)}$ being the smallest. We note that $D_a=D_i$ for $a,i\leq r_0$ by our by indexing convention.
\begin{algorithm}[H]
  \caption{ \texttt{C\_v1\_min}: Calculate a lower bound for the treated group variance estimator }
 \label{alg:c_v1_min}
  \begin{algorithmic}
    \Require $D^{A}=(D_a)_{a\in[n_A]}$, Observed dataset $\mathcal{D}=\left(Y,D^\obs,R\right)$
        \State Create $\mathbb{A}^1(D^\obs,R)$ and set $n_{A,p} = \left|\mathbb{A}^1(D^\obs,R)\right| $
    \State Set $n_{A,1}=\sum_{a=1}^{n_A}D_a$, $r_{0,1}=\sum_{a=1}^{r_0}D_a$, $n_{A,p} = 
    \left|\mathbb{A}^1(D^\obs,R)\right| $.
    \State $v_{\min}^1\gets \infty$
    \For{$k=0:\left(n_{A,p}-n_{A,1}+r_{0,1}\right)$ }
      \State Create set $X_{k}=\{Y_i\}_{i\leq r_0,D_i=1}\bigcup \{\widetilde{Y}_{(s)}\}_{s=k+1}^{s=k+\left(n_{A,1}-r_{0,1}\right)}$
      \State Calculate $v_k=\texttt{var}(X_k)$.
      \If{$v_k \leq v_{\min}^1$}
        \State $v_{\min}^1\gets v_k$
      \EndIf
    \EndFor
    \Return $v_{\min}^1$
  \end{algorithmic}
\end{algorithm}
\begin{algorithm}[H]
  \caption{ \texttt{C\_v0\_min}: Calculate a lower bound for the control group variance estimator }
\label{alg:c_v0_min} 
\begin{algorithmic}
    \Require $D^{A}=(D_a)_{a\in[n_A]}$, Observed dataset $\mathcal{D}=\left(Y,D^\obs,R\right)$
        \State Create $\mathbb{A}^1(D^\obs,R)$ and set $n_{A,p} = \left|\mathbb{A}^1(D^\obs,R)\right| $
    \State Set $n_{A,0}=\sum_{a=1}^{n_A}\left(1-D_a\right)$, $r_{0,1}=\sum_{a=1}^{r_0}\left(1-D_a\right)$, $n_{A,p} = 
    \left|\mathbb{A}^1(D^\obs,R)\right| $.
    \State $v_{\min}^0\gets \infty$
    \For{$k=0:\left(n_{A,p}-n_{A,0}+r_{0,0}\right)$ }
      \State Create set $X_{k}=\{Y_i\}_{i\leq r_0,D_i=0}\bigcup \{\widetilde{Y}_{(s)}\}_{s=k+1}^{s=k+\left(n_{A,0}-r_{0,0}\right)}$
      \State Calculate $v_k=\texttt{var}(X_k)$.
      \If{$v_k \leq v_{\min}^0$}
        \State $v_{\min}^0\gets v_k$
      \EndIf
    \EndFor
    \Return $v_{\min}^0$
  \end{algorithmic}
\end{algorithm}

\begin{algorithm}[H]
  \caption{ \texttt{C\_v1\_max}: Calculate an upper bound for the treated group variance estimator }
 \label{alg:c_v1_max}
  \begin{algorithmic}
    \Require $D^{A}=(D_a)_{a\in[n_A]}$, Observed dataset $\mathcal{D}=\left(Y,D^\obs,R\right)$
        \State Create $\mathbb{A}^1(D^\obs,R)$ and set $n_{A,p} = 
    \left|\mathbb{A}^1(D^\obs,R)\right| $
    \State $n_{A,1}=\sum_{a=1}^{n_A}D_a$, $r_{0,1}=\sum_{a=1}^{r_0}D_a$, $n_{A,p} = 
    \left|\mathbb{A}^1(D^\obs,R)\right| $
    \State $v^1_{\max}\gets -\infty$
    \For{$k=0:\left(n_{A,1}-r_{0,1}\right)$ }
      \State Create set $X_{k}=\{Y_i\}_{i\leq r_0,D_i=1}\bigcup \{\widetilde{Y}_{(s)}\}_{s=1}^{s=k}\bigcup \{\widetilde{Y}_{(s)}\}_{s=n_{A,p}-n_{A,1}+r_{0,1}+k+1}^{s=n_{A,p}}$
      \State Calculate Calculate $v_k=\texttt{var}(X_k)$.
      \If{$v_k \geq v^1_{\max}$}
        \State $v^1_{\max}\gets v_k$
      \EndIf
    \EndFor
    \Return $v^1_{\max}$
  \end{algorithmic}
\end{algorithm}
We note that $\{\widetilde{Y}_{(s)}\}_{s=1}^{s=k}=\emptyset$ if $k=0$ and $\{\widetilde{Y}_{(s)}\}_{s=n_
      {A,p}-n_{A,1}+r_{0,1}+k+1}^{s=n_{A,p}}=\emptyset$ if $k=n_{A,1}-r_{0,1}$. 
\begin{algorithm}[H]
  \caption{ \texttt{C\_v0\_max}: Calculate an upper bound for the control group variance estimator }
 \label{alg:c_v0_max}
  \begin{algorithmic}
    \Require $D^{A}=(D_a)_{a\in[n_A]}$, Observed dataset $\mathcal{D}=\left(Y,D^\obs,R\right)$
        \State Create $\mathbb{A}^1(D^\obs,R)$ and set $n_{A,p} = 
    \left|\mathbb{A}^1(D^\obs,R)\right| $
    \State Set $n_{A,0}=\sum_{a=1}^{n_A}\left(1-D_a\right)$, $r_{0,0}=\sum_{a=1}^{r_0}\left(1-D_a\right)$.
    \State $v^0_{\max}\gets -\infty$
    \For{$k=0:\left(n_{A,0}-r_{0,0}\right)$ }
      \State Create set $X_{k}=\{Y_i\}_{i\leq r_0,D_i=0}\bigcup \{\widetilde{Y}_{(s)}\}_{s=1}^{s=k}\bigcup \{\widetilde{Y}_{(s)}\}_{s=n_{A,p}-n_{A,0}+r_{0,0}+k+1}^{s=n_{A,p}}$
      \State Calculate Calculate $v_k=\texttt{var}(X_k)$.
      \If{$v_k \geq v^0_{\max}$}
        \State $v^0_{\max}\gets v_k$
      \EndIf
    \EndFor
    \Return $v^0_{\max}$
  \end{algorithmic}
\end{algorithm}

\bibliography{ref}
\bibliographystyle{abbrv}

\end{document}